\pdfoutput=1
\documentclass[a4paper,11pt,eqsecnum,twoside,titlepage,portrait]{book}
\usepackage[isolatin]{inputenc}
\usepackage[T1]{fontenc}
\usepackage[english, francais]{babel}
\usepackage[left=2.5cm,right=2.5cm,top=2.5cm,bottom=3.5cm]{geometry}  
\usepackage{graphicx}
\usepackage{epstopdf}
\usepackage{amsmath,amsfonts,amssymb,amscd,mathbbol,stmaryrd,wasysym,manfnt}

\numberwithin{equation}{section} 
\usepackage{chappg} 
\usepackage{tocloft}  

\usepackage{mathpazo}   
\usepackage{fancyhdr} 
\usepackage{xcolor} 
\usepackage{makeidx} 

\renewcommand{\thesubsubsection}{\thesubsection.\alph{subsubsection}}
\let\mysubsubsection\subsubsection
\renewcommand\subsubsection[1]{\stepcounter{subsubsection}\mysubsubsection{\thesubsubsection\ -\   #1}}
\let\myparagraph\paragraph
\renewcommand\paragraph[1]{\myparagraph{\small{\textsl{#1}}}}

\renewcommand{\imath}{\mathrm{i}}
\newcommand{\emath}{\mathrm{e}}
\newcommand{\tr}{\mathrm{tr}}
\renewcommand{\Im}{\mathrm{Im}}
\renewcommand{\Re}{\mathrm{Re}}

\newcommand{\TRUE}{\textsf{TRUE}}
\newcommand{\FALSE}{\textsf{FALSE}}

\fancypagestyle{plain}{
\fancyhf{}
\fancyhead[LE,RO]{\thepage}
\fancyhead[RE]{\leftmark}
\fancyhead[LO]{\rightmark}
\fancyhead[C]{}
\fancyfoot[LE]{François David, 2012}
\fancyfoot[RO]{Introduction to quantum formalism[s]}
\fancyfoot[RE]{{\small \textsl{Lecture notes\ --\ \today}}}
\fancyfoot[LO]{IPhT 2012}
\fancyfoot[C]{}

}
\fancypagestyle{partie}{
\fancyhf{}
\fancyhead[LE,RO]{\thepage}
\fancyhead[RE]{\leftmark}
\fancyhead[LO]{\rightmark}
\fancyhead[C]{}
\fancyfoot[LE]{F. David, 2012}
\fancyfoot[LO]{Introduction to quantum formalism[s]}
\fancyfoot[RO]{Lecture notes\ --\ \today}
\fancyfoot[RE]{IPhT 2012}

}

\makeindex

\begin{document}
\selectlanguage{english}

\frontmatter
\begin{titlepage}
\begin{center}
{\Huge{\ \\ \ \\ \ \\ A short introduction to the quantum formalism[s]}}
\end{center}
\vskip 1.cm
\begin{center}
{\Large{François David}}
\end{center}
\begin{center}
{\large Institut de Physique Th\'eorique\\
CNRS, URA 2306, F-91191 Gif-sur-Yvette, France\\
CEA, IPhT, F-91191 Gif-sur-Yvette, France
}
\end{center}
\begin{center}
{francois.david@cea.fr}
\end{center}
\date{}

\begin{center}
\end{center}

These notes are an elaboration on: (i) a short course that I gave at the IPhT-Saclay in May-June 2012; (ii) 
 a previous letter 

\cite{PhysRevLett.107.180401} on reversibility in quantum mechanics.

They present an introductory, but hopefully coherent, view of the main formalizations of quantum mechanics, of their interrelations and of their common physical  underpinnings: 
causality, reversibility and locality/separability.
The approaches covered are mainly: (ii)  the canonical formalism; (ii)  the algebraic formalism; (iii)  the quantum logic formulation. 
Other subjects: quantum information approaches, quantum correlations, contextuality and non-locality issues, quantum measurements, interpretations and alternate theories, quantum gravity, are only very briefly and superficially discussed.

Most of the material is not new, but is presented in an original, homogeneous and hopefully not technical or abstract  way. I try to define simply all the mathematical concepts used and to justify them physically. 
These notes should be accessible to young physicists (graduate level) with a good knowledge of the standard formalism of quantum mechanics, and some interest for theoretical physics (and mathematics).

These notes do not cover the historical and philosophical aspects of quantum physics.
\vfill

\noindent{
\hfill Preprint IPhT t12/042
}

\end{titlepage}
%

\tableofcontents

\mainmatter

\chapter{Introduction}

\section{Motivation}

Quantum mechanics in its modern form is now more than 80 years old.
It is probably the most successful physical theory that was ever proposed. 
It started as an attempt to understand the structure of the atom and the interactions of matter and light at the atomic scale, and it became quickly the general physical framework, valid from  the presently accessible high energy scales (10 TeV$\simeq 10^{-19} $m) -- and possibly from the Planck scale ($10^{-35}$ m) -- up to  macroscopic scales (from $\ell\sim 1$ nm up to  $\ell\sim 10^5$ m depending of physical systems and experiments). Beyond these scales,  classical mechanics takes over as an effective theory, valid when quantum interferences and non-local correlations effects can be neglected.

Quantum mechanics has fully revolutionized physics (as a whole, from particle and nuclear physics to atomic an molecular physics, optics, condensed matter physics and material science), chemistry (again as a whole), astrophysics, etc. with a big impact on mathematics and of course a huge impact on modern technology, the whole communication technology, computers, energy, weaponry (unfortunately) etc.
In all these domains, and despites the huge experimental and technical progresses of the last decades, quantum mechanics has never been seriously challenged by experiments, and its mathematical foundations are very solid.

Quantum information  has become a important and very active field (both theoretically and experimentally) in the last decades. It has enriched our points of view on the quantum theory, and on its applications (quantum computing). 
Quantum information, together with the experimental tests of quantum mechanics, the theoretical advances in quantum gravity and cosmology, the slow diffusion of the concepts from quantum theory in the general public, etc. have led to a revival of the discussions about the principles of quantum mechanics and its seemingly paradoxical aspects.

Thus one sometimes gets the feeling that quantum mechanics is both: (i) the unchallenged and dominant paradigm of modern physical sciences and technologies, (ii) still (often presented as) mysterious and poorly understood, and waiting for some revolution.

\medskip

These lecture notes present a brief and  introductory (but hopefully coherent) view of the main formalizations of quantum mechanics (and of its version compatible with special relativity, quantum field theory), of their interrelations and of their theoretical foundation. 

The ``standard'' formulation of quantum mechanics (involving the Hilbert space of pure states, self-adjoint operators as physical observables, and the probabilistic interpretation given by the Born rule), and the path integral and functional integral representations of probabilities amplitudes are the standard tools used in most applications of quantum theory in physics and chemistry.
It is important to be aware that there are other formulations of quantum mechanics, i.e. other representations (in the mathematical sense) of quantum mechanics, which allow a better comprehension and justification of the quantum theory. This course will focus on two of them, algebraic QM and the so called ``quantum logic'' approach, that I find the most interesting and that I think I managed to understand (somehow...). 
I shall insist on the algebraic aspects of the quantum formalism.

In my opinion discussing and comparing the various formulations is useful in order to get a  better understanding of the coherence and the strength of the quantum formalism. 
This is important when discussing  which features of quantum mechanics are  basic principles and which ones are just natural consequences of the former. Indeed this depends on the different formulations. 
For instance the Born rule or the projection postulate are postulates in the standard formulation, while in some other formulations they are mere consequences of the postulates.
This is also important for understanding the relation between quantum physics and special relativity through their common roots, causality, locality and reversibility.

Discussing the different formulations is useful to discuss these issues, in particular when considering the relations between quantum theory, information theory and quantum gravity.

\medskip
These notes started from: (i) a spin-off of more standard lecture notes for a master course in quantum field theory and its applications to statistical physics, (ii) a growing interest
\footnote{A standard syndrome for the physicist over 50... encouraged (for useful purpose) by the European Research Council}
 in understanding what was going on in the fields of quantum information, of quantum measurements and of the foundational studies of the quantum formalism, (iii) a course that I was kindly asked to give at the Institut de Physique Théorique (my lab) and at the graduate school of physics of the Paris Area (ED107) in May-June 2012, (iv)  a short Letter \cite{PhysRevLett.107.180401} about reversibility in quantum mechanics that I published last year. These notes can be considered partly as a very extended version of this letter.

\section{Organization}

After this introductory section, the second section is a reminder of the basic concepts of classical physics, of probabilities  and of the standard (canonical) and path integral formulations of quantum physics.
I tried to introduce in a  consistent way the important classical concepts of states, observables and probabilities, which are of course crucial in the formulations of quantum mechanics.  
I discuss in particular the concept of quantum probabilities and the issue of reversibility in quantum mechanics in the last subsection.

The third section is devoted to a presentation and a discussion of the algebraic formulation of quantum mechanics and of quantum field theory, based on operator algebras. Several aspects of the discussion are original.
Firstly I justify the appearance of abstract C$^*$-algebras of observables using arguments based on causality and  reversibility. In particular the existence of a $^*$-involution (corresponding to conjugation) is argued to follow from the assumption of reversibility for the quantum probabilities.
Secondly, the formulation is based on real  algebras, not complex ones as usually done, and I explain why this is more natural. I give the mathematical references which justify that the GNS theorem, which ensures that complex abstract C$^*$-algebras are always representable as algebras of operators on a Hilbert space, is also valid for real algebras. 
The standard physical arguments for the use of complex algebras are only given after the general construction.
The rest of the presentation is shorter and quite standard.

The fourth section is devoted to one of the formulations of the so-called quantum logic formalism. This formalism is much less  popular outside the community interested in the foundational basis of quantum mechanics, and in mathematics, but deserves to be better known.
Indeed, it provides a  convincing justification of the algebraic structure of quantum mechanics, which for an important part is still postulated in the algebraic formalism. Again, if the global content is not original, I try to present the quantum logic formalism in a similar light than the algebraic formalism, pointing out which aspects are linked to causality, which ones to reversibility, and which ones to locality and separability. This way to present the quantum logic formalism is  original, I think.
Finally, I discuss in much more details than is usually done Gleason's theorem, a very important theorem of Hilbert space geometry and operator algebras, which justify the Born rule and is also very important when discussing hidden variable theories.

The final section contains short, introductory and more standard discussions of some other questions about the quantum formalism.
I present  some recent approaches based on quantum information. I discuss  some features of quantum correlations: entanglement, entropic inequalities, the Tisrelson bound for bipartite systems.
The problems with hidden variables, contextuality, non-locality, are  reviewed. Some very basic features of quantum measurements are recalled. Then I stress the difference between 
\begin{itemize}
  \item the various formalizations (representations) of quantum mechanics;
  \item the various possible interpretations of this formalism;
\end{itemize}
I finish this section with a few very standard remarks on the problem of quantum gravity.

\section{What this course is not!}
These notes are (tentatively) aimed at a non specialized audience: graduate students and more advanced researchers. The mathematical formalism is the main subject of the course, but it will be presented and discussed at a not too abstract, rigorous or advanced level.
Therefore these notes \textbf{do not intend to be}:
\begin{itemize}
  \item a real course of mathematics  or of mathematical physics;
  \item a real physics course on high energy quantum physics,  on atomic physics and quantum optics, of quantum condensed matter, discussing the physics of specific systems and their applications;
 \item a course on what is \emph{not} quantum mechanics;
  \item a course on the history of quantum physics;
  \item a course on the present sociology of quantum physics;
\item a course on the  philosophical and epistemological aspects of quantum physics.
\end{itemize} 
But I hope that it could be useful as an introduction to these topics. Please keep in mind that this is not a course made by a specialist, it is rather a course made by an amateur, for amateurs!

\section{Acknowledgements}

I thank Roger Balian, Michel Bauer, Marie-Claude David, Kirone Mallick and Vincent Pasquier for their interest and their advices.

\chapter{Reminders}

I first start by  reminders of classical mechanics, probabilities and standard quantum mechanics. 
This is mostly very standard material, taken from notes of my graduate courses (Master level) in Quantum Field Theory.
The sections on classical and quantum probabilities are a bit more original.

\section{Classical mechanics}
\index{Classical mechanics}
The standard books on classical mechanics are the books by Landau \& Lifshitz \cite{LandauMechanics76} and the book by A. Arnold \cite{ArnoldMechanics89}.

\subsection{ Lagrangian formulation}
\index{Lagrangian}
Consider the simplest system: a non relativistic particle of mass $m$ in a one dimensional space (a line). 
Its coordinate (position) is denoted $q$. It is submitted to a conservative force which derives from a potential $V(q)$. 
The potential is independent of time.
The velocity is $\dot{q}(t)={dq\over dt}$.
The dynamics of the particle is given by Newton's equation
\index{Newton's equation}
\begin{equation}
\label{neweq}
m\,\ddot{q}(t)=-{\partial\over\partial q}\, V(q)
\end{equation}
The equation of motion derives from the least action principle. The classical trajectories extremize the action $S$
\index{Action}
\begin{equation}
\label{actlagr}
S[q]=\int_{t_i}^{t_f} dt\ L(q(t),\dot{q}(t))
\qquad,\qquad
L(q,\dot{q})={m\over 2}\dot{q}^2-V(q)
\end{equation}
$L$ is the lagrangian. 
\index{Lagrangian}
Under the variations  the initial and final positions are fixed $q(t_i)=q_i$, $q(t_f)=q_f$. So one requires that a classical solution $q_c(t)$ satisfies
\begin{equation}
\label{varact0}
q(t)=q_c(t)+\delta q(t)\ ,\quad \delta q(t_i)=\delta q(t_f)=0\quad\implies\quad
S[q]=S[q_c]+\mathcal{O}(\delta q^2)
\end{equation}
This functional derivative equation leads to the Euler-Lagrange equation \index{Euler-Lagrange equation}
\begin{equation}
\label{vareq1}
{\delta S[q]\over\delta q(t)}=0
\qquad\iff\qquad
{d\over dt}\,{\partial L(q,\dot{q})\over\partial \dot{q}(t)}={\partial L(q,\dot{q})\over\partial {q}(t)}
\end{equation}
which leads to \ref{neweq} 
\begin{figure}[h]
\begin{center}
\includegraphics[width=3in]{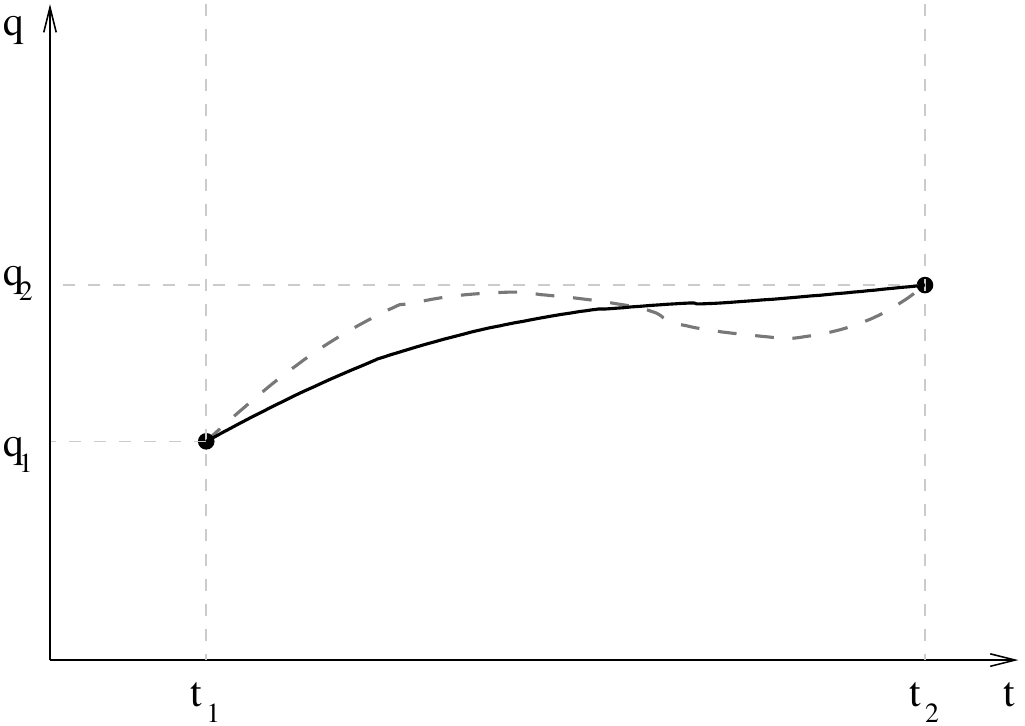}
\caption{The least action principle: the classical trajectoire (full line)  extremizes the action. Under a variation (dashed line) $\delta S=0$. The initial and final positions are kept fixed.}
\label{ }
\end{center}
\end{figure}
This generalizes to many systems: higher dimensional space, many particles, systems with internal degrees of freedom, time-dependent potentials, fields, as long as there is no irreversibility (dissipation).

A good understanding of the origin of the least action principle in classical mechanics comes in fact from the path integral formulation of quantum mechanics.

\subsection{Hamiltonian formulation}
\subsubsection{Phase space and Hamiltonian:} 
\index{Phase space}
\index{Hamiltonian}
The Hamiltonian formulation is in general equivalent, but slightly more general than the Lagrangian formulation.
For a classical system with $n$ degrees of freedom, a state of a system is a point $\boldsymbol{x}$ in the phase space 
$\boldsymbol{\Omega}$ of the system. $\boldsymbol{\Omega}$ is a manifold with even dimension $2n$. 
The evolution equations are flow equations (first order differential equations in time) in the pause space.

For the particle in dimension $d=1$ in a potential there is one degree of freedom, $n=1$ and dim($\Omega$)=2.
The two coordinates  in phase space are the  position $q$ et the momentum $p$.

\begin{equation}
\label{ }
\mathbf{x}=(q,p)
\end{equation}
The Hamiltonian is
\begin{equation}
\label{hamilt1}
H(q,p)={p^2\over 2m}+V(q)
\end{equation}
The equations of motion are the  Hamilton equations
\index{Hamilton equations}
\begin{equation}
\label{hamilt2}
\dot{p}=-{\partial H\over\partial q}
\qquad,\qquad
\dot{q}={\partial H\over\partial p}
\end{equation}
so the relation between the momentum and the velocity $p=m\dot q$ is a dynamical relation.
The Hamilton equations derive also from a variational principle. To find the classical trajectory such that 
$q(t_1)=q_1$, $q(t_2)=q_2$
one extremizes the action functional 
  $\mathcal{S}_H$
\begin{equation}
\label{hamilt3b}
\mathcal{S}_H [q,p]=\int_{t_1}^{t_2} dt\,\left[{p(t)\dot{q}(t)-H(q(t),p(t))}\right]
\end{equation}
with respect to variations of 
$q(t)$ and of  $p(t)$, $q(t)$ being fixed at the initial and final times $t=t_1$ et $t_2$, but $p(t)$ being free at  $t=t_1$ and $t_2$.
Indeed, the functional derivatives of $\mathcal{S}_H$ are
\begin{equation}
\label{hamilt3}
{\delta \mathcal{S}_H \over\delta q(t)}=-\dot p(t)-{\partial H\over\partial q}(q(t),p(t))
\ ,\quad
{\delta \mathcal{S}_H \over\delta p(t)}=\dot q(t)-{\partial H\over\partial p}(q(t),p(t))
\end{equation}
The change of variables $(q,\dot q)\to(q,p)$ and of action  fonctionals  $S(q,\dot q)\to \mathcal{S}_H (q,p)$ between the Lagrangian and the Hamiltonian formalism corresponds to a  Legendre transform.
\begin{figure}[h]
\begin{center}
\includegraphics[width=3in]{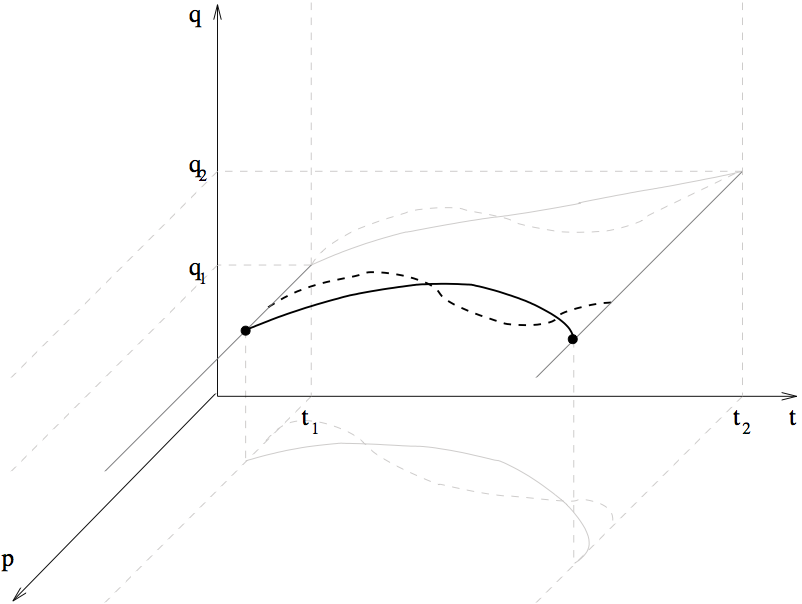}
\caption{Least action principle in phase space: The classical trajectory (full line) extremizes the action $S_H[q,p]$.
The initial and final positions are fixed. The initial and final momenta are free. Their actual value is given by the variational principle as a function of the initial and final positions and times.}
\label{ }
\end{center}
\end{figure}

\subsubsection{Hamilton-Jacobi equation} 

For a classical trajectory $q_\mathrm{cl}(t)$ solution of the equations of motion, the action functionals $\mathcal{S}_H$ and $S$ are equal!
If we fix the initial time $t_1$ and the initial position $q_1$, this classical action can be considered as a function of the final time $t_2={t}$ and of the final position $q(t_2)=q_2={q}$. This function is called the Hamilton-Jacobi action, or the Hamilton function, and I note it $\mathcal{S}(q,t)=\mathcal{S}_{\scriptscriptstyle{\mathrm{HJ}}}({q},{t})$ to be explicit (the initial conditions $q(t_1)=q_1$ being implicit)
\index{Hamilton-Jacobi action}
\begin{align}
\label{ }
\mathcal{S}(q,t)=\mathcal{S}_{\scriptscriptstyle{\mathrm{HJ}}}({q},{t})=
\mathcal{S}[q_{\mathrm{cl}}] \quad&\text{with}\ q_{\mathrm{cl}}\ \text{classical solution such that}\quad {q(t_2)={q}, t_2={t}}\ 
\nonumber\\
&\text{and where}\ t_1\ \text{and}\ {q(t_1)=q_1}\ \text{are kept fixed}
\end{align}

Using the equations of motion it is easy to see that the evolution with the final time $t$ of this function $\mathcal{S(}q,t)$ is given by the differential equation
\index{Hamilton-Jacobi Equation}
\begin{equation}
\label{ }
{\partial\mathcal{S}\over\partial t}=-H\left(q,{\partial\mathcal{S}\over\partial q}\right)
\end{equation}
with $H$ the Hamiltonian function.
This is is a first order differential equation with respect to the final time $t$. It is called the Hamilton-Jacobi equation.

From this equation on can show that (the initial conditions $(t_1,q_1)$ being fixed) the impulsion $p$ and the total energy $E$ of the particle at the final time $t$, expressed as a function of its final position $q$ and of $t$, are 
\begin{equation}
\label{ }
E(q,t)=-{\partial\mathcal{S}\over\partial t}(q,t)\ ,\quad p(q,t)={\partial\mathcal{S}\over\partial q}(q,t)
\end{equation}
These formulas extends to the case of systems with $n$ degrees of freedom and of mote gerenal Hamiltonians.
Positions and momenta are now $n$ components vectors
\begin{equation}
\label{ }
\mathbf{q}=\{q^i\}\quad,\qquad \mathbf{p}=\{p_i\}\qquad i=1,\cdots,n
\end{equation}

\subsubsection{Symplectic manifolds} 

\index{Symplectic manifold} 
\index{Phase space}

A more general situation is a general system whose phase space $\mathbf{\Omega}$ is a manifold with an even dimension $N=2n$,  not necessarily the Euclidean space $\mathbb{R}^{N}$, but with for instance a non trivial topology.
Locally $\mathbf{\Omega}$ is described by local coordinates $\mathbf{x}=\{x^i,i=1,2n\}$ (warning! The $x-i$ are coordinate in phase space, not some position coordinates in physical space).

Hamiltonian dynamics requires a symplectic structure on $\Omega$. This symplectic structure allows to define (or amounts to define) the Poisson brackets. $\Omega$ is a symplectic manifold if it is embodied with an antisymmetric 2-form  $\omega$ (a degree 2 differential form) which is non-degenerate and closed ($d\omega=0$).
This means that to each point  $\mathbf{x}\in\Omega$ is associated (in the coordinate system $\{xî\}$
$$\omega(\mathbf{x})={1\over 2}\omega_{ij}(\mathbf{x}) dx^i \wedge dx^j$$
caracterized by an antisymmetric matrix  $2n\times 2n$ which is invertible
$$w_{ij}(\mathbf{x})=-w_{ji}(\mathbf{x})\quad,\qquad \det(\omega)\neq 0$$
$dx^i\wedge dx^j$ is the antisymmetric product (exterior product) of the two 1-forms $dx^i$ and $dx^j$.
This  form is closed. Its exterior derivative $d\omega$ is zero
$$ d\omega(\mathbf{x})={1\over 3!}\sum_{i,j,k} \partial_i \omega_{jk}(\mathbf{x})\, dx^i\wedge dx^j\wedge dx^k = 0$$
In term of components this means
 $$\forall\  i_1<i_2<i_3\quad,\qquad \partial_{i_1}\omega_{i_2 i_3}+\partial_{i_2}\omega_{i_3 i_1}+\partial_{i_3}\omega_{i_1 i_2}=0$$
The fact that $\omega$ is a differential form means that under a local change of coordinates $\mathbf{x}\to\mathbf{x'}$ (in phase space) the components of the form change as
 $$\mathbf{x}\to\mathbf{x'}\quad,\qquad  \omega=\omega(\mathbf{x})_{ij}\,dx^i\wedge dx^j=\omega'(\mathbf{x}')_{ij}\,d{x'}^i\wedge d{x'}^j$$
that is for the components
 $$  \omega'(\mathbf{x}')_{ij} =  \omega(\mathbf{x})_{kl}\ {\partial  x^k\over \partial {x'}^i} {\partial  x^l\over \partial {x'}^j} $$
The Poisson brackets will be defined in the next subsection.%

For the particule on a ligne $n=1$, $\mathbf{\Omega}=\mathbb{R}^2$, $\mathbf{x}=(q,p)$,
The symplectic form is simply $\omega= dq\wedge dp$. Its components are
\begin{equation}
\label{omegan1}
\omega=(\omega_{ij})=\begin{pmatrix}
    0  &  1  \\
    -1  &  0
\end{pmatrix}
\end{equation}
In $d=n$ dimensions $\mathbf{\Omega}=\mathbb{R}^{2n}$, $\mathbf{x}=(q^i,p^i)$, and $\omega={1\over 2}\sum\limits_i dq^i\wedge dp^i$, i.e.
\begin{equation}
\label{omegaDarb}
( \omega_{ij})=\begin{pmatrix}
   0   &   1 & 0 & 0 & \cdots \\
   -1   &  0 & 0 & 0 & \cdots \\
   0 & 0 & 0 & 1 & \cdots \\
   0 & 0 & -1 & 0 & \cdots \\
   \vdots & \vdots & \vdots & \vdots & \ddots
\end{pmatrix}
\end{equation}

The Darboux theorem 
\index{Darboux theorem}
\index{Darboux coordinates}
shows that for any symplectic manifold $\Omega$ with a symplectic form $\omega$, it is always possible to find local coordinate systems (in the neighborhood of any point) such that the symplectic form takes the form \ref{omegaDarb} ($\omega$ is constant and is a direct sum of antisymmetric symbols). The $(q^i,p^j)$ are local pairs of conjugate variables.

The fact that locally the symplectic form may be written under its generic  constant form means that symplectic geometry is characterized only by global invariants, not by local ones. This is different from Riemaniann geometry, where the metric tensor $g_{ij}$ cannot in general be written in its flat form $h_{ij}=\delta_{ij}$, because of curvature, and where there are local invariants.

 \subsubsection{Observables, Poisson brackets}

\index{Observable}
The observables of the system defined by a symplectic phase espace $\Omega$ may be identified with the (``sufficiently regular'') real functions on $\Omega$. The value of an observable $f$ for the system in the state $x$ is simply $f(\mathbf{x})$.
Of course observables may depend explicitly on the time $t$ in addition on $x$.
\begin{equation}
\label{ }
\text{system in state}\ x\quad\to\quad \text{measured value of}\ f\ =\ f(x)
\end{equation}
For two differentiable functions (observables) $f$ and $g$, the Poisson bracket $\{f,g\}_\omega$ is the function (observable) defined by
\index{Poisson bracket}
\begin{equation}
\label{hamilt4}
{\{f,g\}}_\omega(x)=\omega^{ij}(x)\ \partial_i f(x)\,\partial_j g(x)
\quad\text{with}\quad \partial_i={\partial\over\partial x^i}\quad\text{and}\quad w^{ij}(x)=\left(w^{-1}(x)\right)_{ij}
\end{equation}
the matrix elements of the inverse of the antisymmetric matrix $\omega(x)$.
When no ambiguity are present, I shall omit the subscript $\omega$.
In a canonical local coordinate system (Darboux coordinates) the Poisson bracket is
\begin{equation}
\{f,g\}=\sum_i\ {\partial f \over \partial q^i}{\partial g \over \partial p^i}-{\partial f \over \partial p^i }{\partial g \over \partial q^i}
\qquad\hbox{and}\qquad
\{q^i,p^j\}=\delta_{ij}
\end{equation}
The Poisson bracket is  antisymmetric
\begin{equation}
\label{antisymBr}
\{f,g\}=-\{g,f\}
\end{equation}
The fact that it involves first order derivatives only implies the Leibnitz rule (the Poisson bracket acts as a derivation)
\begin{equation}
\label{LeibnitzBra}
\{f,gh\}\ =\ \{f,g\}h+g \{f,h\}
\end{equation}
The fact that the symplectic form is closed $d\omega=0$ is equivalent to the Jacobi identity
\index{Jacobi identity}
\begin{equation}
\label{JacobiBra}
\{f,\{g,h\}\}+\{g,\{h,f\}\}+\{h,\{f,g\}\}=0
\end{equation}
Knowing the Poisson bracket $\{\ ,\}$ is equivalent to know the symplectic form $\omega$ since
\begin{equation}
\label{ }
\{x^i,x^j\}=\omega^{ij}(\mathbf{x})
\end{equation}

\subsubsection{Dynamics, Hamiltonian flows:}
\index{Hamiltonian flow}
In Hamiltonian mechanics, the dynamics of the system is generated by an Hamiltonian function $H$. The Hamiltonian is a real regular (in general differentiable) function on the phase space $\Omega\to\mathbb{R}$.
The state of the system $\mathbf{x}(t)$ changes with time and the evolution equation for the coordinates $x^i(t)$ in phase space  (the Hamilton équation) take the general form (for a time in dependent Hamiltonian)
\begin{equation}
\label{evolX}
\dot x^i(t)={dx^i(t)\over dt}={\{x^i(t),H\}}
=w^{ij}(\mathbf{x}(t))\,\partial_jH(\mathbf{x}(t))
\end{equation}
This form involves the Poisson Bracket and is covariant under local changes of coordinates in phase espace.
The equations are flow equations of the general form
\begin{equation}
\label{ }
\dot x^i(t)={dx^i(t)\over dt}=F^i(\mathbf{x}(t))
\end{equation}
but the vector field $F^i=\omega^{ij}\partial_jH$ is special and derives from $H$. The flow, i.e. the application
$\phi$: $\Omega\times\mathbb{R}\to \Omega$ is called the Hamiltonian flow associated to $H$.
The evolution functions $\Phi_t(\mathbf{x}$ defined by
\begin{equation}
\label{ }
\mathbf{x}(t=0)=\mathbf{x}\quad\implies\quad \phi_t(\mathbf{x})=\mathbf{x}(t)
\end{equation} 
form a group of transformations (as long as $H$ is independent of the time)
\begin{equation}
\label{ }
\phi_{t_1+t_2}=\phi_{t_1}\circ\phi_{t_2}
\end{equation}
More generally, let us consider a (time independent) observable $f$ (a function on $\Omega$). The evolution of the value of $f$ for a dynamical state $\mathbf{x}(t)$, $f(\mathbf{x},t)=f(\mathbf{x}(t))$ where $\mathbf{x}(t)=\phi_t(\mathbf{x})$, obeys the equation
\begin{equation}
\label{dFdt1}
{\partial f(\mathbf{x},t)\over\partial t}={d f(\mathbf{x}(t))\over dt}={\{f,H\}}(\mathbf{x}(t))
\end{equation}
where the r.h.s. is the Poisson bracket of the observable $f$ and the Hamiltonian $H$.
In particular (when $H$ is independent of $t$) the energy $E(t)=H(\mathbf{x}(t))$ is conserved
\begin{equation}
\label{ }
{\partial E(\mathbf{x},t)\over\partial t}={d H(\mathbf{x}(t))\over dt}=0
\end{equation}

\subsubsection{The Liouville measure}
\index{Liouville measure}
The symplectic form $\omega$ defines an invariant volume element $d\mu$ on the phase space $\Omega$. 
\begin{equation}
\label{ }
d\mu(\mathbf{x})=\omega^n = \prod_{i=1}^{2 n} d x^i\, | \omega | ^{1/2}
\quad,\qquad  | \omega |=|\det(\omega_{ij})|
\end{equation}
This defines the so-called Liouville measure on $\Omega$. 
This mesure is invariant under all the Hamitonian flows.

\subsubsection{Example: the classical spin}
\index{Spin}
\index{Classical spin}
The simplest example of a system with a non trivial phase space is the classical spin (the classical top with constant total angular momentum).
The states of the spin are labelled by unit 3-components vector $\vec n=(n_1,n_2,n_3)$, $|\vec n|=1$ (the direction of the angular momentum). Thus the phase space is the 2-dimensional unit sphere and is compact
$$\Omega=\mathcal{S}_2$$
The classical precession equation
$${d\vec n\over dt}=\vec B\times \vec n $$
can be written in Hamiltonian form.
$\vec B$ is a vector in $\mathbb{R}^3$, possibly a 3-component vector field on the sphere depending on $\vec n$. 

There is a symplectic structure on $\Omega$. It is related to the natural complex structure on $\mathcal{S}_2$ (the Riemann sphere).
The Poisson bracket of two functions $f$ and $g$ on $\mathcal{S}_2$ is defined as
$$\{f,g\}=(\vec \nabla f \times \vec\nabla g )\cdot \vec n\ .$$
The gradient field $\vec\nabla f$ of a function $f$ on the sphere is a vector field  tangent to the sphere, so $\vec \nabla f \times \vec\nabla g$ is normal to the sphere, hence collinear with $\vec n$.
In spherical coordinates
$$  \vec n=(\sin\theta \cos\phi,\sin\theta \sin\phi, \cos\theta) $$
the Poisson bracket is simply
$$
\{f,g\}={1\over\sin\theta}\left({\partial f\over\partial\theta}{\partial g\over\partial\phi}-  {\partial g\over\partial\theta}{\partial f\over\partial\phi}\right)
$$
Admissible local Darboux coordinates $\mathbf{x}=(x^1,x^2)$ such that $\omega= dx^1\wedge dx^2$ must be locally orthogonal, area preserving mappings. 
\index{Darboux coordinates}
Examples are 
\begin{itemize}
  \item  
``action-angle'' variables (the Lambert cylindrical equal-area projection)
$$\mathbf{x}=(\cos\theta,\phi)$$
 \item or plane coordinates (the Lambert azimuthal equal-area projection).
$$\mathbf{x}=(2\sin(\theta/2)\cos\phi,2\sin(\theta/2)\sin\phi)$$
\end{itemize}
\index{Lambert coordinates}
\begin{figure}[h]
\begin{center}
\includegraphics[height=1.2in]{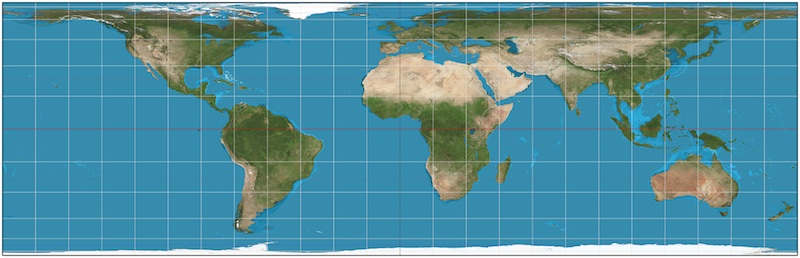}
\qquad
\includegraphics[height=2.in]{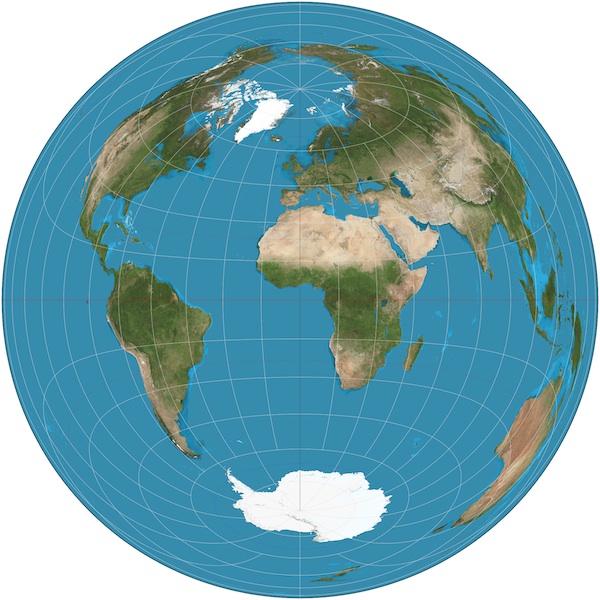}
\medskip
\caption{The Lambert cylindrical and azimuthal coordinates}
\label{ }
\end{center}
\end{figure}
With this Poisson bracket, the Hamiltonian which generates the precession dynamics is simply (for constant $\vec B$)
$$H=\vec B\cdot\vec n$$

\subsubsection{Statistical states, distribution functions, the Liouville equation}
\index{Statistical ensemble}
\index{Statistical distribution}
\index{Mixed state}
\index{Expectation value}
We now consider statistical ensembles. If we have only some partial information on the state of the system, to this information is  associated  a statistical (or mixed) state $\varphi$. This statistical state is described by a probability distribution on the phase space $\Omega$, that we write
\begin{equation}
\label{ }
d\rho_\varphi(\mathbf{x})=d\mu(\mathbf{x})\, \rho_\varphi(\mathbf{x})
\end{equation}
with $d\mu(\mathbf{x})$ the Liouville measure and $\rho_\varphi(\mathbf{x})$ the probability density, a non negative distribution (function) such that
\begin{equation}
\label{ }
\rho_\varphi(\mathbf{x})\ge 0\quad,\qquad \int_\Omega d\mu(\mathbf{x})\, \rho_\varphi(\mathbf{x})=1
\end{equation}
On a given statistical state $\varphi$ the expectation for an observable $f$ (its expectation value, i.e. its mean value if we perform a large number of measurements of $f$ on independent systems in the same state $\varphi$) is 
\begin{equation}
\label{ }
\langle f\rangle_\varphi = \int_\mathbf{\Omega} d\mu(\mathbf{x})\, \rho_\varphi(\mathbf{x})\, f(\mathbf{x})
\end{equation}

When the system evolves according to some Hamiltonian flow $\phi_t$ generated by a Hamiltonian $H$, the statistical state depends on time 
$\varphi\to \varphi(t)$, as well as the distribution function $\rho_\varphi\to\rho_{\varphi(t)}$. 
$\varphi$ being the initial state of the system at time $t=0$, we can denote this function
\begin{equation}
\label{ }
\rho_{\varphi(t)}(\mathbf{x})=\rho_\varphi(\mathbf{x},t)
\end{equation}
This time dependent distribution function is given by
\begin{equation}
\label{ }
\rho_\varphi(\mathbf{x}(t),t)=\rho_\varphi(\mathbf{x}) \quad,\qquad \mathbf{x}(t)=\phi_t(\mathbf{x})
\end{equation}
(using the fact that the Liouville measure is conserved by the Hamiltonian flow).
Using the evolution equation for $\mathbf{x}(t)$  \ref{evolX}, one obtains the evolution equation for the distribution function $\rho_\varphi(\mathbf{x},t)$, called the Liouville equation 
\index{Liouville equation}
\begin{equation}
\label{LiouvilleEq}
{\partial\over \partial t}\rho_\varphi(\mathbf{x},t)=\left\{H,\rho_\varphi\right\}(\mathbf{x},t)
\end{equation}
$\left\{H,\rho_\varphi\right\}$ is the Poisson bracket of the Hamiltonian $H$ and of the density function $\rho_\varphi$, considered of course as a function of $\mathbf{x}$ only (time is fixed in the r.h.s. of \ref{LiouvilleEq}).

With these notations, the expectation of the observable $f$ dépends on the time $t$ , and is given by the two equivalent integrals
\begin{equation}
\label{ }
\langle f\rangle(t)=\int_\mathbf{\Omega} d\mu(\mathbf{x})\, \rho_\varphi(\mathbf{x})\, f(\mathbf{x}(t))= \int_\mathbf{\Omega} d\mu(\mathbf{x})\, \rho_\varphi(\mathbf{x},t)\, f(\mathbf{x})
\end{equation}

Of course when the state of the system is a ``pure state'' ($\varphi_\mathrm{pure}=\mathbf{x}_0$) the distribution function is a Dirac measure $\rho_\mathrm{pure}(\mathbf{x})=\delta(\mathbf{x}-\mathbf{x}_0)$ and the Liouville equation leads to
\begin{equation}
\label{ }
\rho_\mathrm{pure}(\mathbf{x},t)=\delta(\mathbf{x}-\mathbf{x}(t))
\quad,\qquad \mathbf{x}(t)=\phi_t(\mathbf{x}_0)
\end{equation}
\index{Pure state}

\subsubsection{Canonical transformations}
\index{Canonical transformation}
The Hamiltonian flow is an example of canonical transformations.
Canonical transformations $\mathcal{C}$ are (bijective) mappings $\mathbf{\Omega}\to\mathbf{\Omega}$ which preserve the symplectic structure.
Denoting the image of the point $\mathbf{x}\in\Omega$ (by the canonical transformation $\mathcal{C}$) by $\mathbf{X}$
\begin{equation}
\label{defmapC}
\mathbf{x}\  \mathop{\to}^\mathcal{C}\ \mathbf{X}=\mathcal{C}(\mathbf{x})
\end{equation}
This means simply that the symplectic form $\omega^*$ defined by
\begin{equation}
\label{omegastar}
\omega^*(x)=\omega(X)
\end{equation}
is equal to the original form
\begin{equation}
\label{ }
\omega^*=\omega
\end{equation}
$\omega^*$ is called the pullback of the symplectic form $\omega$ by the mapping $\mathcal{C}$ and is also denoted $\mathcal{C}^*\omega$.
In a given coordinate system such that
\begin{equation}
\label{ }
\mathbf{x}=(x^i)\ ,\ \ 
\mathbf{X}=(X^k)
\end{equation}
\ref{omegastar} means that $\omega$ and $\omega^*$ read
\begin{equation}
\label{ }
\omega(x)=w_{ij}(\mathbf{x})\,dx^i\wedge dx^j 
\quad,\qquad
\omega^*(x)=w_{ij}(\mathbf{X})\,dX^i\wedge dX^j 
\end{equation}
so that the components of $\omega^*$  are
\begin{equation}
\label{ }
\omega^*_{ij}(\mathbf{x})=w_{kl}(\mathbf{X})  {\partial X^k\over\partial x^i}{\partial X^l\over\partial x^j}
\end{equation}
$\mathcal{C}$ is a canonical transformation if
\begin{equation}
\label{ }
\omega_{ij}(\mathbf{x})=\omega^*_{ij}(\mathbf{x})
\end{equation}

Canonical transformations are the transformations that preserve the Poisson brackets. Let $f$ and $g$ be two observables (functions $\Omega\to\mathbb{R}$ and $F=f\circ \mathcal{C}^{-1}$ and $G=g\circ \mathcal{C}^{-1}$ their transform by the transformation $\mathcal{C}$
\begin{equation}
\label{ }
f(\mathbf{x})=F(\mathbf{X})\quad,\qquad g(\mathbf{x})=G(\mathbf{X})
\end{equation}
$\mathcal{C}$ is a canonical transformation if
\begin{equation}
\label{ }
\{f,g\}_\omega=\{F,G\}_\omega
\end{equation}
Taking for $f$ and $g$ the coordinate change $x^i\to X^i$ itself, canonical transformations are change of coordinates such that
\begin{equation}
\label{ }
\{X^i,X^j\}=\{x^i,x^j\}
\end{equation}

Canonical transformations are very useful tools in classical mechanics. They are the classical analog of unitary transformations in quantum mechanics.

In the simple example of the classical spin, the canonical transformations are simply the smooth area preserving diffeomorphisms of the 2 dimensional sphere.

\subsubsection{Along the Hamiltonian flow}

As an application, one can treat the Hamiltonian flow $\phi_t$ as a time dependent change of coordinate in phase space (a change of reference frame) and look at the dynamics of the system in this new frame which moves with the system.
In this new coordinates, denoted $\bar{\mathbf{x}}=\{\bar x^i\}$, if at time $t=0$ the system is in an initial  state $\bar{\mathbf{x}}=\mathbf{x}_0$, at time $t$ it is of course still in the same state
$\bar{\mathbf{x}}(t)=\mathbf{x}_0$.
\index{Hamiltonian flow}

It is the observables $f$ which become time dependent.
Indeed, if in the original (time independent) coordinate system one considers a time independent observable $f$ (a function $\mathbf{x}\to f(\mathbf{x})$), in the new coordinate system one must consider the time dependent observable $\bar f$, defined by
\begin{equation}
\label{ }
\bar f(\mathbf{\bar x}, t)=f(\mathbf{x}(t))
\qquad\text{with}\quad \mathbf{x}(t)=\phi_t(\mathbf{\bar x})\quad\text{i.e.}\quad \mathbf{x}(0)=\mathbf{\bar x}
\end{equation}
This time dependent observable $\bar f(\mathbf{\bar x},t)$ describes how the value of the observable $f$ evolves with the time $t$, when expressed as a function of the initial state $\mathbf{\bar x}$.
Of course the time  evolution of $\bar f$ depends on the dynamocs of the system, hence of the Hamiltonian $H$.
The dynamics for the observables is given by evolution equation (similar to the Liouville equation, up to a sign)
\begin{equation}
\label{LEqF}
{\partial\bar f\over \partial t}=- \{H,\bar f\}
\quad\text{i.e.}\qquad
{\partial\bar  f(\mathbf{\bar x},t)\over \partial t}=\{\bar f,H\}(\mathbf{\bar x},t)
\end{equation}
In this dynamical frame the Hamiltonian is still time independent, i.e. $\bar H=H$, since its evolution equation is
\begin{equation}
\label{ }
{\partial H \over \partial t}=- \{H,H\}=0
\end{equation}
The Poisson bracket is always the Poisson bracket for the symplectic form $\omega$, since $\omega$ is conserved by the canonical transformations, in particular this change of reference frame.

This change of frame corresponds to a change from a representation of the dynamics by a flow of the states in phase space, the observables being fixed functions, to a representation where the states do not move, but where there is a flow for the functions.
This is the analog for Hamiltonian flows to what is done in fluid dynamics: going from the Eulerian specification (the fluid moves in a fixed coordinate system) to the Lagrangian specification (the coordinate system moves along the fluid).
But these two representations are of course the classical analog of the Schrödinger picture (vector states evolves, operators are fixed) and of the Heisenberg picture (vector states are fixed, operators depend on time) in quantum mechanics.
\index{Lagrangian specification}
\index{Eulerian specification}

\subsection{The commutative algebra of observables}
\index{Observable}
\index{Commutative algebra}
Let us adopt a slightly more abstract point of view. 
The (real or) complex functions on phase space $f$ $\Omega\to\mathbb{C}$ form a commutative algebra $\mathcal{A}$ with the standard addition and multiplication laws.
\begin{equation}
\label{ }
(f+g)(x)=f(x)+g(x)\quad,\qquad (fg)(x)=f(x) g(x)
\end{equation}
This is more generally true if $\Omega=X$ is simply a locally compact topological space, and $\mathcal{A}$ the algebra of continuous functions with compact support. 

Statistical states (probability distributions on $X$) are then normalized positive linear forms $\varphi$ on $\mathcal{A}$
\begin{equation}
\label{ }
\varphi(\alpha f+\beta g)=\alpha \varphi(f) + \beta \varphi(g)
\ ,\quad
\varphi( f f^*)\ge 0
\ ,\quad
\varphi(1)=1
\end{equation}
The sup or $\mathcal{L}^\infty$ norm, defined as
\begin{equation}
\label{ }
\| f\|^2=\sup_{x\in X} |f(x)|^2 =\sup_{\varphi\,\mathrm{states}} \varphi(|f(x)|^2 )
\end{equation}
has clearly the following properties
\begin{equation}
\label{ }
\| f \|=\| f^* \|
\ ,\quad
\| fg\|\le\| f \| \ \| g\|
\ ,\quad
\| f f^* \| =\| f \|^2
\end{equation}
and $\mathcal{A}$ is complete under this norm. This makes the algebra $\mathcal{A}$ a so-called commutative C$^*$-algebra.
\index{C$^*$-algebra}

For any element $x\in X$, consider the subalgebra of the functions that vanish at $x$
\begin{equation}
\label{ }
\mathcal{I}_x:\{f\in\mathcal{A};\ f(x)=0\}
\end{equation}
They are maximal ideals of $\mathcal{A}$, (left-)ideals $\mathcal{I}$ of an algebra $\mathcal{A}$ being subalgebras of $\mathcal{A}$ such that $x\in\mathcal{I}$ and $y\in\mathcal{A}$ implies $xy\in\mathcal{I}$. It is easy to show that the set of maximal ideals of $\mathcal{A}=C(X)$ is isomorphic to $X$, and that $\mathcal{A}/\mathcal{I}_x=\mathbb{C}$ the target space.

Now a famous theorem by Gelfand and Naimark states that the reciprocal is true. Any commutative C$^*$-algebra is isomorphic to the algebra of continuous functions on some topological (locally compact) space $X$! This seems a formal result (the space $X$ and its topology may be quite wild, and very far from a regular smooth manifold), but it is important to keep in mind that a mathematical object (here a topological space) can be defined intrinsically (by its elements) or equivalently by the abstract properties of some set of functions from this object to some other object (here the commutative algebra of observables). This modern point of view in mathematics (basically this idea is at the basis of the category formalism) is also important in modern physics, as we shall see later in the quantum case.
\index{Gelfand-Naimark}

For the Hamiltonian systems, the algebra of (differentiable) functions on $\Omega$ is equipped with a additional product, the Poisson bracket $\{f,g\}$.
The corresponding algebra, with the three laws (addition, multiplication, Poisson bracket) is now a commutative Poisson algebra.
A Poisson algebra is a (not necessarily commutative) associative algebra with a bracket that satisfies \ref{antisymBr}, \ref{LeibnitzBra} and \ref{JacobiBra}.
\index{Poisson algebra}

\subsection{"Axiomatics"}
The most general formulation for classical Hamiltonian dynamics is that of Poisson manifold. This is a more general formulation that symplectic manifolds, since it encompasses special situations where the symplectic form is degenerate.
Poisson manifolds can in general be split (foliated) into ``symplectic leaves'' embodied with a well defined induced symplectic structure.

The fact that in classical mechanics dynamics are given by Hamiltonian flows on a phase space which is a symplectic or Poisson manifold can be somehow justified, if one assumes that the possible dynamics are flows generated by some smooth vector fields, that these flows are generated by conserved quantities (Hamiltonians) and that these dynamics are covariant under change of frames generated by these flows (existence and invariance of canonical transformations). 
\index{Poisson manifold}

However a real understanding and justification of classical Hamiltonian dynamics comes from quantum mechanics.
Indeed, the Poisson bracket structure is the ``classical limit'' of the commutators of observables (operators)  in quantum mechanics, and the canonical transformations are the classical version of unitary transformations.

\section{Probabilities}
\index{Probabilities}

Probabilities are an important aspect of classical physics and are one of the essential components of quantum physics. 
Without going into any details and any formalism, I think it is important to recall the two main ways to consider and use probabilities in statistics and physics: the frequentist point of view and the Bayesian point of view. At the level considered here, these are  different point of views on the same mathematical formalism, and on its use. As we shall see, in some sense quantum mechanics forces us to treat them on the same footing.
There are of course many different, more subtle and more precise mathematical as well as philosophical points of view on probability theory. I shall not enter in any discussion about the merits and the consistency of objective probabilities versus subjective probabilities.

Amongst  many standard references on the mathematical formalism of probability, there is the book by Kolmogorov
\cite{Kolmogorov-book},
and the book by Feller \cite{Feller-book}. 
See also the quick introduction for and by a physicist by M. Bauer (in french) \cite{BauerCours2009}.
References on Bayesian probabilities are the books by de Finetti \cite{deFinetti-book}, by Jaynes \cite{Jaynes-Book} and the article by Cox \cite{Cox1946}.

\subsection{The frequentist point of view}
\index{Frequentist probabilities}
The frequentist point of view is the most familiar and the most used in statistical physics, dynamical systems, as well as in mathematics (it is at the basis of the formulation of modern probability theory from the beginning of 20th century, in particular for the Kolmogorov axiomatic formulation of probabilities). 
Roughly speaking, probabilities represent a measure of ignorance on the state of a system, coming for instance from: uncertainty on its initial state, uncertainty on its dynamical evolution due to uncertainty on the dynamics, or high sensibility to the initial conditions (chaos). 
Then probabilities are asymptotic frequencies of events (measurements) if we repeat observations on systems prepared by the same initial procedure.
More precisely, one has a set $\Omega$ of samples (the sample space), a $\sigma$-algebra  $\mathcal{F}$ of ``measurable'' subsets of the sample space $\Omega$, and a  measure $P$ on $\mathcal{F}$ (equivalently a probability measure $\mu_\mathcal{F}$ on $\Omega$.
This probability measure is a priori given.

\subsection{The Bayesian point of view}
\index{Bayesian probabilities}
The so called Bayesian point of view is somehow broader, and of use in statistics, game theory, economy, but also in experimental sciences. It is also closer to the initial formulations of probabilities (or ``chance'') in the 18th and 19th centuries. It has been reviewed by statisticians like de Finetti or Jaynes (among others) in the 20th century. 

Probabilities are considered as qualitative estimates for the ``plausibility'' of some proposition (it can be the result of some observation), given some ``state of knowledge'' on a system.
The rules that these probabilities must satisfy are constrained by some logical principles (objectivist point of view where the degree of plausibility is constructed by a ``rational agent''), or may correspond  simply to some ``degree of personal belief'' of  propositions (subjectivist point of view).

\subsection{Conditional probabilities}
\index{Conditional probabilities}
\index{Bayes formula}
The basic rules are the same in the different formulations. A most important concept is conditional probabilities $P(A|B)$ (the probability of $A$,  $B$ being given), and the Bayes (or conditional probability) relation 
\begin{equation}
\label{ }
P(A|B)={P(B|A) P(A)\over P(B)}
\end{equation}
where $P(A)$ and $P(B)$ are the initial probabilities for $A$ and $B$ (the prior), and $P(A|B)$ $P(B|A)$ the conditional probabilities. 

\paragraph{Frequentist:} In the frequentist formulation $P(A|B)$ is the frequency of $A$, once we have selected the samples such that $B$ is true. Bayes formula has the simple representation with Venn diagrams in the set of samples

\begin{figure}[h]
\begin{center}
\includegraphics[width=2in]{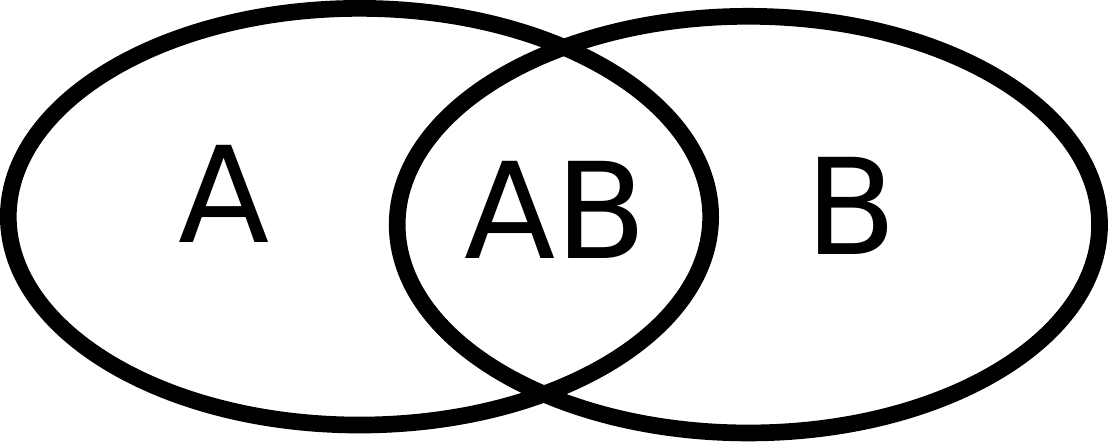}
\caption{Venn representation of the conditional probabilities formula}
\label{ }
\end{center}
\end{figure}

\paragraph{Bayesian:} In the Baysian formulation (see for instance the book by Jaynes), one may consider every probabilities as  conditional probabilities. For instance $P_C(A)=P(A|C)$, where the proposition $C$ corresponds to  the ``prior knowledge'' that leads to the probability assignment $p_C(A)$  for $A$ (so $P_C$ is the probability distribution). If $AB$ means the proposition ``$A$ and  $B$'' ($A\wedge B$ or $A+B$), Bayes formula follows from  the ``product rule''
\begin{equation}
\label{ }
P(AB|C)=P(A|BC) P(B|C)=P(B|AC) P(A|C)
\end{equation}
whose meaning is the following: given $C$, if I already know the plausibility for $AB$ of being true ($P(AB|C)$), and the plausibility for $B$ of being true (the prior $P(B|C)$), the formula tells me how I should modify the plausibility for $A$ of being true, if I learn that $B$ is true ($P(A|BC)$). Together with the ``sum rule''
\begin{equation}
\label{ }
P(A|C)+P(\neg A|C)=1
\end{equation}
($\neg$ is the negation), these are the basic rules of probability theory in this framework.

\section{Quantum mechanics: ``canonical'' formulation}
Let us recall the so called ``canonical formalism'' of quantum mechanics, as it is presented in textbooks. This is the standard presentation when one uses the ``correspondence principle'' to quantize a classical non-relativistic system, or simple field theories.
\index{Canonical quantization}

There is of course an enormous number of good books on quantum mechanics and quantum field theory. 
Among the very first books on quantum mechanics, those  of P. A. Dirac  (1930) \cite{Dirac30} and J. von Neumann (1932) \cite{vonNeumann32G} (\cite{vonNeumann32} for the english traduction of 1955) are still very useful and valuable.
Modern books with a contemporary view and treatment of the recent developments are the books by Cohen-Tanoudji, Laloe \& Diu \cite{Cohen-Tannoudji:1977ys} , by M. le Bellac \cite{LeBellac-2011}, by Auletta, Fortunato and Parisi \cite{Auletta-book-2009}. 

Some standard references on quantum field theory are the books by J. Zinn-Justin \cite{ZinnJustin-book}, by A. Zee \cite{Zee03} (in a very different style). Refernce more oriented towards mathematical physics will be given later.

Amongst the numerous references on the questions of the foundation and the interpretation of quantum mechanics, one may look at the encyclopedic review by Auletta \cite{Auletta}, and at the recent shorter book by F. Laloe \cite{Laloe-book} (see also \cite{Laloe-book-fr,laloe-2001-69}). More later.
\subsection{Principles}

\subsubsection{Pure states:}
\index{Phase space}\index{Hilbert space}
\index{Bra-ket notation}
\index{Scalar product}
\index{Dirac}
The phase space  $\mathbf{\Omega}$ of classical mechanics is replaced by the complex Hilbert space ${\mathcal H}$ of states. 
Elements of $\mathcal{H}$ (vectors) are denoted $\psi$ or $|\psi\rangle$ (``kets'' in Dirac notations). 
The scalar product of two vectors $\psi$ and $\psi'$ in ${\mathcal H}$ is denoted
$\psi^*\!\!\cdot\!\psi'$ or $\langle\psi|\psi'\rangle$. The $\psi^*=\langle\psi\vert$  are the ``bra'' and belong to the dual $\mathcal{H}^*$ of $\mathcal{H}$. 
Note that in the mathematical litterature the scalar product is often noted in the opposite order $\langle\psi\vert\psi'\rangle=\psi'\!\!\cdot\!\psi^*$. We shall stick to the physicists notations.

Pure quantum states are rays of the Hilbert space, i.e. 1 dimensional subspaces of $\mathcal{H}$. They correspond to unit norm vectors $|\psi\rangle$, such that $\|{\psi}\|^2=\langle\psi |\psi\rangle=1$, and modulo an arbitrary  phase $|\psi\rangle\equiv\emath^{\imath\theta}|\psi\rangle$.

\subsubsection{Observables:} 
\index{Observable}\index{Self-adjoint operator}\index{Physical observable}
The physical observables $A$ are the self-adjoint operators on  $\mathcal{H}$ (Hermitian or symmetric operators), such that $A=A^\dagger$, where the conjugation is defined by $\langle A^\dagger \psi'|\psi\rangle=\langle \psi'|A\psi\rangle$.
Note that the conjugation $A^\dagger$ is rather denoted $A^*$ in the mathematical literature, and in some chapter we shall use this notation, when dealing with general Hilbert spaces not necessarily complex.

The operators on $\mathcal{H}$ form an associative, but non commutative operator algebra.
Any set of of commuting operators $\{A_i\}$ corresponds to a set of classically compatible observables, which can be measured independently.
\index{Non commutative algebra}
\index{Compatible observables}

\subsubsection{Measurements, Born principle:} 
\index{Measurement}
\index{Born principle}
\index{Expectation value}
The outcome of the measurement of an observable $A$ on a state $\psi$ is in general not deterministic. 
Quantum mechanics give only probabilities, and in particular the expectation value of the outcomes. 
This expectation value is given by the Born rule
\begin{equation}
\label{mq1}
\langle A \rangle_{\psi}=\langle\psi|A|\psi\rangle=\langle\psi|A\psi\rangle
\end{equation}
For compatible (commuting) observables the probabilities of outcome obey the standard rule of probabilities and these measurements can be repeated and performed independently.

This implies in particular that the possible outcomes of the measurement of $A$ must belong to the spectrum of $A$, i.e. can only be eigenvalues of $A$ (I consider the simple case where $A$ has a discrete spectrum).
Moreover the probability to get the outcome $a_i$ ($a_i$ being the eigenvalue of $A$ and $|i\rangle$ the corresponding eigenvector) is the modulus squared of the probability amplitude $\langle i|\psi\rangle$
$$  \text{probability of outcome of }A = a_i \text{\ in the state } |\psi\rangle \  =\  p_i\ =\   | \langle i|\psi\rangle |^2$$
\index{Eigenvector}\index{Eigenvalue}\index{Spectrum}

It follows also that quantum measurements are irreversible process. In ideal measurements or non destructive measurements which can be repeated, if the outcome of $A$ was $a_i$, after the measurement the system is found to be in the eigenstate $|i\rangle$. This is the projection postulate.
\index{Irreversible process}
\index{Ideal measurement}
\index{Non-destructive measurement}
\index{Projection postulate}
\index{Orthogonal projector}
In the more general situation where the eigenspace of $A$ associated to the eigenvalue $a_i$ is a higher dimensional subspace $V_i$, the state of the system is obtained by applying the orthogonal projector $P_i$ onto $V_i$ to the initial state $|\psi\rangle$.
Things are more subtle in the case of a continuous spectrum and non normalizable eigenstates.

At that stage I do not discuss what it means to ``prepare a system in a given state'', what ``represents'' the state vector, what is really a measurement process (the problem of quantum measurement) and what means the projection postulate. We shall come back to some of these questions along the course.
\index{Projection postulate}
\index{Measurement}
\index{Preparation}

\subsubsection{Unitary dynamics}
\index{Closed system}
\index{Unitary transformation}
\index{Hamiltonian}
\index{Evolution operator}

For a closed system, the time evolution of the states is linear and it must preserve the probabilities, hence the scalar product $\langle.|.\rangle$. Therefore is given by unitary transformations $U(t)$ such that $U^{-1}=U^\dagger$.
Again if the system is isolated the time evolution form a multiplicative group acting on the Hilbert space and its algebra of observables, hence it is generated by an Hamiltonian self-adjoint operator $H$

$$ U(t)=\exp\left({t\over\imath\hbar}H\right) $$
The evolution equations for states and observables are discussed below.

\subsubsection{Multipartite systems:}
\index{Multipartite system}
\index{Bipartite system}
\index{Tensor product}
\index{Entangled state}
\index{Entanglement}

Assuming that it is possible to perform independent measurements on two independent (causally) subsystems $S_1$ and $S_2$ implies (at least in the  finite dimensional case) that the Hilbert space $\mathcal{H}$ of the states of the composite system $S=``S_1\cup S_2''$ is the tensor product of the Hilbert spaces $\mathcal{H}_1$ and $\mathcal{H}_2$ of the two subsystems.
$$\mathcal{H}=\mathcal{H}_1\otimes \mathcal{H}_2$$
This implies the existence for the system $\mathcal{S}$ of generic ``entangled  states'' between the two subsystems
$$|\Psi\rangle =c { |\psi\rangle}_1\otimes {|\phi\rangle}_2 +  c'  {|\psi'\rangle}_1\otimes {|\phi'\rangle}_2 $$
Entanglement is one of the most important feature of quantum mechanics, and has no counterpart in classical mechanics. 
It is entanglement that leads to many of the counter-intuitive features of quantum mechanics, but  it  leads also to many of its interesting aspects and to some of its greatest success.

\subsubsection{Correspondence principe, canonical quantization}

\index{Canonical quantization}
\index{Correspondence principe}
\index{Conjugate variables}
\index{Commutation relation}
The correspondence principle has been very important in the elaboration of quantum mechanics. Here by correspondence principle I mean that when quantizing a classical system, often one can associate to canonically conjugate variables $(q_i,p_i)$ self-adjoint operators $(Q_i, P_i)$ that satisfy the canonical commutation relations
\begin{equation}
\label{mq2}
\{q_i,p_i\}=\delta_{ij}\ \implies\ [Q_i,P_j]=\imath\hbar\delta_{ij}
\end{equation}
and to take has Hamiltonian the operator obtained by replacing in the classical Hamiltonian the variables $(q_i,p_i)$ by the corresponding operators.
\index{Hamiltonian}
\index{Position}
\index{Momentum}

For instance, for the particle on a line in a potential, one takes as $(Q,P)$ the position and the momentum and for the Hamiltonian
\begin{equation}
\label{HamQV}
H={P^2\over 2m}+V(Q)
\end{equation}
The usual explicit representation is $\mathcal{H}=\mathcal{L}^2(\mathbb{R})$, the states $|\psi\rangle$  correspond to the wave functions $\psi(q)$, and the operators are represented as
\begin{equation}
\label{ }
Q=q\quad,\qquad P={\hbar\over\imath}{\partial\over\partial q}
\end{equation}
\index{Wave function}

\subsection{Representations of quantum mechanics}
The representation of states and observables as vectors and operators is invariant under global unitary transformations (the analog of canonical transformations in classical mechanics). These unitary transformations may depend on time. Therefore there are different representations of the dynamics in quantum mechanics. I recall the two main representations.
La représentation des états et des observables étant invariante par des transformations unitaires globales, pouvant dépendre du temps (l'équivalent des transformations canoniques classiques), les états et la dynamique du système peuvent se représenter de plusieurs façon équivalentes. Je rappelle ici les deux principales.

\subsubsection{The Schrödinger picture}
\index{Schrödinger picture}
It is the most simple, and the most used in non relativistic quantum mechanics, in canonical quantization and to formulate the path integral.
In the Schrödinger picture the states $\psi$ (the kets $|\psi\rangle$) evolve with time and are noted $\psi(t)$. The observables are represented by time independent operators. The evolution is given by the Schrödinger equation
\index{Schrödinger equation}\index{State}\index{Observable}
\begin{equation}
\label{mq3}
\imath\hbar{d\psi\over dt}=H\psi
\end{equation}
The expectation value of an observable $A$ measured at time $t$ for a system in the state $\psi$ is thus
\begin{equation}
\label{evAPsi}
\langle A\rangle_{\psi(t)}=
\langle\psi(t)|A|\psi(t)\rangle
\end{equation}
\index{Evolution operator}\index{Evolution equation}
The evolution operator $U(t)$ is defined by 
\begin{equation}
\label{ }
\psi(t=0)=\psi_0\quad \to\quad \psi(t)=U(t)\psi_0
\end{equation}
It is given by
\begin{equation}
\label{ }
U(t)=\exp\left({{t\over\imath\hbar}H}\right)
\end{equation}
and obeys the evolution equation 
\begin{equation}
\label{mq5}
\imath\hbar\,{d \over d t}U(t)=H\ U(t)
\qquad;\qquad
U(0)=\mathbf{1}
\end{equation}
This generalizes easily to the case where the Hamiltonian depends explicitly of the time $t$. Then 
\begin{equation}
\imath\hbar\,{d \over d t}U(t,t_0)=H(t)\ U(t,t_0)
\qquad;\qquad
U(t_0,t_0)=\mathbf{1}
\end{equation}
and
\begin{equation}
\label{mq6b}
U(t,t_0)
=T\left[\exp\left({{1\over\imath\hbar}\int_{t_0}^{t} dt\,H(t)}\right)\right]
=\sum_{k=0}^\infty (\imath\hbar)^{-k}\ \int_{t_0<t_1<\cdots<t_{k}<t} \hskip -4 em dt_1\cdots dt_k\ H(t_{k})\cdots H(t_1)
\end{equation}
where $T$ means the time ordered product (more later).
\index{Time ordered product}

\subsubsection{The Heisenberg picture}
\index{Heisenberg picture}
This representation is the most useful in relativistic quantum field theory. It is in fact the best mathematically fully consistent formulation, since the notion of state in more subtle, in particular it depends on the reference frame.
It is required for building the relation between critical systems and Euclidean quantum field theory (statistical field theory).

In the Heisenberg representation, the states are redefined as a function of time via the unitary transformation  $U(-t)$ on $\mathcal{H}$, where $U(t)$ is the evolution operator for the Hamiltonian $H$. They are denoted
 \begin{equation}
\label{mq7}
|\psi;t\rangle=U(-t)|\psi\rangle
\end{equation}
The unitary transformation redefines the observables $A$. They becomes time dependent and are denoted $A(t)$
\begin{equation}
\label{mq9}
A(t)_{}=U(-t)AU(t)
\end{equation}
The dynamics given by the Schrödinger equation is reabsorbed by the unitary transformation. The dynamical states are independent of time!
\index{Unitary transformation}
\begin{equation}
\label{mq8}
|\psi(t);t\rangle = U(-t)U(t)|\psi\rangle =  |\psi\rangle
\end{equation}
\index{Expectation value}
The expectation value of an observable $A$ on a state $\psi$ at time $t$ is in the Heisenberg representation
\begin{equation}
\label{mq10}
{\langle A(t)\rangle}_{\psi}
=\langle\psi(t);t|A(t)|\psi(t;,t\rangle
=\langle\psi|A(t)|\psi\rangle
\end{equation}
The Schrödinger and Heisenberg representation are indeed equivalent, since they give the same result for the physical observable (the expectation values)
\begin{equation}
\label{evHevS}
\langle A \rangle_{\psi(t)}={\langle A(t)\rangle}_{\psi}
\end{equation}
\index{Hamiltonian}
In the Heisenberg representation the Hamitonian $H$ remains independent of time (since it commutes with $U(t)$
\begin{equation}
\label{ }
H(t)=H
\end{equation}
The time evolution of the operators is given by the evolution equation
\begin{equation}
\label{mq13}
\imath\hbar{d \over d t}A(t)=[A(t),H]
\end{equation}
\index{Liouville equation}
This is the quantum version of the classical Liouville equation \ref{LEqF}. Of course the Schrödinger and the Heisenberg representations are the quantum analog of the two ``Eulerian'' and` ``Lagrangian''  representations of classical mechanics discussed above.
\index{Eulerian specification}\index{Lagrangian specification}

For the particle in a potential the equations for $Q$ and $P$ are the quantum version of the classical Hamilton equations of motion
\index{Hamilton equations}
\begin{equation}
\label{mq14}
{d \over d t}Q(t)={1\over m} P(t)
\qquad,\qquad 
{ d\over d t}P(t)=-V'(Q(t))
\end{equation} 
For an observable $A$ which depends explicitly of time (in the Schrödinger picture), the evolution equation becomes
\begin{equation}
\label{mq13a}
\imath\hbar{d \over d t}A(t)=\imath\hbar{\partial \over \partial t}A(t)+[A(t),H]
\end{equation}
and taking its expectation value in some state $\psi$ one obtains Ehrenfest theorem
\begin{equation}
\label{mq13b}
\imath\hbar{d \over d t}\langle A\rangle (t)=\imath\hbar{\partial \over \partial t}\langle A\rangle (t)+\langle[A,H]\rangle(t)
\end{equation}\index{Ehrenfest's theorem}

\subsection{Quantum statistics and the density matrix}
\index{Quantum statistics}\index{Density matrix}\index{Probabilities}\index{Information}
\subsubsection{The density matrix}
As in classical physics, in general on has only a partial information on the physical system one is interested in.
Its state has to be described by a concept of statistical or mixed state. But in quantum mechanics all the information one can get on a system is provided by the expectation values of the observable of the system. Statistics is already there! The pure quantum states $|\psi\rangle$ are the special ``mixed states'' with the property that a maximal amount of information can be extracted by appropriate sets of compatible measurements on the (ensemble of) state. The difference with classical physics is that different maximal sets of information can be extracted from the same state if one chose to perform different incompatible sets of measurements.

The mathematical concept that represents a mixed state is that of density matrix.
But before discussing this, one can start by noticing that, as in classical physics, an abstract statistical state $\omega$ is fully characterized by the ensemble of the expectation values $\langle A\rangle_\omega$ of all the observables $A$ of the system, measured over the state $\omega$.
\index{Mixed state}\index{Statistical state}\index{Expectation value}\index{Observable}
\begin{equation}
\label{mqst1}
\langle\mathbf{A}\rangle_{\omega}
\quad=\ \text{expectation value of  $\mathbf{A}$ measured over the state $\omega$}
\end{equation}
I denote general statistical states by Greek letters (here $\omega$) and pure states by the bra-ket notation when there is a ambiguity.
The $\omega$ here should not be confused for the notation for the symplectic form over the classical phase space of a classical system. We are dealing with quantum systems and there is no classical phase space anymore.

\index{Operator}\index{Algebra of operator}\index{Linear form}
From the fact that the observables may be represented as an algebra of operators over the Hilbert space $\mathcal{H}$, it is natural to consider that statistical states $\omega$ corresponds to linear forms over the algebra of operators hence applications  $\mathbf{A}\to\langle \mathbf{A}\rangle_\omega$, with the properties
\index{Positivity}
\begin{align}
\label{mqst2}
&\langle a\mathbf{A}+b\mathbf{B}\rangle_\omega=a\langle\mathbf{A}\rangle_\omega+b\langle\mathbf{B}\rangle_\omega
\qquad\text{linearity}\\
&\langle\mathbf{A}^\dagger\rangle_\omega=\overline{\langle\mathbf{A}\rangle}_\omega
\qquad \text{reality}\\
&\langle \mathbf{A}^\dagger \mathbf{A}\rangle_{\omega}\ge 0\qquad 
\text{and}\qquad \langle \mathbf{1}\rangle_{\omega}=1
\qquad\text{positivity and normalization}
\end{align}
For finite dimensional Hilbert spaces and for the most common infinite dimensional cases (for physicists) this is equivalent to state that to any statistical state is associated a normalized positive self-adjoint matrix $\rho_\omega$ such that
\begin{equation}
\label{mqst3}
\langle \mathbf{A}\rangle_{\omega}=\tr (\rho_\omega\, \mathbf{A})
\end{equation}
This is the density matrix or density operator. It was introduced by J. von Neumann (and L. Landau  and F. Bloch) in 1927. \ref{mqst3} is the generalization of Born rule for statistical states.
\index{Density operator}\index{von Neumann J.}\index{Born rule}\index{Projection operator}

For pure states $|\psi\rangle$  the density operator is simply the projection operator onto the state
\begin{equation}
\label{ }
\rho_{\psi}=|\psi\rangle\langle\psi|
\end{equation}

Before discussing some properties and features of the density matrix, let me just mention that in the physics literature, the term ``state'' is usually reserved to pure states, while in the mathematics literature the term `state'' is used for general statistical states. The denomination ``pure state'' or ``extremal state'' is used for vectors in the Hilbert state and the associated projector.
There are in fact some good mathematical reasons to use this general denomination of state.
\index{State}\index{Pure state}

\subsubsection{Interpretations}
Let us consider a system whose Hilbert space is finite dimensional ($\dim(\mathcal{H})=N$), in a state given by a density matrix ${\rho}_\omega$.
${\rho}_\omega$  is a $N\times N$ self-adjoint positive matrix. It is diagonalizable and its eigenvalues are $\ge 0$. If it has $1\le K\le N$ orthonormal eigenvectors labeled by  $|n\rangle$ ($n=1,\cdots K$) associated with $K$  non-zero eigenvalues $p_n$ ($n=1,\cdots K$)
one can write
\begin{equation}
\label{ }
\rho_\omega=\sum_{n=1}^K p_n\, |n\rangle\langle n|
\end{equation}
with
\begin{equation}
\label{ }
0 < p_n\le 1\ ,\quad \sum_n p_n=1
\end{equation}
The expectation value of any observable $\mathbf{A}$ in the state $\omega$ is 
\begin{equation}
\label{mqst3}
\langle \mathbf{A}\rangle_{\omega}=
\sum_n p_n \langle n| \mathbf{A} |n\rangle
\end{equation}
The statistical state $\omega$ can therefore be viewed as a classical statistical mixture of the $K$ orthonormal pure states $|n\rangle$, $n=1,\cdots K$, the probability of the system to be in the pure state $|n\rangle$ being equal to $p_n$.

\index{Preparation}\index{Measurement}\index{Statistical ensemble}\index{Information}
This point of view is usually sufficient if one wants to think about results of measurements on a single instance of the system. But it should not be used to infer statements on how the system has been prepared. 
One can indeed build a statistical ensemble of independently prepared copies of the system corresponding to the state $\omega$ by picking at random, with probability $p_n$ the system in the state $|n\rangle$.
But this is not the only way to build a statistical ensemble corresponding to $\omega$.
More precisely, there are many different ways to prepare a statistical ensemble of states for the system, by picking with some probability $p_\alpha$ copies of the system in different states among a pre chosen set $\{|\psi_\alpha\rangle\}$ of (a priori not necessarily orthonormal) pure states, which give the same density matrix $\rho_\omega$. 

This is not a paradox. The difference between the different preparation modes is contained in the quantum correlations between the (copies of the) system and the devices used to do the preparation. These quantum correlations are fully inaccessible if one performs measurements on the system alone. The density matrix contains only the information about the statistics of the measurement on the system alone (but it encodes the maximally available information obtainable by measurements on the system only).

Another subtle point is that an ensemble of copies of a system is described by a density matrix $\rho$ for the single system if the different copies are really independent, i.e. if there are no quantum correlations between different copies in the ensemble. Some apparent paradoxes 
arise if there are such correlations. One must then consider the matrix density for several copies, taken as a larger composite quantum system.

\subsubsection{The von Neumann entropy}
\index{Entropy}\index{von Neumann entropy}
The ``degree of disorder'' or the ``lack  of information'' contained in a mixed quantum state $\omega$  is given by the von Neumann entropy
\begin{equation}
\label{mqst4}
S(\omega)=-\ \tr(\rho_\omega\,\log \rho_\omega) =-\sum_n  p_n\log p_n
\end{equation}
It is the analog of the Boltzman entropy for a classical statistical distribution. It shares also some deep relation with Shannon entropy in  information theory (more later).
\index{Boltzmann entropy}\index{Shannon entropy}

The entropy of a pure state is minimal and zero. Conversely, the state of maximal entropy is the statistical state where all quantum pure states are equiprobable. It is given by a density matrix proportional to the identity, and the entropy is the logarithm of the number of accessible different (orthogonal) pure quantum state, i.e. of the dimension of the Hilbert space (in agreement with the famous Boltzmann formula $W=k_B\log N$).
\begin{equation}
\label{ }
\rho={1\over N} \mathbf{1}\quad,\quad S=\log N
\quad,\qquad N= \mathrm{dim}\mathcal{H}
\end{equation}

\index{Entanglement}\index{Entanglement entropy}
\subsubsection{Application: Entanglement entropy}
An important context where the density matrix plays a role is the context of open quantum systems and multipartite quantum systems.
Consider a bipartite system $\mathcal{S}$ composed of two distinct subsystems $\mathcal{A}$ and $\mathcal{B}$.
The Hilbert space $\mathcal{H}_\mathcal{S}$ of the pure states of $\mathcal{S}$ is the tensor product of the Hilbert space of the two subsystems
\index{Tensor product}\index{Bipartite system}
\begin{equation}
\label{ }
\mathcal{H}_\mathcal{S}=\mathcal{H}_\mathcal{A} \otimes\mathcal{H}_\mathcal{B}
\end{equation}
Let us assume that the total system is in a statistical state given by a density matrix $\rho_\mathcal{S}$, but that one is interested only in the subsystem $\mathcal{A}$ (or $\mathcal{B}$). In particular one can only perform easement on observables relative to $\mathcal{A}$ (or $\mathcal{B}$). 
\index{Reduced density matrix}\index{Partial trace}
Then all the information on $\mathcal{A}$ is contained in the reduced density matrix $\rho_\mathcal{A}$; obtained by taking the partial trace of the density matrix for the whole system $\rho_\mathcal{S}$ over the (matrix indices relative to the) system $\mathcal{B}$.
\begin{equation}
\label{ }
\rho_\mathcal{A}=\tr_{\mathcal{B}}\left[\rho_\mathcal{S}\right]
\end{equation}
This is simply the quantum analog of taking the marginal of a probability distribution $p(x,y)$ with respect to one of the random variables $\rho_x(x)=\int dy\, \rho (x,y)$).
\index{Marginal probability distribution}

If the system $\mathcal{S}$ is in a pure state $|\psi\rangle$, but if this state is entangled between $\mathcal{A}$ and $\mathcal{B}$, the reduced density matrix $\rho_\mathcal{A}$ is that of a mixed state, and its entropy is $S_A(\rho_\mathcal{A})>0$. Indeed when considering $\mathcal{A}$ only the quantum correlations between  $\mathcal{A}$ and $\mathcal{B}$ have been lost.
If $\mathcal{S}$ is in a pure state the entropies  $S_A(\rho_\mathcal{A})=S_B(\rho_\mathcal{B})$. This entropy  is then called the entanglement entropy. Let us just recall that this is precisely one of the context where the concept of von Neumann entropy was introduced around 1927.
More properties of features  of quantum entropies will be given later.

\subsubsection{Gibbs states}
\index{Gibbs state}\index{Temperature}\index{Partition function}
A standard example of density matrix is provided by considering an quantum system $\mathcal{S}$ which is (weakly) coupled to a large thermostat, so that it is at equilibrium, exchanging freely energy (as well as other quantum correlations) with the thermostat, and at a finite temperature $T$.
Then the mixed state of the system is  a Gibbs state (or in full generality called a Kubo-Martin-Schwinger or KMS state).
If the spectrum of the Hamiltonian $H$ of the system is discrete, with the eigenstates $|n\rangle$, $n\in\mathbb{N}$ and eigenvalues (energy levels) by $E_n$ (with $E_0<E_1<E_2\cdots$), the density matrix is
\begin{equation}
\label{mqtf2}
\rho_\beta\ =
\ {1\over Z(\beta)}\,\exp(-\beta H)
\end{equation}
with $Z(\beta)$ the partition function
\begin{equation}
\label{mqtf3}
Z(\beta)\ =\ \mathrm{tr}\left[\exp(-\beta H)\right]
\end{equation}
and 
\begin{equation}
\label{ }
\beta={1\over k_BT}
\end{equation}
In the energy eigenstates basis the density matrix reads
 \begin{equation}
\label{mqtf2}
\rho_\beta\ =\ \sum_n p_n\,|n\rangle\langle n|
\end{equation}
with $p_n$ the standard Gibbs probability \index{Gibbs probability}
\begin{equation}
\label{mqtf1}
p_n\ =\ {1\over Z(\beta)}\,\exp(-\beta E_n)
\qquad;\qquad
Z(\beta)=\sum_n\, \exp(-\beta E_n) 
\end{equation}
The expectation value of an observable $A$ in the thermal state at temperature $T$ is
\index{Thermal state}
\begin{equation}
\label{mqtf4}
\langle A\rangle_\beta\ =
\ \sum_n p_n\,\langle n | A | n \rangle 
\ =\ \ {\mathrm{tr}\left[A\, \exp(-\beta H) \right]\over\tr\left[\exp(-\beta H)\right]}
\end{equation}
For infinite systems with an infinite number of degrees of freedom, such that several equilibrium macroscopic states may coexist, the density matrix formalism is not sufficient and must be replaced by the formalism of KMS states ((Kubo-Martin-Schwinger). This will be discussed a bit more later in connection with superselection sectors in the algebraic formalism.
\index{KMS state}

\subsubsection{Imaginary time formalism}
\index{Imaginary time}
Let us come back to the simple case of a quantum non-relativistic system, whose energy spectrum is bounded below (and discrete to make things simple), but unbounded from above.
The evolution operator \index{Evolution operator}
\begin{equation}
\label{mqtim1}
U(t)=\exp\left({t\over\imath\hbar}\,H\right)
\end{equation} 
considered as a function of the time $t$, may be extended from ``physical'' real time $t\in\mathbb{R}$ to complex time variable, provided that 
\begin{equation}
\label{mqtf5}
\mathrm{Im}(t)\le 0
\end{equation}
More precisely, $U(t)$ as an operator, belongs to the algebra $\mathcal{B}(\mathcal{H})$ of bounded operators on the Hilbert space $\mathcal{H}$.
\index{Bounded operator}
A bounded operator $A$ on $\mathcal{H}$ is an operator whose $L^\infty$ norm, defined as
\begin{equation}
\label{ }
\| A \|^2=\sup_{\psi\in\mathcal{H}} {\langle\psi | A^\dagger A |\psi\rangle \over \langle\psi | \psi\rangle}
\end{equation}
is finite.
This is clear in the simple case where
\begin{equation}
\label{mqtim2}
U(t)=\sum_n\exp\left({t\over \imath\hbar}E_n\right) \,|n\rangle\langle n|
\qquad, \quad \|U(t)\|=
\begin{cases}
      \exp\left({\Im(t)\over\hbar} E_0\right)& \text{if}\ \Im{(t)}\le 0, \\
      +\infty& \text{otherwise}.
\end{cases}
\end{equation}
The properties of the algebras of bounded operators and of their norm will be discussed in more details  in the next section on the algebraic formulation of quantum mechanics.

\begin{figure}[h]
\begin{center}
\includegraphics[width=6.cm]{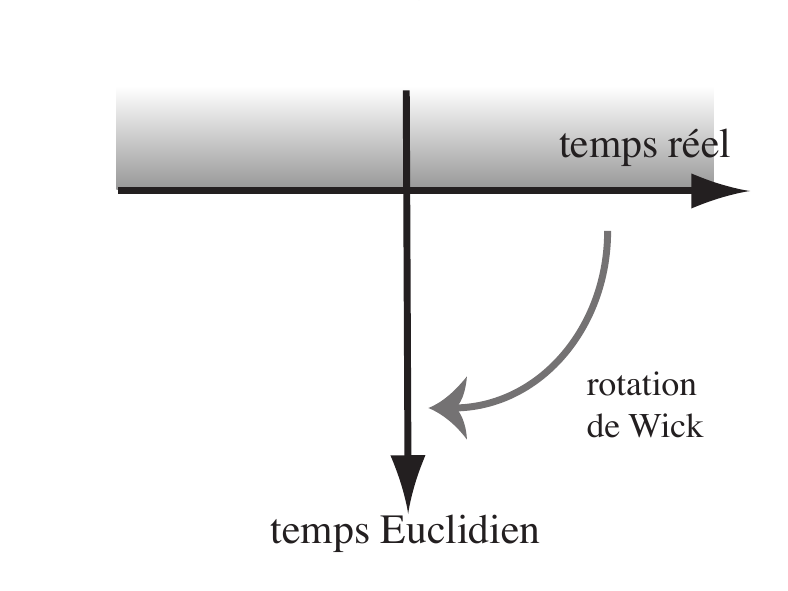}
\caption{Real time $t$ and imaginary (Euclidean) time $\tau=\imath t$: Wick rotation }
\label{mqtf6}
\end{center}
\end{figure}

Consider now the case where $t$  is purely imaginary
\begin{equation}
\label{mqtim3}
t=-\imath\,\tau\quad,\quad\tau>0\quad\text{real}
\qquad\,\quad
U(-\imath\tau)=\exp\left(-{\tau\over\hbar} H\right)
\end{equation}
The evolution operator has the same form than the density matrix for the system in a Gibbs state at temperature $T$
\begin{equation}
\label{rhovsU}
\rho_\beta\ =\ {1\over Z(\beta)}\,  U(-\imath\tau)
\qquad,\quad \beta={1\over k_{\mathrm{B}}\,T}={\tau\over\hbar}={\imath}\ {t\over\hbar}
\end{equation}
For relativistic quantum field theories, time became an ``Euclidean coordinate'' $\tau=x^0$, and Minkowski space time becomes Euclidean space.
\index{Euclidean space}\index{Minkowski space}
There is deep analogy
\begin{center}
imaginary time $=$\ finite temperature
\end{center}
This analogy has numerous applications. It is at the basis of many applications of quantum field theory to statistical physics (Euclidean Field Theory).
Reciprocally, statistical physics methods have found applications in quantum physics and high energy physics (lattice gauge theories).
 Considering quantum theory for imaginary time is also very useful in high energy physics, in quantum gravity. Finally this relation between Gibbs (KMS) states and the unitary evolution operator extends to a more general relation between states and automorphisms of some operator algebras (Tomita-Takesaki theory), that we shall discuss (very superficially) in the next chapter.

\section{Path and functional integrals formulations}

\subsection{Path integrals}
\index{Path integral}\index{Feynman}
It is known since Feynman that a very useful, if not always rigorous, way to represent matrix elements of the evolution operator of a quantum system (transition amplitudes, or ``propagators'') is  provided by path integrals (for non-relativistic systems with a few degrees of freedom) and functional integrals (for relativistic or non relativistic systems with continuous degrees of freedoms, i.e. fields).

Standard references on path integral methods on quantum mechanics and quantum field theory are the original book by Feynman \& Hibbs \cite{FeynmanHibbs}, and the books by J. Zinn-Justin \cite{ZinnJustin-book}, \cite{Zinn-Justin:2010fk}.

For a single particle in an external potential this probability amplitude $K$ for propagation from $q_i$ at time $t_i$ to $q_f$ at time $t_f$\
\begin{equation}
\label{ }
\langle q_f|U(t_f-t_i)|q_i\rangle= \langle q_f,t_f|q_i,t_i\rangle\qquad U(t)=\exp\left({t\over i \hbar} H\right)
\end{equation}
(the first notation refers to the Schrödinger picture, the second one to the Heisenberg picture) 
\index{Schrödinger picture}\index{Heisenberg picture}
can be written as a sum of histories $q(t)$
\begin{equation}
\label{ }
\int_{q(t_i)=q_I,\ q(t_f)=
q_f}  \mathcal{D}[q]\ \exp\left({i\over \hbar} S[q]\right)
\end{equation}
where $S[q]$ is the classical action.
\index{Action}

\begin{figure}[h!]
\begin{center}
\includegraphics[width=4in]{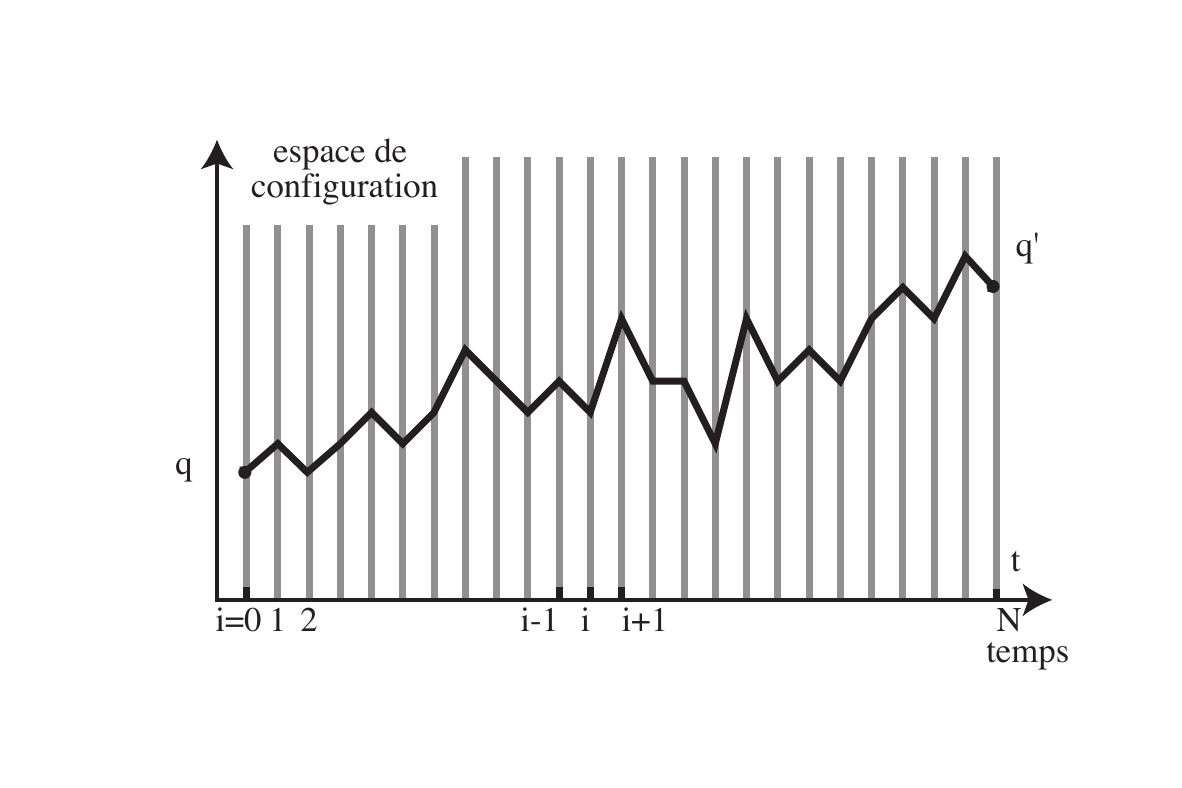}
\caption{Path integral: time discretization}
\label{ }
\end{center}
\end{figure}

The precise derivation of this formula, as well as its proper mathematical definition, is obtained by decomposing the evolution of the system in a large number $N$ of evolutions during  elementary time step $\Delta t=\epsilon=t/N$, at arbitrary intermediate positions $q(t_n=n\epsilon)$, $n\in\{1,\cdots ,N-1\}$, using the superposition principle.
One then uses the explicit formula for the propagation kernel at small time (the potential $V(q)$ may be considered as constant locally)
\begin{equation}
\label{ }
K(q_f,\epsilon, q_i,0)\simeq \left({2 i \pi \hbar\epsilon\over m}\right)^{-1/2}\ \exp\left({i\over \hbar}\left({m\over 2} {(q_f-q_i)^2\over \epsilon}-\epsilon V\left({q_f+q_i\over 2}\right)\right)\right) 
\end{equation}
and one then takes the continuous time limit $\epsilon\to 0$.
The precise definition of the measure over histories or paths  is (from the prefactor)
\begin{equation}
\label{ }
\mathcal{D}[q]=\prod_{n=1}^{N-1} \left(dq(t_n)\left({2 i \pi \hbar \epsilon\over m}\right)^{-1/2}\right)
\end{equation}

\index{Lagrangian}\index{Hamiltonian}
The  ``Lagrangian'' path integral has a ``Hamiltonian'' version (path integral in phase space)
\begin{equation}
\label{ }
\int_{q(t_i)=q_I,\ q(t_f)=
q_f}  \mathcal{D}[q,p]\ \exp\left({i\over \hbar} \int dt\left(p \dot q -H(q,p))\right)\right)
\end{equation}
But one must be very careful on the definition of this path integral (discretization and continuum time limit) and on the measure in order to obtain a consistent quantum theory. 

\subsection{Field theories, functional integrals}

\index{Functional integral}
\index{Quantum field theory}
Path integral representations extend to the case of relativistic quantum field theories. For instance for the scalar field, whose classical action (giving the Klein-Gordon equation) is
\index{Scalar field}\index{Klein-Gordon equation}
\begin{equation}
\label{ }
S[\phi]=\int dt \int d^3\vec x\ {1\over 2} \left(\left({\partial\phi\over\partial t}\right)^2- \left({\partial\phi\over\partial \vec x}\right)^2 - m^2 \phi^2\right)
=\int d^4x {1\over 2} \left(-\partial^\mu\phi\, \partial_\mu \phi - m^2 \phi^2\right)  
\end{equation}
a path integral involves an integral over field configurations over space-time of the form
$$\int \mathcal{D}[\phi)\ \emath^{{\imath\over\hbar} S[\phi]}$$
and is usually denoted a functional integral.

\index{Time ordered product}
More precisely, the vacuum expectation value of time ordered product of local field operators $\boldsymbol{\phi}$  in this quantized  field theory can be expressed as a functional integral
\begin{equation}
\label{ }
\langle\Omega|T \boldsymbol{\phi}(x_1)\cdots \boldsymbol{\phi}(x_N)|\Omega\rangle={1\over Z}
\int \mathcal{D}[\phi)\ \emath^{{\imath\over\hbar} S[\phi]}\phi(x_1)\cdots\phi(x_N)
\end{equation}
with $Z$ the partition function or vacuum amplitude
\begin{equation}
\label{ }
{Z}=
\int \mathcal{D}[\phi)\ \emath^{{\imath\over\hbar} S[\phi]}
\end{equation}
The factor $Z$ means that the functional integral is normalized so that the vacuum to vacuum amplitude is $$\langle\Omega|\Omega\rangle=1$$

The path integral and functional integral formulations are invaluable tools to formulate many quantum systems and quantum field theories, and perform calculations. 
They give a very simple and intuitive picture of the semiclassical regimes. It explains why the laws of classical physics can be formulated via variational principles, since classical trajectories are just the stationary phase trajectories (saddle points) dominating the sum over trajectories in the classical limit $\hbar\to 0$.
In many cases it allows to treat and visualize quantum interference effects when a few semi-classical trajectories dominates (for instance for trace formulas).

Functional integral methods are also very important conceptually for quantum field theory: from the renormalization of QED to the quantization and proof of renormalisability of non abelian gauge theories, the treatment of topological effects and anomalies in QFT, the formulation of the Wilsonian renormalization group, the applications of QFT methods to statistical mechanics, etc.
They thus provides a very useful way to quantize a theory, at least in semiclassical regime where one expect that the quantum theory is not too strongly coupled and quantum correlations and interference effects can be kept under control.

I will not elaborate further here. When discussing the quantum formalism, one should keep in mind that the path and functional integrals represent a very useful and powerful (if usually not mathematically rigorous) way to visualize, manipulate and compute transition amplitudes, i.e. matrix elements of operators. They  rather represent an application of the standard canonical formalism, allowing to construct the Hilbert space (or part of it) and the matrix elements of operators of a quantum theory out of a classical theory via a quick and efficient recipe.

{
\section{Quantum mechanics and reversibility}
}
\label{S:reversibility}
\index{Reversibility}
\index{Irreversibility}
\subsection{Is quantum mechanics reversible or irreversible?}
An important property of quantum  (as well as classical) physics is reversibility: 
the general formulation of the physical laws is the same under time reversal. 
This  is often stated as: 
\index{Time reversal}\index{Time arrow}
\begin{center}
``There is no microscopic time arrow.''
\end{center}
This does not mean that the fundamental interactions (the specific physical laws that govern our universe) are invariant under time reversal.
It is known that (assuming unitarity, locality and Lorentz invariance) they are invariant only under CPT, the product of charge conjugation, parity and time reversal. This reversibility statement means that the dynamics, viewed forward in time (press key \fbox{${\RHD}$} ), of any given state of a system is similar to the dynamics, viewed backward in time (press key \fbox{$\LHD$} ), of some other state.

This reversibility statement is of course also different from the macroscopic irreversibility that we experience in everyday life (expansion of the universe, second principle of Thermodynamics, quantum measurement, 
Parkinson's laws \cite{Parkinson1955}
, etc.).
\index{Second principle of thermodynamics}
\index{Parkinson's law}
\index{Measurement}

In classical mechanics reversibility is an obvious consequence of the Hamiltonian formulation. In quantum mechanics things are more subtle. Indeed if the evolution of a ``closed system'' (with no interaction with its environment or some observer) is unitary and reversible (and in particular possible quantum correlations between the system and its ``outside'' are kept untouched), quantum measurements are irreversible processes.
However it is known since a long time that microscopic reversibility is not really in contradiction with this irreversibility.
See for instance the '64 paper by Aharonov, Bergmann \& Lebowitz \cite{Aharonov1964}.
Since this will be very important in these following lectures, especially in the presentation of the quantum logic formalism, let us discuss it on a simple, but basic example, with the usual suspects involved in quantum measurements.

\subsection{Reversibility of quantum probabilities}
\label{ssRevQP}
We consider two observers, Alice and Bob. 
Each of them can  measure a different observable (respectively $A$ and $B$) on a given quantum system $\mathcal{S}$ (for simplicity $\mathcal{S}$ can be in a finite number of states, i.e. its Hilbert space is finite dimensional). We take these observations to be perfect (non demolition) test measurements, i.e. yes/no measurements, represented by some selfadjoint projectors $\mathbf{P}_A$ and $\mathbf{P}_B$ such that $\mathbf{P}_A^2=\mathbf{P}_A$ and $\mathbf{P}_B^2=\mathbf{P}_B$, but not necessarily commuting.
The eigenvalues of these operators are $1$ and $0$, corresponding to the two possible outcomes $1$ and $0$ (or $\TRUE$ and $\FALSE$ ) of the  measurements of the observables $A$ and of the observable $B$. 
\index{Observable}\index{Eigenvalue}\index{Eigenvector}

Let us consider now the two following protocols.

\paragraph{Protocol 1:}

\index{Alice}\index{Bob}\index{Measurement}
Alice gets the system $\mathcal{S}$ (in a state  she knows nothing about). She measures $A$ and if she finds $\TRUE$, then she send the system to Bob, who measures $B$. What is the plausibility
\footnote{In a Bayesian sense. 
}
 for Alice that Bob will find that $B$ is $\TRUE$? Let us call this the conditional probability for $B$ to be found true, $A$ being known to be true, and denote it $P(B{\,\mapsfrom\hskip -2.ex\vert\hskip 1.ex} A)$.
The arrow $\mapsfrom$ denotes the causal ordering between the measurement of $A$ (by Alice) and of $B$ (by Bob).
\index{Conditional probabilities}
\index{Causal ordering}
\index{Bayesian probabilities}

\begin{figure}[h]
\begin{center}
\includegraphics[width=2.in]{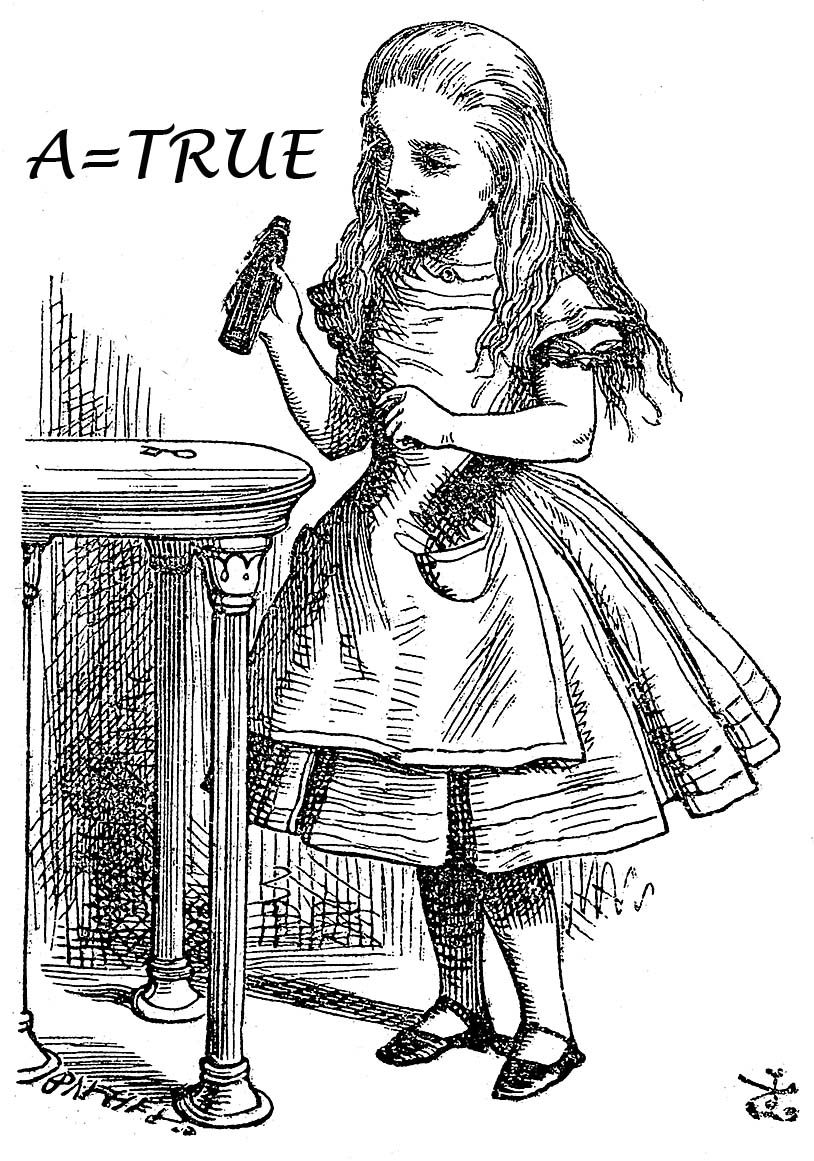}\qquad
\includegraphics[width=2.in]{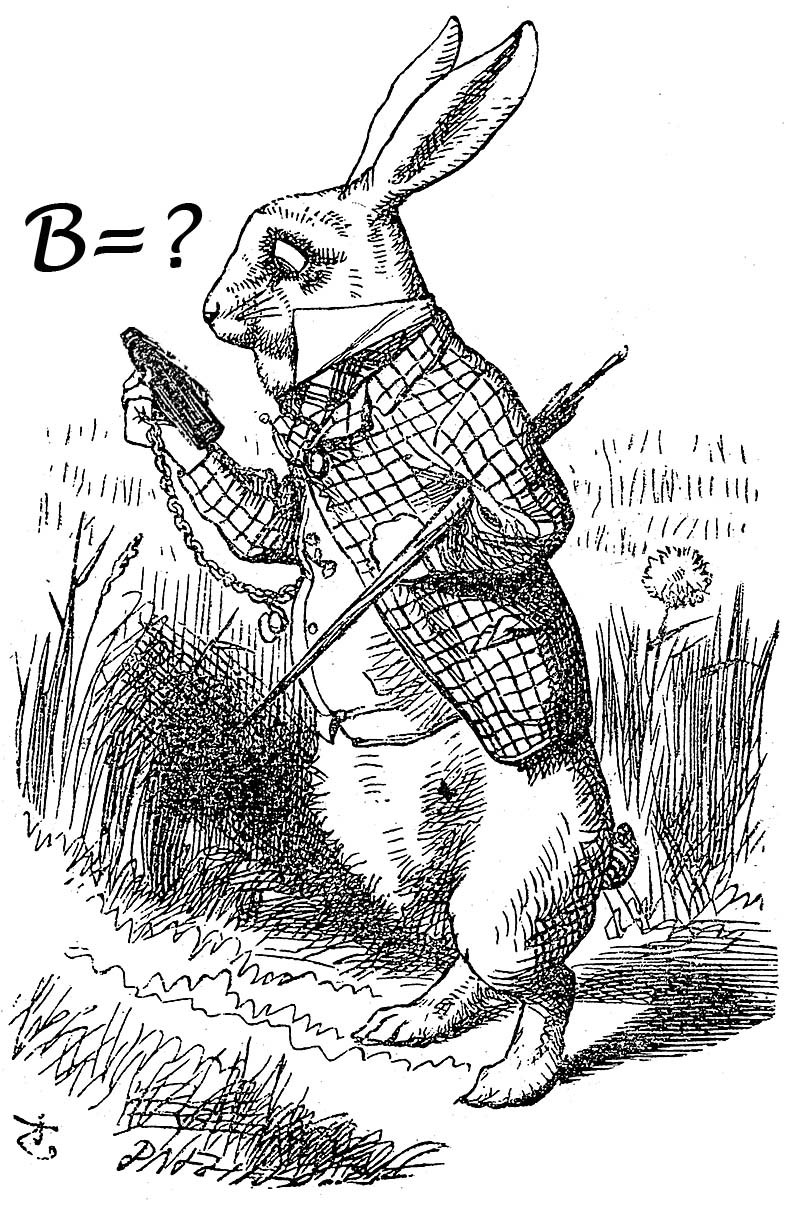}
\caption{Protocol 1: Alice wants to guess what Bob will measure. This defines the conditional probability  $ P(B{\,\mapsfrom\hskip -2.ex\vert\hskip 1.ex} A)$.}
\label{ }
\end{center}
\end{figure}

\paragraph{Protocol 2:}

  Alice gets the system $\ {S}$ from Bob, and knows nothing else about $\mathcal{S}$. Bob tells her that he has measured $B$, but does not tell her the result of his measurement, nor how the system was prepared before he performed the measurement (he may know nothing about it, he just measured $B$). Then Alice measures $A$ and (if) she finds $\TRUE$ she asks herself the following question: what is the plausibility 
(for her, Alice) that Bob had found that $B$ was $\TRUE$? 
\footnote{This question makes sense if for instance, Alice has made a bet with Bob. Again, and especially for this protocol, the probability has to be taken in a Bayesian sense.}
Let us call this the conditional probability for $B$ to have been found true, $A$ being known to be true, and denote it by 
$P(B{\,\mapsto\hskip -2.3ex\vert\hskip 1.ex} A)$.
The arrow $\mapsto$ denotes the causal ordering between the measurement of $A$ (by Alice) and of $B$ (by Bob).

  \begin{figure}[h!]
\label{ }
\begin{center}
\includegraphics[width=2.in]{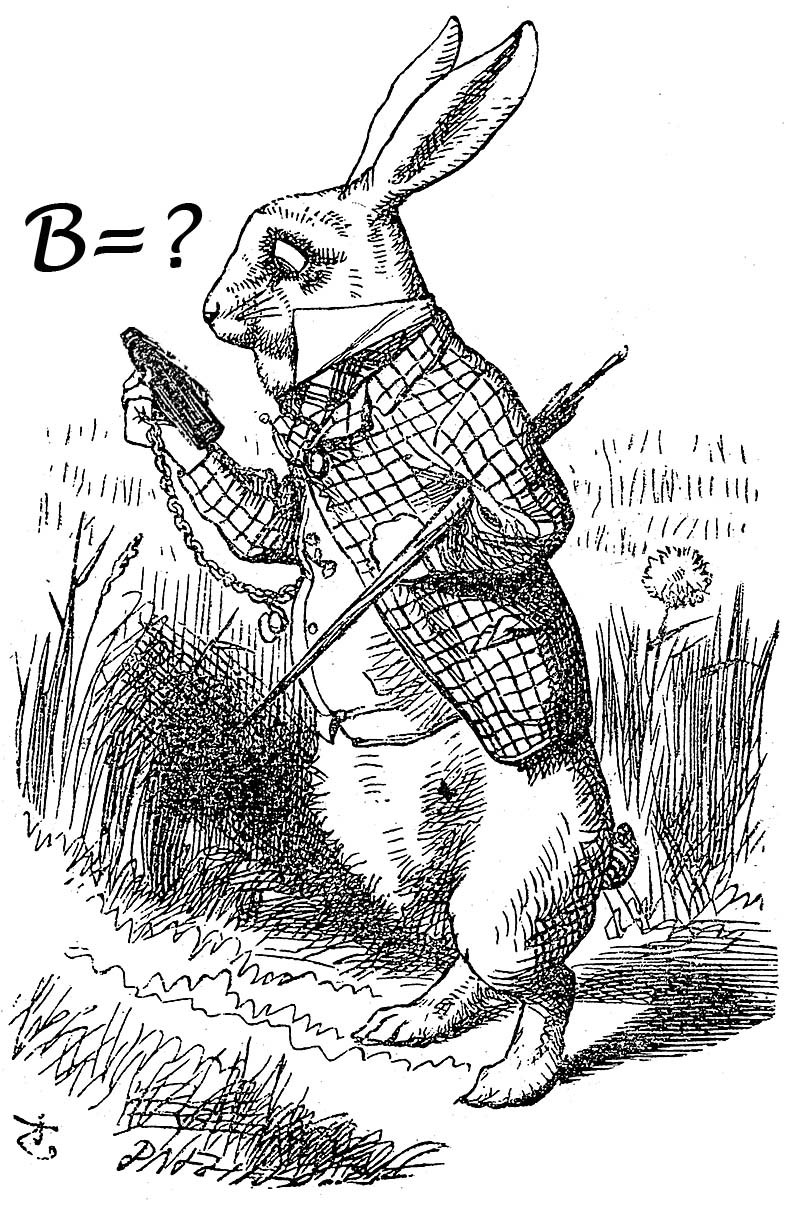}\qquad
\includegraphics[width=2.in]{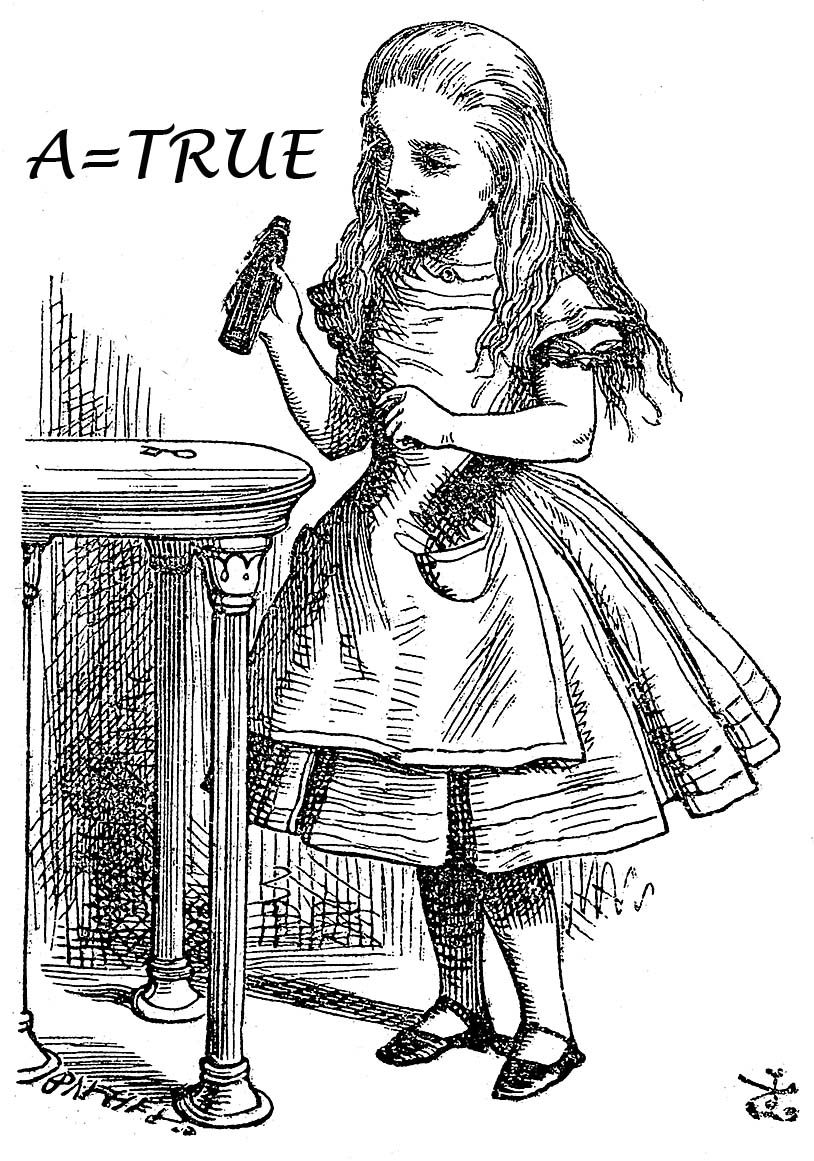}\nonumber\\
\caption{Protocol 2: Alice wants to guess what Bob has measured. This defines the conditional probability  \quad $P(B{\,\mapsto\hskip -2.ex\vert\hskip 1.ex} A)$.}
\end{center}
\end{figure}

If $\mathcal{S}$ was a classical system, and the mesurements  were classical measurements which do not change the state of $\mathcal{S}$,  then the two protocols are equivalent and  the two quantities  equal  the standard conditional probability (Bayes formula)
$$\mathcal{S}\ \text{classical system}\,:  \ \ P(B{\,\mapsfrom\hskip -2.ex\vert\hskip 1.ex} A) =P(B{\,\mapsto\hskip -2.3ex\vert\hskip 1.ex} A)=P(B|A)=P(B\cap A)/P(A)\ .$$
\index{Bayes formula}

In the quantum case, at a purely logical level, knowing only that the measurement process may perturb the system $\mathcal{S}$, $P(B{\,\mapsfrom\hskip -2.ex\vert\hskip 1.ex} A)$ and $P(B{\,\mapsto\hskip -2.3ex\vert\hskip 1.ex} A)$ may be different. 
A crucial and remarkable property of quantum mechanics is that they are still equal. Indeed in the first protocol $P(B{\,\mapsfrom\hskip -2.ex\vert\hskip 1.ex} A)$ is given by the Born rule; if Alice finds that $A$ is $\TRUE$ and knows nothing more, her best bet is that the state of $\mathcal{S} 
$ is given by the density matrix 
$$\rho_A=\mathbf{P}_A/\mathrm{Tr}(\mathbf{P}_A)$$
Therefore the probability for Bob to find that $B$ is $\TRUE$ is 
$$P(B{\,\mapsfrom\hskip -2.ex\vert\hskip 1.ex} A)=\tr(\rho_A\mathbf{P}_B) .$$

In the second protocol the best guess for Alice is to assume that before Bob measures $B$ the state of the system is given by the equidistributed density matrix $\rho_{\mathbf{1}}=\mathbf{1}/\tr(\mathbf{1})$.  
In this case the probability that Bob finds that $B$ is $\TRUE$, then that Alice finds that $A$ is $\TRUE$, is 
$$p_1=\tr(\mathbf{P}_B)/\tr(\mathbf{1})\times \tr(\rho_B\mathbf{P}_A)
\quad\text{ with}\quad
\rho_B=\mathbf{P}_B/\mathrm{Tr}(\mathbf{P}_B).$$
Similarily the probability that  Bob finds that $B$ is $\FALSE$, then that Alice finds that $A$ is $\TRUE$ is 
$$p_2=\tr(\mathbf{1}-\mathbf{P}_{B})/\tr(\mathbf{1})\times \tr(\rho_{\overline B}\mathbf{P}_A)=(\tr(\mathbf{P}_A)-\tr(\mathbf{P}_A\mathbf{P}_B))/\tr(\mathbf{1})$$
 where $\rho_{\overline B}=(\mathbf{1}-\mathbf{P}_B)/\tr(\mathbf{1}-\mathbf{P}_B)$. 
 The total probability is then 
 $$ P(B{\,\mapsto\hskip -2.3ex\vert\hskip 1.ex} A)= p_1+p_2 = \tr(\rho_A\mathbf{P}_B) .$$
 
Therefore, even if $A$ and $B$ are not compatible, i.e. if  $\mathbf{P}_A$ and $\mathbf{P}_B$ do not commute, we obtain in  both case the standard result for quantum conditional probabilities
\begin{equation}
\label{ }
\mathcal{S}\ \text{quantum system}\, :\ \ 
P(B{\,\mapsfrom\hskip -2.ex\vert\hskip 1.ex} A)=P(B{\,\mapsto\hskip -2.3ex\vert\hskip 1.ex} A)=\mathrm{Tr}[\mathbf{P}_A \mathbf{P}_B]/ \mathrm{Tr}[\mathbf{P}_A]
\end{equation}

This reversibility property (that I denote here causal reversibility, in order not to confuse it with time reversal invariance) is very important, as we shall see later.\index{Causal reversibility}

\chapter{Algebraic quantum formalism}
\label{s:AlgQM}
\section{Introduction}
In this formulation, quantum mechanics is constructed from the classical concepts of observables and states, assuming  that observables are not  commuting quantities anymore 
but still form an algebra, and using the concepts of causality and reversibility. Of course such ideas go back to the matrix mechanics of Heisenberg, but the precise formulation relies on the mathematical theory of operator algebras, initiated by F. J. Murray and J. von Neumann 
\index{von Neumann J.}
in the end of the thirties (one motivation of J. von Neumann was precisely to understand quantum mechanics). It was developped by Segal (Segal 47), and then notably by Wightman, Haag, Kastler, Ruelle, etc.

The standard and excellent reference on the algebraic and axiomatic approaches to quantum field theory is the book by R. Haag, Local Quantum Physics,  especially the second edition (1996) \cite{Haag96}. 
Another older reference is the book by N. N. Bogoliubov; A. A. Logunov, A.I. Oksak and  I.T. Todorov (1975, 1990) \cite{Bogo-Todo90}. 
Another useful reference is the famous book by R. F. Streater and A. S. Wightman (1964, 1989) \cite{StreaterWightmanBook} .

Standard references on operator algebras in the mathematical litterature are the books by J. Dixmier (1981, 1982)  \cite{Dixmier:1969vn}, Sakai (1971) \cite{Sakai71}, P. de la Harpe and V. Jones (1995) \cite{delaHarpeJones95}. References more oriented towards the (mathematical) physics community are Bratteli and Robinson(1979) \cite{BratteliRobinson2002}, A. Connes (1994) \cite{Connes94} and A. Connes and M. Marcolli \cite{ConnesMarcolli07}.
 I shall need also some results on real $\mathrm{C}^*$-algebras and the only good reference I am aware of is Goodearl (1982) \cite{Goodearl82}.
 
 I shall give here a very brief and crude presentation of the algebraic formulation of quantum theory. 
 It will stay at a very heuristic level, with no claim of precision or of mathematical rigor.
 However the starting point will be a bit different from the usual presentation, and was presented in \cite{PhysRevLett.107.180401}. 
 I shall start from the general concepts of observables and states, and derive why \emph{abstract real C$^*$-algebras} are the natural framework to formulate quantum theories. 
 Then I shall explain which mathematical results ensure that the theory can always be represented by algebras of operators on Hilbert spaces. 
 Finally I shall explain why locality and separability enforce the use of complex algebras and of complex Hilbert spaces.

{
\section{The algebra of observables}
}

\subsection{The mathematical  principles}
\label{AxiomsAlg}
A quantum system is described by its observables, its states and a causal involution acting on the observables and enforcing constraints on the states. Let us first give the axioms and motivate them later on.

\subsubsection{Observables}
 \index{Observable}\index{Real algebra}\index{Associatice algebra}\index{Unital algebra}
 The physical observables of the system generate  a \emph{real associative unital algebra $\mathcal{A}$} (whose elements will still be denoted ``observables'' ) .
 $\mathcal{A}$ is a linear vector space
  \begin{equation*}
\label{ }
\mathbf{a},\,\mathbf{b}\in \mathcal{A}\quad \lambda,\, \mu\in\mathbb{R}
\qquad
\lambda \mathbf{a} +\mu \mathbf{b} \in A 
\end{equation*}
\index{Associativity}\index{Distributivity}
with an associative product (distributive w.r.t the addition)
 \begin{equation}
\label{algebra}
\mathbf{a},\,\mathbf{b},\,\mathbf{c}\in \mathcal{A}
\qquad\mathbf{ab}\in A
\qquad
(\mathbf{ab})\mathbf{c}=\mathbf{a}(\mathbf{bc})
\end{equation}
and an unity
\begin{equation}
\label{ }
\mathbf{1 a}=\mathbf{a 1}=\mathbf{a}\ ,\ \ \forall \mathbf{a}\in\mathcal{A}
\end{equation}
We shall precise later what are ``physical observables''.

\subsubsection{The $^*$-conjugation}
\index{Conjugation}\index{Involution}\index{Anti-automorphism}
There is an involution $^\star$ on $\mathcal{A}$ (denoted conjugation). It is an anti-automorphism whose square is the identity. This means that
  \begin{equation*}
\label{ }
(\lambda \mathbf{a}+\mu \mathbf{b})^\star = \lambda \mathbf{a}^\star  + \mu \mathbf{b}^\star  
\end{equation*}
 \begin{equation}
\label{involution}
{(\mathbf{a}^*)}^*=\mathbf{a}
\qquad (\mathbf{ab})^\star =\mathbf{b}^\star  \mathbf{a}^\star 
\end{equation}

\subsubsection{States}
\index{State}\index{Expectation value}
   Each $\varphi$ associates to an observable $\mathbf{a}$ its expectation value $\varphi(\mathbf{a})\in\mathbb{R}$ in the state $\varphi$.
The states satisfy
  \begin{equation*}
\label{ }
\varphi(\lambda \mathbf{a} +\mu \mathbf{b})=\lambda \varphi(\mathbf{a}) +\mu \varphi(\mathbf{b})
\end{equation*}
  \begin{equation}
\label{defsta}
\varphi(\mathbf{a}^\star )=\varphi(\mathbf{a}) \qquad \varphi(\mathbf{1})=1 \qquad \varphi(\mathbf{a}^\star \mathbf{a})\ge 0
\end{equation}
The set of states is denoted $\mathcal{E}$. It is natural to assume that it allows to discriminate between observables, i.e.
\begin{equation}
\label{ }
\forall \ \mathbf{a}\neq\mathbf{b}\in\mathcal{A} \ (\text{and}\ \neq \mathbf{0}), \exists\  \varphi\in\mathcal{E}\ \text{such that}\ \varphi(\mathbf{a})\neq\varphi(\mathbf{b})
\end{equation}

I do not discuss the concepts of time and dynamics at that stage. This will be done later. I first discuss the relation between these ``axioms'' and the physical concepts of causality, reversibility and probabilities.

\subsection{Physical discussion}
\subsubsection{Observables and causality}
\index{Causality}
{
In quantum physics, the concept of physical observable corresponds both to an operation on the system (measurement) and to the response on the system (result on the measure), but I shall not elaborate further. 
We already discussed why in classical physics observables form a real commutative algebra. The removal of the commutativity assumption is the simplest modification imaginable compatible with the uncertainty principle (Heisenberg 1925).

Keeping the mathematical structure of an associative but non commutative algebra reflects the assumption that there is still some concept of \og causal ordering\fg\   
 between observables (not necessarily physical), in a formal but  loose sense. 
Indeed the multiplication and its associativity means that we can \og combine\fg{} successive observables, e.g. $\mathbf{ab}\simeq (\mathbf{b}$ then $\mathbf{a}$), in a linear process such that (($\mathbf{c}$ then $\mathbf{b}$) then $\mathbf{a}) \simeq (\mathbf{c}$ then ($\mathbf{b}$ then $\mathbf{a}$)). This ``combination'' is different from the concept of ``successive measurement".
\index{Non commutative algebra}

Without commutativity the existence of an addition law is already a non trivial fact, it means that we can \og combine\fg{} two non compatible observations into a new one whose mean value is always the sum of the first two mean values.

Both addition and mutiplication of observables are in fact more natural in the context of relativistic theories, via the analyticity properties of correlation functions and the short time and short distance expansions.
 
 \subsubsection{The $^*$-conjugation and reversibility}
 \index{Conjugation}\index{Irreversibility}
 The existence of the involution $^*$ (or conjugation) is the second and very important feature of quantum physics. It implies that although the observables do not commute, there is no favored arrow of time (or causal ordering) in the formulation of a physical theory, in other word this is reversibility. To any causal description of a system in term of a set of  observables $\{\mathbf{a},\,\mathbf{b},\,\ldots\}$ corresponds an equivalent \og anti-causal\fg{} description it terms of  conjugate observables $\{\mathbf{a}^*,\,\mathbf{b}^*,\,\ldots\}$. Although there is no precise concept of time or dynamics yet, the involution $^*$ must not be confused with the time reversal operator $\mathbf{T}$ (which may or may not be a symmetry of the dynamics).
 
\subsubsection{States, mesurements and probabilities}
\index{State}\index{Measurement}\index{Probabilities}
The states $\varphi$ are the simple generalisation of the classical concept of statistic (or probabilistic) states describing our knowledge of a system through the expectation value of the outcome of measurements for each possible observables. At that stage we do not assume anything about whether there are states such that all the values of the observables can be determined or not. Thus a state can be viewed also as the characterization of all the information which can be extracted from a system through a measurement process (this is the point of view often taken in quantum information theory).
We do not consider how states are prepared, nor how the measurements are performed (this is the object of the subpart of quantum theory known as the theory of quantum measurement) and just look at the consistency requirements on the outcome of measurements.

The \og expectation value\fg{} $\varphi(\mathbf{a})$ of an observable $\mathbf{a}$ can be considered as well as given by the average of the outcome of measurements $\mathbf{a}$  over many realisations of the system in the same state (frequentist view) or as the sum over the possible outcomes $a_i$ times the plausibility for the outcomes  in a given state (Bayesian view).
\index{Measurement}\index{Frequentist probabilities}\index{Bayesian probabilities}
In fact both point of views have to be considered, and are somehow unified, in the quantum formalism.

The linearity of the $\varphi$'s follows from (or is equivalent to) the assumptions that the observables form a linear vector space on $\mathbb{R}$.

The very important condition $$\varphi(\mathbf{a}^*)=\varphi(\mathbf{a})$$
 for any $\mathbf{a}$ follows from the assumption of reversibility. If this were not the case, there would be observables which would allow to favor one causal ordering, irrespective of the dynamics and of the states of the system.
 \index{Reversibility}

The positivity condition $\varphi(\mathbf{a}^*\mathbf{a})\ge 0$ ensures that the states have a probabilistic interpretation, so that on any state the expectation value of a positive observable is positive, and that there are no  negative probabilities, in other word it will ensure unitarity.
It is the simplest consistent positivity condition compatible with reversibility, and in fact the only possible without assuming more structure on the observables. 
Of course the condition $\varphi(\mathbf{1})=1$ is the normalisation condition for probabilities.
\index{Positivity}

\subsection{Physical observables and pure states}
Three important concepts follow from these principles.
\subsubsection{Physical (symmetric) observables:}
An observable $\mathbf{a}\in\mathcal{A}$ is symmetric (self adjoint, or self conjugate) if $\mathbf{a}^*=\mathbf{a}$.
Symmetric observables correspond to the physical observables, which are actually measurable. 
Observable such that $\mathbf{a}^*=-\mathbf{a}$ are skew-symmetric (anti-symmetric or anti conjugate). They do not correspond to physical observables but must be included in order to have a consistent algebraic formalism.
\index{Physical observable}

\subsubsection{Pure states:}
The set of states $\mathcal{E}$ is a convex subset of the set of real linear forms on $\mathcal{A}$ (the dual of  $\mathcal{A}$). Indeed if $\varphi_1$ et $\varphi_2$ are two states and $0\le x\le 1$, $\varphi=x \varphi_1+(1-x) \varphi_2$ is also a state.
This corresponds to the fact that any statistical mixture of two statistical mixtures is a statistical mixture.
Then the extremal points in  $\mathcal{E}$, i.e. the states which cannot be written as a statistical mixture of two differents states in $\mathcal{E}$, are called the pure states. Non pure states are called mixed states. If a system is in a pure state one cannot get more information from this system than what we have already.
\index{Pure state}

\subsubsection{Bounded observables}
We just need to impose two additional technical and natural assumptions:
(i)  for any observable $\mathbf{a}\neq 0$, there is a state $\varphi$ such that $\varphi(\mathbf{a}^*\mathbf{a})>0$, if this is not the case, the observable $\mathbf{a}$ is indistinguishable from the observable $0$ (which is always false); 
(ii) $\sup_{\varphi\in\mathcal{E}} \varphi(\mathbf{a}^*\mathbf{a})<\infty$, i.e. we restrict $\mathcal{A}$ to the algebra of bounded observables, this will be enough to characterize the system.
\index{Bounded observable}

\section{The C$^*$-algebra of observables}
\label{RCstarAlg}

The involution $^*$ et the existence of the states $\varphi\in \mathcal{E}$ on $\mathcal{A}$  strongly constrain  the structure of the algebra of observables and of its representations. 
Indeed this allows to associate to $\mathcal{A}$ a unique norm $\| \cdot \|$  with some specific properties. This norm makes $\mathcal{A}$ a $\mathrm{C}^*$-algebra, and more precisely a real abstract $\mathrm{C}^*$-algebra.
\index{C$^*$-algebra}
\index{Abstract C$^*$-algebra}
This structure  justifies the standard representation of quantum mechanics where pure states are elements of an Hilbert space and physical observables are  self-adjoint operators.

\subsection{The norm on observables, $\mathcal{A}$ is a Banach algebra}
\index{Norm}
Let us consider the  function $\mathbf{a}\to ||\mathbf{a}||$  from $\mathcal{A}\to\mathbb{R}^+$ defined by
\begin{equation}
\label{normstate}
||\mathbf{a}||^2=\sup_{\mathrm{states}\,\varphi\in\mathcal{E}}\varphi(\mathbf{a}^\star \mathbf{a})
\end{equation}
We have assumed that 
$||\mathbf{a}||<\infty$, $\forall a\in\mathcal{A}$ and that $||\mathbf{a}||=0\iff\mathbf{a}=0$ (this is equivalent to $\mathbf{a}\neq 0\implies\exists \varphi\in\mathcal{E}$ such that $\varphi(\mathbf{a^*a})\neq 0$). It is easy to show that  $||\cdot ||$ is a norm on $\mathcal{A}$, such that
\begin{equation}
\label{ }
||\lambda \mathbf{a}||=|\lambda|\,||\mathbf{a}||
\qquad
||\mathbf{a}+\mathbf{b}||\le ||\mathbf{a}||+||\mathbf{b}||
\qquad
||\mathbf{ab}||\le ||\mathbf{a}||\,||\mathbf{b}||
\end{equation}
If  $\mathcal{A}$ is not closed for this norm, we can take  its completion  $\overline{\mathcal{A}}$.  The algebra of observables is therefore a real Banach algebra.
\index{Banach algebra}

\paragraph{Derivation:}\ 

The first identity comes from the definition and the linearity of states.

Taking $\mathbf{c}=x\mathbf{a}+(1-x)\mathbf{b}$ and using the positivity of $\varphi(c^*c)\ge 0$ for any $x\in\mathbb{R}$ we obtain Schwartz inequality
$\varphi(a^*b)^2=\varphi(a^*b)\varphi(b^*a)\le \varphi(a^*a)\varphi(b^*b)$, $\forall\, a,\,b\in\mathbf{A}$. This implies the second inequality.
\index{Positivity}
\index{Schwartz inequality}

The third inequality comes from the fact that if
$\varphi \in\mathcal{E}$ and $b\in\mathcal{A}$ are such that $\varphi(b^*b)>0$, then $\varphi_b$ defined by
$\varphi_b(a)={\varphi(b^*ab)\over \varphi(b^*b)}$
is also a state for $\mathcal{A}$.
Then
$
||ab||^2= \sup_\varphi \varphi(b^*a^*ab)=\sup_\varphi \varphi_b(a^*a) \varphi(b^*b)\le \sup_g g(a^*a)\ \sup_\varphi \varphi(b^*b)=|| a||^2\, ||b||^2
$.

\subsection{The observables form a real $\mathrm{C}^*$-algebra}
Moreover the norm satisfies the two non-trivial properties.
\begin{equation}
\label{C*cond}
||\mathbf{a}^* \mathbf{a}||=||\mathbf{a}||^2=|| \mathbf{a}^*||^2
\end{equation}
and
\begin{equation}
\label{invcond}
\mathbf{1}+\mathbf{a}^*\mathbf{a}\quad\text{is invertible}\quad\forall\, \mathbf{a}\in \mathcal{A}
\end{equation}
These two properties are equivalent to state that  $\mathcal{A}$ is a real C$^*$-algebra.
\index{Real C$^*$-algebra}
\footnote{The  first condition on the norm and the involution \ref{C*cond} is sometimes called the C$^*$ condition. The \og C\fg{} letter in the denomination C$^*$-algebra originally comes from term  \og closed\fg{}, the closure condition specific to subalgebras of the algebra of bounded operators on a Hilbert space which defines also C$^*$-algebras. 
The second condition \ref{invcond} is specific to real algebras.}.
For a definition of real C$^*$-algebras and the properties used below see the book by Goodearl \cite{Goodearl82}.

\paragraph{Derivation:}\ 

One has $||a^*a||\le ||a||\,||a^*||$.
Schwartz inequality implies that  
$\varphi(a^*a)^2\le \varphi\left( (a^*a)^2\right) \varphi(1)$, hence $||a||^2\le ||a^*a||$. This implies (\ref{C*cond}).

To obtain (\ref{invcond}), notice that if  $1+a^*a$ is not inversible, there is a $b\neq 0$ such that $(1+a^*a)b=0$, hence $b^*b+(ab)^*(ab)=0$. Since there is a state $\varphi$ such that $\varphi(b^*b)\neq 0$, either $\varphi(b^*b)<0$ or $\varphi((ab)^*(ab)<0$, this contradicts the positivity of states.

\medskip

The full consequences will be discussed in next subsection. Before that we can introduce already the concept of spectrum of an observable.

\subsection{Spectrum of observables and results of measurements}
Here I discuss in a slightly more precise way the relationship between the spectrum of observables and results of measurements.
The spectrum 
\footnote{The exact definition is slightly different for a general real Banach algebra.}
 of an element $\mathbf{a}\in\mathcal{A}$ is defined as  $$\mathrm{Sp}^{\scriptscriptstyle{\mathbb{C}}}(\mathbf{a})=\{z\in\mathbb{C}:\ (z-\mathbf{a})\ \text{not inversible in}\ \mathcal{A}_{\mathbb{C}}\ \text{the complexified of}\ \mathcal{A}\}\ \ .$$
The spectral radius of $\mathbf{a}$ is defined as
\index{Spectrum}\index{Spectral radius}
$$
r^{\scriptscriptstyle{\mathbb{C}}}(\mathbf{a})=\sup( |z|;\ z\in
\mathrm{Sp}^{\scriptscriptstyle{\mathbb{C}}}(\mathbf{a}))
$$
For a real C$^*$-algebra it is known that the norm  $||\cdot||$ defined by \ref{normstate} is
$$ ||a||^2=r^{\scriptscriptstyle{\mathbb{C}}}(\mathbf{a}^*\mathbf{a})$$
that the spectrum of  any physical observable (symetric) is real
$$
\mathbf{a}=\mathbf{a}^*\ \implies\ \mathrm{Sp}^{\scriptscriptstyle{\mathbb{C}}}(\mathbf{a})\subset\mathbb{R}
$$
and that for any $\mathbf{a}$, the product  $\mathbf{a^*a}$ is a symmetric positive element of $\mathcal{A}$, i.e. its spectrum is real and positive
$$
\mathrm{Sp}^{\scriptscriptstyle{\mathbb{C}}}(\mathbf{a^*a})\subset\mathbb{R}^+
$$
Finally, for any (continuous) real function $F$ $\mathbb{R}\to\mathbb{R}$ and any $\mathbf{a}\in\mathcal{A}$ one can define the observable $F(\mathbf{a})$. Now consider a physical observable $\mathbf{a}$.
Physically, measuring $F(\mathbf{a})$ amounts to measure $\mathbf{a}$ and when we get the real number $A$ as a result, return $F(A$) as a result of the measure of $F(\mathbf{a})$ (this is fully consistent with the algebraic definition of $F(\mathbf{a})$ since $F(\mathbf{a})$ commutes with $\mathbf{a}$).
Then is can be shown easily that the spectrum of $F(\mathbf{a})$ is the image by $F$ of the spectrum of $\mathbf{a}$, i.e. 
$$\mathrm{Sp}^{\scriptscriptstyle{\mathbb{C}}}(F(\mathbf{a}))=F(\mathrm{Sp}^{\scriptscriptstyle{\mathbb{C}}}(\mathbf{a}))$$
In particular, assuming that the spectrum is a discrete set of points, let us  choose for $F$ the function 
$$F[\mathbf{a}]=1/(z\mathbf{1}-\mathbf{a})$$
For any state $\varphi$, the expectation value of this observable on the state $\varphi$ is 
$$E_\varphi(z)=  \varphi(1/(z\mathbf{1}-\mathbf{a})$$ and is an analytic function of $z$ away from the points of the spectrum $\mathrm{Sp}^{\scriptscriptstyle{\mathbb{C}}}(\mathbf{a}))$. (Assuming that the singularity at each $z_p$ is  a single pole) the residue of $E_\varphi(z)$ at $z_p$ is nothing but
\begin{align}
\label{Probzp}
    Res_{z_p}E_\varphi &  =\ \varphi(\delta(\mathbf{a}-z_p\mathbf{1}))  \nonumber \\
    &=\   \text{probabiliy to obtain}\ z_p\ \text{when measuring}\  \mathbf{a}\  \text{on the state}\  \varphi  
\end{align}
with $\delta(z)$ the Dirac distribution.
\index{Dirac distribution}\index{Physical observable}\index{Measurement}\index{Output of a measurement}

This implies that for any physical observable $\mathbf{a}$, its spectrum is the set of all the possible real numbers $z_p$ returned by a measurement of $\mathbf{a}$. This is one of the most important axioms of the standard formulation of quantum mechanics, and we see that it is a consequence of the axioms in this formulation. Of course the probability to get a given value $z_p$ (an element of the spectrum) depends on the state $f$ of the system, and it is given by \ref{Probzp} which is nothing but  some kind of Born rule for the abstract definiton of states.
\index{Born rule}

\subsection{Complex C$^*$-algebras}
The theory of operator algebras (C$^*$-algebras and W$^*$-algebras) and their applications (almost) exclusively deal with complex algebras, i.e. algebras over $\mathbb{C}$. 
In the case of quantum physics we shall see a bit later why  quantum (field) theories must be represented by complex C$^*$-algebras. I give here some definitions.
\index{C$^*$-algebra}\index{Abstract C$^*$-algebra}\index{Complex C$^*$-algebra}
\index{Associative algebra}\index{Involution}

Abstract complex $C^*$-algebras and complex states $\phi$ are defined as in \ref{AxiomsAlg}.
A complex C$^*$-algebra  $\mathfrak{A}$ is a complex associative involutive algebra.
The involution is now anti-linear
\index{Antilinear application}
$$(\lambda \mathbf{a}+\mu \mathbf{b})^\star = \bar\lambda \mathbf{a}^\star  + \bar\mu \mathbf{b}^\star
\qquad \lambda,\,\mu\in\mathbb{C}$$ 
$\bar z $ denotes the complex conjugate of $z$.
$\mathfrak{A}$ has a norm $\mathbf{a}\to ||\mathbf{a}||$ which still satisfy the C$^*$ condition  \ref{C*cond}, 
\begin{equation}
\label{C*condC}
||\mathbf{a}^* \mathbf{a}||=||\mathbf{a}||^2=|| \mathbf{a}^*||^2
\end{equation}
and it is closed under this norm.
The condition \ref{invcond} is not necessary any more (it follows from \ref{C*condC} for complex algebras).

The states are defined now as the complex linear forms $\phi$ on $\mathfrak{A}$ which satisfy
\begin{equation}
\label{ }
\phi(\mathbf{a}^*)=\overline{\phi(\mathbf{a})}
\qquad
\phi(\mathbf{1})=1
\qquad
\phi(\mathbf{a}^*\mathbf{a})\ge 0
\end{equation}
Any complex C$^*$-algebra $\mathfrak{A}$ can be considered as a real C$^*$-algebra $\mathcal{A}_{\mathbb{R}}$ (by considering $\imath=\sqrt{-1}$ as an element $\mathbf{i}$ of the center of $\mathcal{A}_{\mathbb{R}}$) but the reverse is not true in general. 
\index{Center of an algebra}

However if a real algebra $\mathcal{A}_{\mathbb{R}}$ has an element (denoted $\mathbf{i}$) in its center $\mathcal{C}$ that is isomorphic to $\sqrt{-1}$, i.e. $\mathbf{I}$ is such that
\begin{equation}
\label{ }
\mathbf{i}=-\mathbf{i}\ ,\  \mathbf{i}^2=-\mathbf{1}\ ,\ \ \mathbf{i a}=\mathbf{a i}\ \ \forall\ \mathbf{a}\in \mathcal{H}_{\mathbb{R}}
\end{equation}
then the algebra $\mathcal{A}_{\mathbb{R}}$ is isomorphic to a complex algebra $\mathcal{A}_{\mathbb{C}}=\mathfrak{A}$. One identifies $x\mathbf{1}+y\mathbf{i}$ with the complex scalar $z=x+\imath y$.
The conjugation $^*$ (linear on $\mathcal{A}_{\mathbb{R}}$) is now anti-linear on $\mathcal{A}_{\mathbb{C}}$.
One can associate to each  $\mathbf{a}\in\mathcal{A}_{\mathbb{R}}$ its real and imaginary part
\begin{equation}
\label{ }
\Re(\mathbf{a})={\mathbf{a}+\mathbf{a}^*\over 2}
\ ,\ \ 
\Im(\mathbf{a})=\mathbf{i}{\mathbf{a}^*-\mathbf{a}\over 2}
\end{equation}
and write in $\mathcal{A}_{\mathbb{C}}$
\begin{equation}
\label{ }
\mathbf{a}=\Re(\mathbf{a})+\imath\ \Im(\mathbf{a})
\end{equation}
To any real state (and in fact any real linear form) $\varphi_{\mathbb{R}}$ on $\mathcal{H}_{\mathbb{R}}$ one associates the complex state (the complex linear form) $\phi_{\mathbb{C}}$ on $\mathcal{A}_{\mathbb{C}}$ defined as
\begin{equation}
\label{ }
\phi_{\mathbb{C}}(\mathbf{a})=\varphi_{\mathbb{R}}(\Re(\mathbf{a}))+\imath\, \varphi_{\mathbb{R}}(\Im(\mathbf{a}))
\end{equation}
It has the expected properties for a complex state on the complex algebra $\mathfrak{A}$.

\section{The GNS construction, operators and Hilbert spaces}
General theorems show that abstract C$^*$-algebras can always be represented as algebra of operators on some Hilbert space. This is the main reason why pure states are always represented by vectors in a Hilbert space and observables as operators. Let us briefly consider how this works.

\subsection{Finite dimensional algebra of observables}
Let us first consider the case of finite dimensional algebras, which corresponds to quantum system with a finite number of independent quantum states.
This is the case considered in general in quantum information theory.
\index{Finite dimensional algebra}

If $\mathcal{A}$ is a finite dimensional real algebra, one can show by purely algebraic methods that $\mathcal{A}$ is a direct sum of matrix algebras over $\mathbb{R}$, $\mathbb{C}$ or $\mathbb{H}$ (the quaternions).
See \cite{Goodearl82} for details. The idea is to show that the C$^*$-algebra conditions  implies that the real algebra $\mathcal{A}$ is semi-simple (it cannot have a nilpotent two-sided ideal) and to use the Artin-Wedderburn theorem. 
\index{Artin-Wedderburn theorem}
\index{Quaternion}
One can even relax the positivity condition $\varphi(\mathbf{a}^*\mathbf{a})\ge 0$ for any $\mathbf{a}$ to the condition $\varphi(\mathbf{a}^2)\ge 0$ for physical observables $\mathbf{a}=\mathbf{a}^*$, which is physically somewhat more satisfactory (F. David unpublished, probably known in the math litterature...).
Thus the algebra is of the form
\begin{equation}
\label{ }
\mathcal{A}=\bigoplus_{i} M_{n_i}(K_i)\qquad K_i=\mathbb{R},\ \mathbb{C},\ \mathbb{H}
\end{equation}
The index $i$ label the components of the center of the algebra.
Any observable reads
$$
\mathbf{a}=\oplus_i \mathbf{a}_i\ ,\ \ \mathbf{a}_i  \in\mathcal{A}_i=M_{n_i}(K_i) 
$$
The multiplication corresponds to the standard matrix multiplication and the involution $^*$ to the standard conjugation (transposition, transposition+complex conjugation and transposition+conjugation respectively for real, complex and quaternionic matrices). 
One thus recovers the familiar matrix ensembles of random matrix theory.
\index{Random matrix}

Any state $\omega$ can be written as
\index{State}
$$\omega(\mathbf{a})=\sum_i p_i\,\tr(\boldsymbol{\rho}_i\mathbf{a}_i)  
\qquad p_i\ge 0,\quad \sum_i p_i=1
$$
and the $\boldsymbol{\rho}_i$'s some symmetric positive normalised matrices in each $\mathcal{A}_i$
$$
\boldsymbol{\rho}_i\in\mathcal{A}_i=M_{n_i}(K_i) \ ,\quad \boldsymbol{\rho}_i= \boldsymbol{\rho}_i^*\ ,\quad \tr(\boldsymbol{\rho}_i)=1\ ,\quad \boldsymbol{\rho}_i\ge 0
$$
The algebra of observables is indeed a subalgebra of the algebra of operators on a finite dimensional real Hilbert space $\mathcal{H}=\bigoplus_i K_i^{n_i}$ ($\mathbb{C}$ and $\mathbb{H}$ being considered as 2 dimensional and 4 dimensional real vector spaces respectively).
But it is not necessarily the whole algebra $\mathcal{L}(\mathcal{H})$.
The system corresponds to a  disjoint collection of standard quantum systems described by their Hilbert space $\mathcal{H}_i=K_i^{n_i}$ and their algebra of observables $\mathcal{A}_i$. 
This decomposition is (with a bit of abuse of language) a decomposition into superselection sectors\footnote{For many authors the term of superselection sectors is reserved to infinite dimensional algebras which do have inequivalent representations.}.
The  $\boldsymbol{\rho}_i$ are the quantum density matrices corresponding to the state. The $p_i$'s correspond to the classical probability to be in a given sector, i.e. in a state described by $(\mathcal{A}_i,\mathcal{H}_i)$. 

A pure state is (the projection onto a) single vector $|\psi_i\rangle$ in a single sector $\mathcal{H}_i$.
\index{Pure state}
Linear superpositions of pure states in different sectors $|\psi\rangle=\sum_i c_i |\psi_i\rangle$ do not make sense, since they do not belong to the representation of $\mathcal{A}$.
No observable $\mathbf{a}$ in $\mathcal{A}$ allows to discriminate between the seemingly-pure-state $|\psi\rangle\langle\psi|$ and the mixed state
$\sum_i |c_i|^2 |\psi_i\rangle\langle \psi_i|$.
Thus the different sectors can be viewed as describing completely independent systems with no quantum correlations, in other word really parallel universes with no possible interaction or communication between them.
\index{Superselection sector}

\subsection{Infinite dimensional real algebra of observables}
\index{Infinite dimensional algebra}
This result generalizes to the case of infinite dimensional real C$^*$-algebras, but it is much more difficult to prove, analysis and topology enter in the game and the fact that the algebra is closed under the norm is crucial (for a physicist this is a natural requirement).

\paragraph{Theorem (Ingelstam NN \cite{Ingelstam64,Goodearl82}):} For any real C$^*$-algebra, there exists a real Hilbert space $\mathcal{H}$ such that $\mathcal{A}$ is isomorphic to a real symmetric closed real sub-algebra of the algebra $B(\mathcal{H})$ of bounded operators on $\mathcal{H}$.
\index{Bounded operator}

\paragraph{}Now any real algebra of symmetric operators on a real Hilbert space $\mathcal{H}$ may be extended (by standard complexification) into a complex algebra of self-adjoint operator on a Hilbert space $\mathcal{H}_{\mathbb{C}}$ on $\mathbb{C}$ and thus one can reduce the study of real algebra to the study of complex algebra. In particular the theory of representations of real C$^*$-algebra is not really richer than that of complex C$^*$-algebra and mathematicians usuallyI  considers only the later case.

I will discuss later why in quantum physics one should restrict oneself also to complex algebras.
But note that in physics real (and quaternionic) algebra of observables do appear as the subalgebra of observables of some system described by a complex Hilbert space, subjected to some additional symmetry constraint (time reversal invariance $\mathbf{T}$ for real algebra, time reversal and an additional SU(2) invariance for quaternionic algebras).

\subsection{The complex case, the GNS construction}

\index{Complex C$^*$-algebra}\index{GNS construction}
Let us discuss  more the case of complex C$^*$-algebras, since their representation in term of Hilbert spaces are simpler to deal with.
The famous GNS construction (Gelfand-Naimark-Segal \cite{GeN43,Seg47-1}) allows to construct the representations of the algebra of observables in term of its pure states. It is interesting to see the basic ideas, since this allows to understand how the Hilbert space of physical pure states emerges from the abstract\footnote{in the mathematic sense: they are not defined with reference to a given representation such as operators in Hilbert space, path integrals, etc.} concepts of observables and mixed states.

The idea is somewhat simple. To every state $\phi$ we associate a representation of the algebra $\mathcal{A}$ in a Hilbert space $\mathcal{H}_\phi$. This is done as follows. 
\index{Representation}
The state
$\phi$ allows to define a bilinear form $\langle\  |\  \rangle$ on $\mathcal{A}$, considered as a vector space on $\mathbb{C}$, through
\begin{equation}
\label{ }
{\langle \mathbf{a}|\mathbf{b}\rangle}_\phi =\phi(\mathbf{a}^\star \mathbf{b})
\end{equation}
This form is $\ge 0$ but is not $>0$, since there are in general isotropic (or null)  vectors such that   $\langle \mathbf{a}|\mathbf{a}\rangle_\phi=0$. Thus $\mathcal{A}$ with this norm is a per-Hilbert space. \index{Pre-Hilbert space}
However, thanks to the C$^*$-condition, these vectors form a linear subspace $\mathcal{I}_\phi$  
of $\mathcal{A}$. 
\begin{equation}
\label{ }
\mathcal{I}_\phi=\{\mathbf{a}\in\mathcal{A}:\,\langle \mathbf{a}|\mathbf{a}\rangle_\phi=0\}
\end{equation}
Taking the (completion of the) quotient space one obtains the vector space
\begin{equation}
\label{ }
\mathcal{H}_\phi=\overline{\mathcal{A}/\mathcal{I}_\phi}
\end{equation}
When there is no ambiguity, if $\mathbf{a}$ is an element of the algebra $\mathcal{A}$ (an observable), we denote by $|a\rangle$ the corresponding vector in the Hilbert space $\mathcal{H}_\phi$, that is the equivalent class of $\mathbf{a}$ in $\mathcal{H}_\phi$ \begin{equation}
\label{ }
|a\rangle=\{\mathbf{b}\in\mathcal{A}:\ \mathbf{b}-\mathbf{a}\in\mathcal{I}_\phi\}
\end{equation}

On this space the scalar product $\langle a|b\rangle$ is $>0$ (and $\mathcal{H}_\phi$ is closed) hence $\mathcal{H}_\phi$ is a Hilbert space. \index{Hilbert space}

Now the algebra $\mathcal{A}$ acts linearily on $\mathcal{H}_\phi$ through the representation $\pi_\phi$ (in the space of bounded linear operators ${B}(\mathcal{H}_\phi$ on $\mathcal{H}_\phi$) defined as 
\begin{equation}
\label{ }
\pi_\phi(\mathbf{a})  |b\rangle=|ab\rangle
\end{equation}

Moreover, if we consider the vector $|\xi_\phi\rangle=|1\rangle\in\mathcal{H}_\phi$ (the equivalence class of the operator identity $\mathbf{1}\in\mathcal{A}$), it is of norm $1$ and such that 
\begin{equation}
\label{ }
\phi(\mathbf{a})=\langle \xi_\phi  | \pi_\phi(\mathbf{a})  | \xi_\phi \rangle
\end{equation}
(this follows basically from the definition of the representation).
Moreover this vector $\xi_\phi \rangle$ is cyclic, this means that the action of the operators on this vector allows to recover the whole Hilbert space  $\mathcal{H}_\phi$, more precisely
\begin{equation}
\label{ }
\overline{\pi_\phi(\mathcal{A})|\xi_\phi\rangle}=\mathcal{H}_\phi
\end{equation}

However this representation is in general neither faithful (different observables may be represented by the same operator, i.e. the mapping $\pi_\phi$ is not injective), nor irreducible ($\mathcal{H}_\phi$ has invariant subspaces). 
The most important result of the GNS construction is:

\paragraph{Theorem (Gelfand-Naimark 43):} The representation $\pi_\phi$ is irreducible if and only if $\phi$ is a pure state.
\index{Pure state}\index{Irreducible representation}

\paragraph{Proof:} The proof is standard and may be found in \cite{delaHarpeJones95}

\medskip
This theorem has far reaching consequences. First it implies that the algebra of observables $\mathcal{A}$ has always a faithful representation in some big Hilbert space $\mathcal{H}$. Any irreducible representation $\pi$ of $\mathcal{A}$ in some Hilbert space $\mathcal{H}$ is unitarily equivalent to the GNS representation $\pi_\phi$
constructed from a unit vector $|\xi\rangle \in\mathcal{H}$ by considering the state 
$$\phi(\mathbf{a})=\langle\xi | \pi(\mathbf{a}) | \xi\rangle  $$

\paragraph{Equivalent pure states}
Two pure states $\phi$ and $\psi$ are equivalent if their GNS representations $\pi_{\phi}$ and $\pi_\psi$ are equivalent. Then $\phi$ and $\psi$ are unitarily equivalent, i.e. there is a unitary element $\mathbf{u}$ of $\mathcal{A}$ ($\mathbf{u}^*\mathbf{u}=\mathbf{1}$) such that 
$\phi(\mathbf{a})=\psi(\mathbf{u}^*\mathbf{au})$ for any $\mathbf{a}$.
As a consequence, to this pure state $\psi$ (which is unitarily equivalent to  $\phi$) is associated a unit vector $|\psi\rangle=\pi_\phi(\mathbf{u})|\xi_\phi\rangle$ in the Hilbert space $\mathcal{H}=\mathcal{H}_\phi$, and we have the representation
\begin{equation}
\label{BornEquivSt}
\psi(\mathbf{a})=\langle\psi| A |\psi\rangle\quad,\qquad A=\pi_\phi(\mathbf{a})
\end{equation}
\index{Unitary transformation}

In other word, all pures states which are equivalent can be considered as projection operators $|\psi\rangle\langle\psi|$ on some vector $|\psi\rangle$ in the same Hilbert space $\mathcal H$. Any observable $\mathbf{a}$ is represented by some bounded operator $A$ and the expectation value of this observable in the state $\psi$ is given by the Born formula \ref{BornEquivSt}. Equivalent classes of equivalent pure states are in one to one correspondence with the irreducible representations of the algebra of observables $\mathcal{A}$.

The standard formulation of quantum mechanics in terms of operators and state vectors is thus recovered!

\section{Why complex algebras?}
\index{Complex C$^*$-algebra}

In the mathematical presentation of the formalism that I give here, real algebras play the essential role.
However it is known that quantum physics is described by complex algebras.
There are several arguments (besides the fact that it actually works) that point towards the necessity of complex algebras. 
Indeed one must take into account some essential physical features of the quantum word: time, dynamics and locality.

\subsection{Dynamics:}
Firstly, if one wants the  quantum system to have a ``classical limit'' corresponding to a classical Hamiltonian system, one would like to have conjugate observables $P_i,Q_i$ whose classical limit are conjugate coordinates $p_i$, $q_i$ with a correspondence between the quantum commutators and the classical Poisson brackets
\index{Correspondence principle}
\index{Poisson bracket}
\begin{equation}
\label{ }
[Q,P]\ \to\ \imath\, \{p,q\}
\end{equation}
Thus anti-symmetric operators must be in one to one correspondence with symmetric ones. This is possible only if the algebra of operators is a complex one, i.e. if it contains an $\mathbf{i}$ element in its center.

Another (but related) argument is that if one wants a time evolution group of inner automorphism acting on the operators (and the states), it is given by unitary evolution operators $U(t)$ of the form 
\begin{equation}
\label{ }
U(t)=\exp(t A) \quad,\qquad A=-A^*
\end{equation}
This corresponds to an Hamiltonian dynamics with a physical observable corresponding to a conserved energy (and given by a Schrödinger equation) only if the algebra is complex, so that we can write
\begin{equation}
\label{ }
A=- \imath H
\end{equation}
\index{Unitary transformation}\index{Schrödinger equation}

There has been various attempts to construct realistic quantum theories of particles or fields based on strictly real Hilbert spaces, most notably by Stueckelberg and his collaborators in the '60. 
See \cite{Stueckelberg1960}.
None of them is really satisfying.

\subsection{Locality and separability:}
Another problem with real algebras comes from the requirement of locality in quantum field theory, and to the related concept of separability of subsystems. Locality will be discussed a bit more later on. But there is already a problem with real algebras when one wants to characterize the properties of a composite system out of those of its subconstituents.
As far as I know, this was first pointed out by Araki, and recovered by various people, for instance by Wooter \cite{} (see Auletta \cite{} page 174 10.1.3).\index{Locality}\index{Separability}

Let us considers a system $\mathcal{S}$ which consists of two separated subsystem $\mathcal{S}_1$ and $\mathcal{S}_2$. 
Note that in QFT a subsystem is defined by its subalgebra of observables and of states. These are for instance the \og system\fg{} generated by the observables in two causally separated regions. 
Then the algebra of observables $\mathcal{A}$  for the total system $1+2$ is the tensor product of the two algebras $\mathcal{A}_1$ and $\mathcal{A}_2$
\begin{equation}
\label{ }
\mathcal{A}=\mathcal{A}_1\otimes\mathcal{A}_2
\end{equation}
which means that  $\mathcal{A}$ is generated by the linear combinations of the elements $\mathbf{a}$ of the  form $\mathbf{a}_1\otimes\mathbf{a}_2$.

Let us now assume that the algebras of observables $\mathcal{A}_1$ and $\mathcal{A}_2$ are (sub)algebras of the algebra of operators on some \emph{real} Hilbert spaces $\mathcal{H}_1$ and $\mathcal{H}_2$. The Hilbert space of the whole system is the tensor product $\mathcal{H}=\mathcal{H}_1\otimes\mathcal{H}_2$. Observables are represented by operators $A$, and physical (symmetric ) operators $\mathbf{a}=\mathbf{a}^*$ correspond to symmetric operators $A=A^T$.
Now it is easy to see that the physical (symetric) observables of the whole system are generated by the products of pairs of observables$(A_1,A_2)$ of the two subsystems which are of the form
\begin{equation}
\label{ }
A_1\otimes A_2\quad\text{such that\quad}\begin{cases}
      & A_1\ \text{and}\ A_2\  \text{are both symmetric,  or} \\
      & A_1\ \text{and}\ A_2\  \text{are both skew-symmetric}
\end{cases}
\end{equation}
In both case the product is symmetric, but these two cases do not generate the same observables. This is different from the case of algebras of operators on \emph{complex} Hilbert spaces, where all symmetric operators on $\mathcal{H}=\mathcal{H}_1\otimes\mathcal{H}_2$ are generated by the tensor products of the form
\begin{equation}
\label{ }
A_1\otimes A_2\quad\text{such that\ }
       A_1\ \text{and}\ A_2\  \text{are symmetric} 
\end{equation}
In other word, if a  quantum system is composed of two independent subsystems, and the physics is described by a real Hilbert space, there are physical observables of the big system  which cannot be constructed out of the physical observables of the two subsystems! 
This would turn into a problem with locality, since one could not characterize the full quantum state of a composite system by combining the results of separate independent measurements on its subparts.
Note that this is also related to the idea of quantum tomography.

\subsection{Quaternionic Hilbert spaces:}
There has been also serious attempts to build quantum theories (in particular of fields) based on quaternionic Hilbert spaces, both in the '60 and more recently by S. Adler \cite{Adler1995}. One idea  was that the SU(2) symmetry associated to quaternions could be related to the symmetries of the quark model and of some gauge interaction models. These models are also problematic. In this case there are less physical observables for a composite system that those one can naively construct out of those of the subsystems, in other word there are many non trivain constraints to be satisfied. A far as I know, no satisfying theory based on $\mathbb{H}$, consistent with locality and special relativity, has been constructed.
\index{Quaternionic Hilbert space}

\section{Superselection sectors}
\index{Superselection sector}
\subsection{Definition}

In the general infinite dimensional (complex) case the decomposition of an algebra of observables $\mathcal{A}$ along its center $Z(\mathcal{A})$ goes in a similar way as in the finite dimensional case. One can write something like
\begin{equation}
\label{ }
\mathcal{A}=\int_{c\in\mathcal{A}'} \,\mathcal{A}_c
\end{equation}
where each $\mathcal{A}_c$ is a simple C$^*$-algebra.

A very important difference with the finite dimensional case is that an infinite dimensional C$^*$-algebra $\mathcal{A}$ has in general many inequivalent irreducible representations in a Hilbert space.
Two different irreducible representations $\pi_1$ and $\pi_2$ of $\mathcal{A}$ in two subspaces $\mathcal{H}_1$ and $\mathcal{H}_2$ of a Hilbert space $\mathcal{H}$ are generated by two unitarily inequivalent pure states $\varphi_1$ and $\varphi_2$ of $\mathcal{A}$.
\index{Irreducible representation}
Each irreducible representation $\pi_i$ and the associated Hilbert space $\mathcal{H}_i$ is called a \emph{superselection sector}.
The great Hilbert space $\mathcal{H}$ generated by all the unitarily inequivalent pure states on $\mathcal{A}$ is the direct sum of all superselections sectors.
The operators in $\mathcal{A}$  do not mix the  different superselection sectors. 
It is however often very important to consider the operators in $\mathcal{B}(\mathcal{H})$ which mixes the different superselection sectors of $\mathcal{A}$ while respecting the structure of the algebra $\mathcal{A}$  (i.e. its symmetries). Such operators are called \textsl{intertwinners}.
\index{Intertwinner}

\subsection{A simple example: the particle on a circle}

One of the simplest examples is the ronrelativistic particle on a one dimensional circle. 
Let us first consider the particle on a line.
The two  conjugate operators $\mathbf{Q}$ and $\mathbf{P}$ obey the canonical commutation relations
\index{Commutation relation}
\begin{equation}
\label{ }
[\mathbf{Q},\mathbf{P}]=\imath
\end{equation}
They are unbounded, but their exponentials
\begin{equation}
\label{ }
\mathbf{U}(k)=\exp(\imath k\mathbf{Q})\quad,\qquad \mathbf{V}(x)=\exp(\imath x\mathbf{Q})
\end{equation}
generates a C$^*$-algebra. Now a famous theorem by Stone and von Neumann states that all representations of their commutation relations are unitary equivalent.
\index{Stone-von Neumann theorem}
 In other word, there is only one way to quantize the particle on the line, given by  canonical quantization and the standard representation of the operators acting on the Hilbert space of functions on $\mathbb{R}$.
\begin{equation}
\label{ }
\mathbf{Q}=x\quad,\qquad \mathbf{P}={1\over\imath}{\partial\over\partial x}
\end{equation}

Now, if the particle is on a  circle with radius 1, the position $x$ becomes an angle $\theta$  defined mod. $2\pi$. The operator $\mathbf{U}(k)$ is defined only for integer momenta $k=2\pi n$, $n\in\mathbb{Z}$. The corresponding algebra of operators has now inequivalent irreducible representations, indexed by a number $\Phi$. Each representation $\pi_{\Phi}$ corresponds to the representation of the $\mathbf{Q}$ and $\mathbf{P}$ operators acting on the Hilbert space $\mathcal{H}$ of functions $\psi(\theta)$ on the circle as
\begin{equation}
\label{ }
\mathbf{Q}=\theta\quad,\qquad \mathbf{P}={1\over\imath}{\partial\over\partial \theta}+A
\quad,\qquad A={\Phi\over 2\pi}
\end{equation}
So each superselection sector describes the quantum dynamics of a particle with unit charge $e=1$ on a circle with a magnetic flux $\Phi$.
No global unitary transformation (acting on the Hilbert space of periodic functions on the circle) can map one superselection sector onto another one. Indeed this would correspond to the unitary transformation
\begin{equation}
\label{ }
\psi(\theta)\to\psi(\theta)\, \emath^{\imath\,\theta\,\Delta A}
\end{equation}
and there is a topological obstruction if $\Delta A$ is not an integer.
Here the different superselection sectors describe different ``topological phases'' of the same quantum system.

This is of course nothing but the famous Aharonov-Bohm effect.

\subsection{General discussion}

The notion of superselection sector  was first introduced by Wick, Wightman and Wigner in 1952. They observed (and proved) that is is meaningless in a quantum field theory like QED  to speak of the superposition of two states $\psi_1$ and $\psi_2$ with integer and half integer total spin respectively, since a rotation by $2 \pi$ changes by $(-1)$ the relative phase between these two states, but does not change anything physically. This apparent paradox disappear when one realizes that this is a similar situation than above. No physical observable allows to distinguish a linear superposition of two states in different superselection sectors, such as $|1\ \mathrm{fermion}\rangle +|1\  \mathrm{boson}\rangle$ from a statistical mixture of these two states $|1\ \mathrm{fermion}\rangle  \langle 1\ \mathrm{fermion}|$  and  $|1\ \mathrm{boson}\rangle  \langle 1\ \mathrm{boson}|$.
Indeed, any operator creating or destroying just one fermion is not a physical operator (bur rather an intertwining operator), but of course an operator creating or destroying a pair of fermions (or rather a pair fermion-antifermion) is physical.

Superselection sectors are an important feature of  the mathematical formulation of quantum field theories, but they have also a physical significance. 
One encounters superselection sectors in quantum systems with an infinite number of states (non-relativistic or relativistic) as soon as 
\begin{itemize}
  \item the system may be in different phases (for instance in a statistical quantum system with spontatneous symmetry breaking);
  \index{Macroscopic phase}
  \item the system has global or local gauge symmetries and sectors with different charges $Q_a$ (abelian or non abelian);
  \index{Global charge}
  \item the system contain fermions;
  \index{Fermion}
  \item the system may exhibit different inequivalent topological sectors, this includes the simple case of a particle on a ring discussed above (the Aharonov-Bohm effect), but also gauge theories with $\theta$-vacua;
  \item{Topological sector}
  \item more generally, a given QFT for different values of couplings or masses of particle may corresponds to different superselection sectors of the same  algebra.
  \item superselection sectors have also been used to discuss measurements in quantum mechanics and the quantum-to-classical transition.
\end{itemize}
Thus one should keep in mind that the abstract algebraic formalism contains as a whole the different possible states, phases  and dynamics of a quantum system, while a given representation describes a subclass of states or of possible dynamics.

\section{von Neumann algebras}
\index{von Neumann algebra}\index{W$^*$-algebra}
A special class of C$^*$-algebras, the so-called von Neumann algebras or W$^*$-algebras, is of special interest in mathematics and for physical applications. As far as I know these were the algebras of operators originally studied by Murray and von Neumann (the ring of operators). Here I just give some definitions and some motivations, without details or applications.

\subsection{Definitions}
There are several equivalent definitions, I give here three classical definitions. The first two  refer to an explicit representation of the algebra as an algebra of operators on a Hilbert space, but the definition turns out to be independent of the representation. The third one depends only on the abstract definition of the algebra.

\paragraph{Weak closure:}
$\mathcal{A}$ a unital $^*$- sub algebra of the algebra of bounded operators $\mathcal{L}(\mathcal{H})$ on a complex Hilbert space $\mathcal{H}$ is a W$^*$-algebra iff $\mathcal{A}$ is closed under the weak topology, namely if for any sequence ${A}_n$ in $\mathcal{A}$, if the individual matrix elements $\langle x| A_n| y\rangle$ converge towards some matrix element $A_{xy}$, this defines an operator in the algebra
\begin{equation}
\label{ }
\forall\  x,y\, \in\mathcal{H}\quad \langle x| A_n|y\rangle\to A_{xy}\qquad \implies\qquad A\in\mathcal{A}
\quad\text{such that}\quad
\langle x| A|y\rangle=A_{xy}
\end{equation}
\noindent\textbf{NB:} The weak topology considered here can be replaced in the definition by stronger topologies on $\mathcal{L}(\mathcal{H})$. 
In the particular case of commutative algebras, one can show that W$^*$-algebras correspond to the set of measurable functions $L^\infty(X)$ on some measurable space $X$, while C$^*$-algebras corresponds to the set $C_0(Y)$ of continuous functions on some Hausdorff space $Y$. Thus, as advocated by A. Connes, W$^*$-algebras corresponds to non-commutative measure theory, while C$^*$-algebras to non-commutative topology theory. 

\paragraph{The bicommutant theorem:}
A famous theorem by von Neumann states that $\mathcal{A}\subset L(\mathcal{H})$ is a W$^*$-algebra iff it is a C$^*$-algebra and it is equal to its bicommutant
\begin{equation}
\label{ }
\mathcal{A}=\mathcal{A}''
\end{equation}
(the commutant $\mathcal{A}'$ of $\mathcal{A}$ is the set of operators that commute with all the elements of $\mathcal{A}$, and the bicommutant the commutant of the commutant).

\noindent\textbf{NB:} The equivalence of this ``algebraic'' definition with the previous ``topological'' or ``analytical'' one illustrate the deep relation between algebra and analysis at work in operator algebras and in quantum physics. It is often stated that this property means that a W$^*$-algebra $\mathcal{A}$ is a symmetry algebra (since  $\mathcal{A}$ is the algebras of symmetries of $\mathcal{B}=\mathcal{A}'$).
But one can also view this as the fact that a W$^*$-algebra is a ``causally complete'' algebra of observables, in analogy with the notion of causally complete domain (see the next section on algebraic quantum field theory).

\paragraph{The predual property}
It was shown by Sakai that W$^*$-algebras can also be defined as C$^*$-algebras that have a predual, i.e. when considered as a Banach vector space, $\mathcal{A}$ is the dual of another Banach vector space $\mathcal{B}$  ($\mathcal{A}=\mathcal{B}^\star$). 

\noindent\textbf{NB:} This definition is unique up to isomorphisms, since $\mathcal{B}$ can be viewed as the set of all (ultra weak) continuous linear functionals on $\mathcal{A}$, which is generated by the positive normal linear functionals on $\mathcal{A}$ (i.e. the states) with adequate topology. So W$^*$-algebras are also algebras with special properties for their states.

\subsection{Classification of factors}
\index{Factor}
A word on the famous classification of factors.
Factors are W$^*$-algebras with trivial center $C=\mathbb{C}$ and any W$^*$-algebra can be written as an integral sum over factors.
W$^*$-algebra have the property that they are entirely determined by their projectors elements (a projector is such that $\mathbf{a}=\mathbf{a}^*=\mathbf{a}^2$, and corresponds to orthogonal projections onto closed subspaces $E$ of $\mathcal{H}$).
The famous classification result of Murray and von Neumann states that there are basically three different classes of factors, depending on the properties of the projectors and on the existence of a trace.

\paragraph{Type I:} A factor is of type I if there is a minimal projector $E$ such that there is no other projector $F$ with $0 < F < E$.
Type I factors always corresponds to the whole algebra of bounded operators $L(\mathcal{H})$ on some (separable) Hilbert space $\mathcal{H}$. Minimal projector are projectors on pure states (vectors in $\mathcal{H}$). This is the case usually considered by ``ordinary physicists''. They are denoted I$_n$ if $\dim(\mathcal{H})=n$ (matrix algebra)  and I$_\infty$ if $\dim(\mathcal{H})=\infty$.

\paragraph{Type II:} Type II factors have no minimal projectors, but finite projectors, i.e. any projector $E$ can be decomposed into $E=F+G$ where $E$, $F$ and $G$ are equivalent projectors. 
The type II$_1$ hyper finitefactor has a unique finite trace $\omega$ (a state such that $\omega(\mathbf{1})=1$ and $\omega(\mathbf{aa}^*)=\omega(\mathbf{a}^*\mathbf{a})$), while type $\mathrm{II}_{\infty}=\mathrm{II}_{1}\otimes \mathrm{I}_{\infty}$. They play an important role in non-relativistic statistical mechanics of infinite systems, the mathematics of integrable systems and CFT.

\paragraph{Type III:} This is the most general class. Type III factors have no minimal projectors and no trace. They are more complicated. Their classification was achieved by A. Connes. These are the general algebras one must consider in relativistic quantum field theories.

\subsection{The Tomita-Takesaki theory}
\index{Tomita-Takesaki theory}
Let me say a few words on a important feature of von Neumann algebras, which states that there is a natural ``dynamical flow'' on these algebras induced by the states.
This will be very sketchy and naive.
We have seen that in ``standard quantum mechanics'' (corresponding to a type I factor), the evolution operator $U(t)=\exp(-\imath t H)$ is well defined in the lower half plane $\Im(t)\le 0$.

This correspondence ``state $\leftrightarrow$ dynamics'' can be generalized to any von Neumann algebra, even when the concept of density matrix and trace is not valid any more.
Tomita and Takesaki showed that to any state $\phi$ on $\mathcal{A}$ (through the GNS construction $\phi(\mathbf{a})=\langle \Omega|\mathbf{a}\Omega\rangle$ where $\Omega$ is a separating cyclic vector of the Hilbert space $\mathcal{H}$) one can associate a one parameter family of modular automorphisms $\sigma_t^\Phi$: $\mathcal{A}\to\mathcal{A}$, such that
$\sigma_t^\Phi(\mathbf{a})=\Delta^{\imath t} \mathbf{a}\Delta^{-\imath t}$, where $\Delta$ is positive selfadjoint modular operator in $\mathcal{A}$. 
This group depends on the choice of the state $\phi$ only up to inner automorphisms, i.e. unitary transformations $u_t$ such that
$\sigma_t^\Psi(\mathbf{a})= u_t \sigma_t^\Phi(\mathbf{a}) u_t^{-1}$ , with the 1-cocycle property $u_{s+t}=u_s\sigma_s(u_t)$.

As advocated by A. Connes, this means that there is a ``global dynamical flow'' acting on the von Neumann algebra $\mathcal{A}$ (modulo unitaries reflecting the choice of initial state).
This Tomita-Takesaki theory is a very important tool in the mathematical theory of operator algebras.
It has been speculated by some authors that there is a deep connection between statistics and time (the so called ``thermal time hypothesis''), with consequences in quantum gravity. Without going to this point, this comforts the point of view that operator algebras have a strong link with causality.

\section{Locality and algebraic quantum field theory}

Up to now I have not really discussed the concepts of time and of  dynamics, and  the role of  relativistic invariance and locality in the quantum formalism.
One should remember that the concepts of causality and of reversibility are already incorporated within the formalism from the start.

It is not really meaningful to discuss these issues if not in a fully relativistic framework.
This is the object of algebraic and axiomatic quantum field theory. 
Since I am not a specialist I give only a very crude and very succinct account of this formalism and refer to the excellent book by R. Haag \cite{Haag96} for all the details and the mathematical concepts. 

\subsection{Algebraic quantum field theory in a dash}
\label{ssAQFTdsh}
 \index{Algebraic quantum field theory}
In order to make the quantum formalism compatible with special relativity, one needs three things.

\paragraph{Locality:}
\index{Locality}
Firstly the observables must be built on the local observables, i.e. the observables attached to bounded domains $\mathcal{O}$ of Minkovski space-time $M=\mathbb{R}^{1,d-1}$. 
They corresponds to measurements made by actions on the system in a finite region of  space, during a finite interval of time.
Therefore one associate to each  domain $\mathcal{O}\subset M$ a subalgebra $\mathcal{A}(\mathcal{O})$ of the algebra of observables.
\begin{equation}
\label{ }
\mathcal{O}\to\mathcal{A}(\mathcal{O})\subset\mathcal{A}
\end{equation}
This algebra is such that is
\begin{equation}
\label{ }
\mathcal{A}(\mathcal{O}_1\cup\mathcal{O}_2)=\mathcal{A}(\mathcal{O}_1)\vee\mathcal{A}(\mathcal{O}_2)
\end{equation}
where $\vee$ means the union of the two subalgebras (the intersection of all subalgebras containing $\mathcal{A}(\mathcal{O}_1)$ and $\mathcal{A}(\mathcal{O}_2)$.
\begin{figure}[h]
\begin{center}
\includegraphics[width=2in]{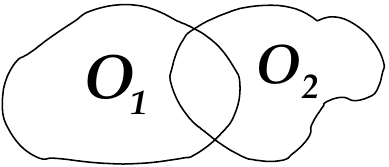}
\caption{The union of two domains}
\label{ }
\end{center}
\end{figure}
\par\noindent Note that this implies
\begin{equation}
\label{ }
\mathcal{O}_1\subset \mathcal{O}_2\ \implies\ \ \mathcal{A}(\mathcal{O}_1)\subset \mathcal{A}(\mathcal{O}_2)
\end{equation}
The local operators are obtained by taking the limit when a domain reduces to a point (this is not a precise or rigorous definition, in particular in view of the UV divergences of QFT and the renormalization problems).
\medskip\par\noindent
$\textdbend$ Caution, the observables of two disjoint domains are not independent if these domains are not causally independent (see below) since they can be related by dynamical/causal evolution.

\begin{figure}[h]
\begin{center}
\includegraphics[width=3in]{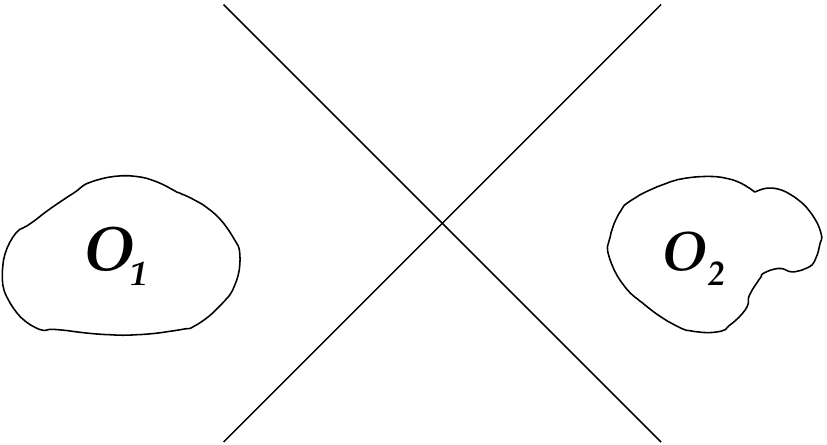}
\caption{For two causally separated domains, the associated observables must commute}
\label{ }
\end{center}
\end{figure}

\paragraph{Causality:}
\index{Causality}
Secondly causality and locality must be respected, this implies that physical local observables which are causally independent must always commute. Indeed the result of measurements of causally independent observables is always independent of the order in which they are performed, independently of the state of the system.
Were this not the case, the observables would not be independent and through some measurement process information could be manipulated and transported at a faster than light pace.
If $\mathcal{O}_1$ and $\mathcal{O}_2$ are causally separated (i.e. any $x_1-x_2$, $x_1\in\mathcal{O}_1$, $x_2\in\mathcal{O}_2$ is space-like)) then  any pair of  operators $A_1$ and $A_2$ respectively in $\mathcal{A}(\mathcal{O}_1)$ and $\mathcal{A}(\mathcal{O}_2)$ commutes
\begin{equation}
\label{ }
\mathcal{O}_1\ \raisebox{1.ex}{$\bigvee$} \hskip -.77em \raisebox{-.8ex}{$ \bigwedge$}  \    \mathcal{O}_2\ ,\quad
A_1\in\mathcal{A}(\mathcal{O}_1) \ ,\quad
A_2\in\mathcal{A}(\mathcal{O}_2) \quad
\implies \quad
[A_1,A_2]=0
\end{equation}
This is the crucial requirement to enforce locality in the quantum theory.

\noindent{NB:} As already discussed, in theories with fermion, fermionic field operators like $\psi$ and $\bar\psi$ are not physical operators, since they intertwin different sectors (the bosonic and the fermionic one) and hence the anticommutation of fermionic operators does not contradict the above rule.

\paragraph{Causal completion:} 
\index{Causal completion}
One needs also to assume causal completion, i.e.
\begin{equation}
\label{ }
\mathcal{A}(\mathcal{O})=\mathcal{A}(\widehat{\mathcal{O}})
\end{equation}
where the domain $\widehat{\mathcal{O}}$ is the causal completion of the domain $\mathcal{O}$ ($\widehat{\mathcal{O}}$ is defined as the set of points $\mathcal{O}''$ which are causally separated from the points of $\mathcal{O}'$, the set of points causally separated from  the points of $\mathcal{O}$, see fig.\ref{f:CausComp} for a self explanating illustration).

\begin{figure}[h]
\begin{center}
\includegraphics[width=1.5in]{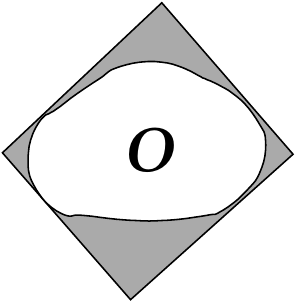}
\caption{A domain $\mathcal{O}$ and its causal completion $\widehat{\mathcal{O}}$ (in gray)}
\label{f:CausComp}
\end{center}
\end{figure}

\par\noindent This implies in particular that  the whole algebra $\mathcal{A}$ is the (inductive) limit of the subalgebras generated by an increasing sequence of bounded domains whose union is the whole Minkovski space
\begin{equation}
\label{ }
\mathcal{O}_i\subset \mathcal{O}_j\ \text{if}\ i<j\ \text{and}\ \ \bigcup_i\mathcal{O}_i=\mathbb{M}^4\quad\implies\qquad \lim_{\longrightarrow}\mathcal{A}(\mathcal{O}_i)=\mathcal{A}
\end{equation}
and also that it is equal to the algebra associated to ``time slices'' with arbitrary small time width.
\begin{equation}
\label{ }
\mathcal{S}_\epsilon=\{\mathbf{x}=(t,\vec x):\ t_0<t<T_0+\epsilon\}
\end{equation}
\begin{figure}[h]
\begin{center}
\includegraphics[width=5in]{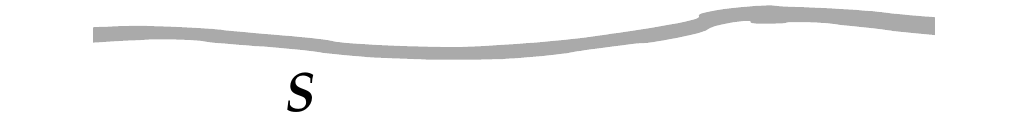}
\caption{An arbitrary thin space-like slice of space-time is enough to generate the algebra of observables $\mathcal{A}$}
\label{f:CausalSlice}
\end{center}
\end{figure}

This indicates also why one should concentrate on von Neumann algebras.
The set of local subalgebras $\mathcal{L}= \{\mathcal{A}(\mathcal{O}):\ \mathcal{O}\ \text{subdomains of}\ M \}$ form an orthocomplemented lattice with interesting properties.

\paragraph{Poincaré invariance:}
\index{Poincaré invariance}
The Poincaré group $\mathfrak{P}(1,d-1)=\mathbb{R}^{1,d-1}\rtimes \mathrm{O}(1,d-1)$ must act on the space of local observables, so that it corresponds to a symmetry of the theory (the theory must be covariant under translations in space and time and Lorentz transformations).
When $\mathcal{A}$ is represented as an algebra of operators on a Hilbert space, the action is usually represented by unitary\footnote{Unitary with respect to the real algebra structure, i.e. unitary or antiunitary w.r.t. the complex algebra structure.} transformations $U(a,\Lambda)$ ($a$ being a translation and $\Lambda$ a Lorentz transformation).
This implies in particular that the algebra associated to the image of a domain by a Poincaré transformation is the image of the algebra under the action of the Poincaré transformation.
\begin{equation}
\label{ }
U(a,\Lambda)\mathcal{A}(\mathcal{O})U^{-1}(a,\Lambda)=\mathcal{A}(\Lambda\mathcal{O}+a)
\end{equation}
\begin{figure}[h]
\begin{center}
\includegraphics[width=3in]{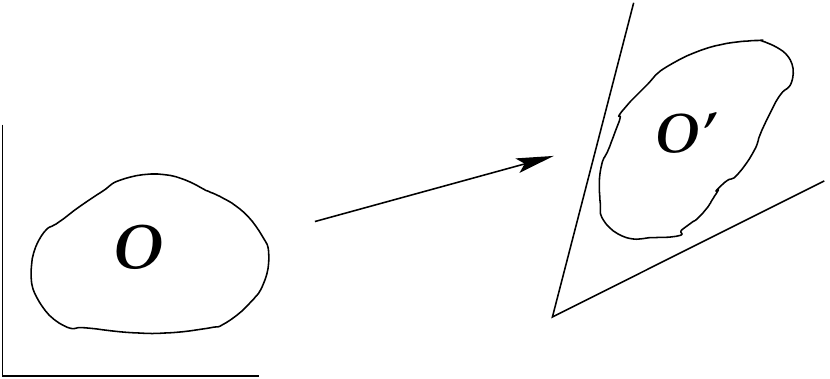}
\caption{The Poincaré group acts on the domains and on the associated algebras}
\label{ }
\end{center}
\end{figure}

The generator of time translations will be the Hamiltonian $P_0=H$, and time translations acting on observables corresponds to the dynamical evolution of the system in the Heisenberg picture, in a given Lorentzian reference frame.

\paragraph{The vacuum state:}
\index{Vacuum}
Finally one needs to assume the existence (and the uniqueness, in the absence of spontaneous symmetry breaking) of a special state,  the vacuum state $|\Omega\rangle$. 
The vacuum state must be invariant under the action of the Poincaré transformations, i.e. 
$U(a,\Lambda)|\Omega\rangle = |\Omega\rangle$.
 At least in the vacuum sector, the spectrum of $\mathbf{P}=(E,\vec P)$ (the generators of time and space translations) must lie in the future cone.
 \begin{equation}
\label{ }
E^2-\vec p^2>0\quad,\qquad E>0
\end{equation}
 This is required since the dynamics of the quantum states must respect causality. In particular, the condition $E>0$ (positivity of the energy) implies that dynamical evolution is compatible with the modular automorphisms on the algebra of observables constructed by the Tomita-Takesaki theory.

\subsection{Axiomatic QFT}
\index{Axiomatic quantum field theory}

\subsubsection{Wightman axioms}
\index{Wightman axioms}

One approach to implement the program of algebraic  local quantum field theory is the so-called axiomatic field theory framework (Wightman \& G\aa rding).
Actually the axiomatic field theory program was started before the algebraic one.
In this formalism, besides the axioms of local, AQFT, the local operators are realized as ``local fields''. 
These local fields $\Phi$ are represented as distributions (over space-time $M$) whose values, when applied to some $C^\infty$ test function with compact support $f$ (typically inside some $\mathcal{O}$) are operators $\mathbf{a}=\langle \Phi{\cdot} f\rangle$.  Local fields are thus ``operator valued distributions''. 
\index{Local field}
They must satisfy the Wightman's axioms (see Streater and Wightman's book \cite{StreaterWightmanBook}  and R. Haag's book, again), which enforce causality, locality, Poincaré covariance, existence (and uniqueness) of the vacuum (and eventually in addition asymptotic completeness, i.e. existence of a scattering S-matrix).

\subsubsection{CPT and spin-statistics theorems}
\index{CPT theorem}
\index{Spin-statistics theorem}

The axiomatic framework is very important for the definition of quantum theories.
It is within this formalism that one can derive the general and fundamental properties of relativistic quantum theories
\begin{itemize}
  \item Reconstruction theorem: reconstruction of the Hilbert space of  states from the vacuum expectation values of product of local fields (the Wightman functions, or correlation functions),
  \item Derivation of the analyticity properties  of the correlation functions with respect to space-time $\mathbf{x}=(t,\vec x)$ and impulsion $\mathbf{p}=(E,\vec p)$ variables,
  \item Analyticity of the S matrix (an essential tool),
  \item The CPT theorem: locality, Lorentz invariance and unitarity imply CPT invariance,
  \item The spin statistics theorem,
  \item Definition of quantum field theories in Euclidean time (Osterwalder-Schrader axioms) and rigorous formulation of the mapping between Euclidean theories 
  and 
   Lorentzian quantum theories.
   \index{Euclidean space}\index{Minkowski space}
\end{itemize}

\section{Discussion}
I gave here a short introduction to the algebraic formulation of quantum mechanics and quantum field theory.
I did not aim at mathematical rigor nor completeness. 
I have not mentioned recent developments  and applications in the direction of gauge theories, of two dimensional conformal field theories, of quantum field theory in non trivial (but classical) gravitational background.

However I hope to have conveyed the idea that the ``canonical structure of quantum mechanics'' -- complex Hilbert space of states, algebra of operators, Born rule for probabilities -- is quite natural and is a representation of an underlying more abstract structure: a real algebra of observables  $+$ states, consistent with the physical concepts of causality, reversibility and locality/separability.

\chapter{The quantum logic formalism}
\label{s:QuanLog}
\section{Introduction:  measurements as logic}
\index{Quantum logic}
The quantum logic formalism is another interesting, albeit more abstract, way to formulate quantum physics.
The bonus of  this approach is that one does not have to assume that the set of observables of a physical system is embodied with the algebraic structure of an associative unital algebra. 
As we have discussed in the previous section, the fact that one can ``add'' and ``multiply'' observables is already a highly non trivial assumption.
This algebraic structure is natural in classical physics since observables form a commutative algebra, coming from the action of  adding and multiplying results of different measurements. 
In quantum physics this is not equivalent, and we have seen for instance that the GNS construction relates the algebra structure of observables to the Hilbert space structure of pure states. 
In particular to the superposition principle for states comes from the addition law for observables.
In the quantum logic formulations  this algebraic structure itself comes out somehow naturally from the symmetries of the measurement operations considered on the physical system.

The ``quantum logic'' approach was initiated by G. Birkhoff \footnote{An eminent mathematician, not to be confused with his father, the famous G. D. Birkhoff of the ergodic theorem} and J. von Neumann (again!) in \cite{BirkVNeumann36}.
\index{von Neumann J.}
It was then (slowly) developped, notably  by physicists like G. Mackey \cite{Mackey63}, J.~M. Jauch \cite{Jauch68} and C. Piron \cite{Piron64,Piron76}, and mathematicians like Varadarajan\cite{Varadarajan1985}.
A good reference on the subject (not very recent but very valuable) is the book by E. Beltrametti and G. Cassinelli \cite{BeltCassi81}.

The terminology ``quantum logic'' for this approach is historical and is perhaps  not fully adequate, since it does not mean that a new kind of logic is necessary to understand quantum physics. It is in fact not a ``logic'' in the mathematical sense, and it relies on the standard logics used in mathematics and exact sciences. It could rather be called ``quantum propositional calculus'' or ``quantum propositional geometry'', where the term ``proposition'' is to be understood as ``test'' or ``projective measurement'' on a quantum system.
The mathematics underlying the quantum logic formalism have  applications in various areas of mathematics, logic and computer sciences. 
The quantum logic approaches do not form a unified precise and consistent framework like algebraic quantum field theory. It has several variants, most of them insisting on propositions, but some older one relying more on the concept of states (the so called convex set approaches). Some recent formulations of quantum physics related to quantum logic have some grandiose categorial formulations.

In this course  I shall give a short, partial presentation of this approach, from a personal point of view\footnote{with the usual reservation on the lecturer's qualifications}. 
I shall try to stress where the physical concepts of causality, reversibility and locality play a role, in parallel to what I tried to do for the algebraic formalism.
My main reference and source of understanding is the review by Beltrametti and Cassinelli \cite{BeltCassi81}.

The idea at the root of this approach goes back to J. von Neumann's book \cite{vonNeumann32,vonNeumann32G}.
It starts from the observation that the observables given by projectors, i.e. operators $\mathbf{P}$ such that $\mathbf{P}^2=\mathbf{P}=\mathbf{P}^\dagger$, correspond to propositions with YES or NO (i.e. TRUE or FALSE) outcome in a logical system. 
\index{Projection operator}
An orthogonal projector $\mathbf{P}$ onto a linear subspace $P\subset\mathcal{H}$ is indeed the operator associated to an observable that can take only the values $1$ (and always $1$ if the state $\psi\in P$ is in the subspace $P$) or $0$ (and always $0$ if the state $\psi\in P^\perp$ belongs to the orthogonal subspace to $P$).
Thus we can consider that measuring the observable $\mathbf{P}$ is equivalent to perform a test on the system, or to check the validity of a logical proposition $\textbf{p} $ on the system. 
\begin{equation}
\label{ }
\mathbf{P}\ =\ \text{orthogonal projector onto}\ P\quad \leftrightarrow\quad\text{proposition}\ \mathbf{p}
\end{equation}
If the result is $1$ the proposition \textbf{p} is found to be TRUE, and if the result is $0$ the proposition \textbf{p} is found to be FALSE.
\begin{equation}
\label{ }
\langle\psi|\mathbf{P}|\psi\rangle=1 \implies  \textbf{p}\ \text{always TRUE on}\ |\psi\rangle
\end{equation}
The projector $\mathbf{1}-\mathbf{P}$ onto the orthogonal subspace $P^\perp$ is associated to the proposition  $\textbf{not\ p}$, meaning usually that $\textbf{p}$ is false (assuming the law of excluded middle)
\begin{equation}
\label{ }
\langle\psi|\mathbf{P}|\psi\rangle=0 \implies  \textbf{p}\ \text{always FALSE on}\ |\psi\rangle
\end{equation}
so that
\begin{equation}
\label{ }
\mathbf{1-P}\ =\ \text{orthogonal projector onto}\ P^\perp\quad \leftrightarrow\quad\text{proposition}\ \mathbf{not\ p}
\end{equation}

In classical logic the negation $\textbf{not}$ is denoted in various ways
\begin{equation}
\label{ }
\textbf{not}\ \mathbf{a}\ = \ {\boldsymbol{\neg}}\mathbf{a} \ ,\ \mathbf{a}'\ ,\ \bar{\mathbf{a}}\ ,\ \tilde{\mathbf{a}}\ ,\ \sim\! \mathbf{a}
\end{equation}
I shall use the first two notations.

Now if two projectors $\mathbf{A}$ and $\mathbf{B}$ (on two subspaces $A$ and $B$) commute, they correspond to classically compatible observables $A$ and $B$ (which can be measured independently), and to a pair of propositions $\mathbf{a}$ and $\mathbf{b}$ of standard logic.
The projector $\mathbf{C}=\mathbf{AB}=\mathbf{BA}$ on the intersection of the two subspaces $C=A\cap B$ corresponds to the proposition $\mathbf{c}= ``\mathbf{a} \ \text{and}\ \mathbf{b}''=\mathbf{a}\wedge \mathbf{b}$.
Similarly the projector $\mathbf{D}$ on the linear sum of the two subspaces  $D=A+ B$ corresponds to the proposition $\mathbf{d}$=``$\mathbf{a}$ or ${\mathbf{b}}$'' =$\mathbf{a}\vee \mathbf{b}$.
\begin{equation}
\label{AandorB}
A\cap B\quad\leftrightarrow \quad\mathbf{a}\wedge\mathbf{b}=\mathbf{a\,and\,b}
\qquad,\qquad
A+ B\quad\leftrightarrow \quad\mathbf{a}\vee\mathbf{b}=\mathbf{a\,or\,b}
\end{equation}
Finally the fact that for subspaces $A\subset B$, i.e. for projectors $\mathbf{AB}=\mathbf{BA}=\mathbf{A}$, is equivalent to state that $\mathbf{a}$ implies $\mathbf{b}$
\begin{equation}
\label{AimpliesB}
A\subset B\quad\leftrightarrow\quad \mathbf{a}\implies\mathbf{b}
\end{equation}

This is easily  extended to a general (possibly infinite) set of \emph{commuting} projectors.
Such a set generates a commuting algebra of observables $\mathcal{A}$, which corresponds to the algebra of functions on some classical space $X$. 
The set of corresponding subspaces, with the operations of  linear sum, intersection and orthocomplementation ($+,\cap,\perp$), is isomorphic to a Boolean algebra of propositions with ($\vee,\wedge,\neg$), or to the algebra of characteristic functions on subsets of $X$.
\index{Boolean algebra} 
Indeed, this is just a reformulation of ``ordinary logic''\footnote{In a very loose sense, I am not discussing mathematical logic theory.} where characteristics functions  of measurable sets  (in a Borel $\sigma$-algebra over some set $X$) can be viewed as logical propositions. 
Classically all the observables of some classical system  (measurable functions over its phase space $\Omega$) can be constructed out of the classical propositions on the system (the characteristic functions of measurable subsets of $\Omega$) . 
\index{Boolean logic}

\begin{figure}[h]
\begin{center}
\includegraphics[width=2in]{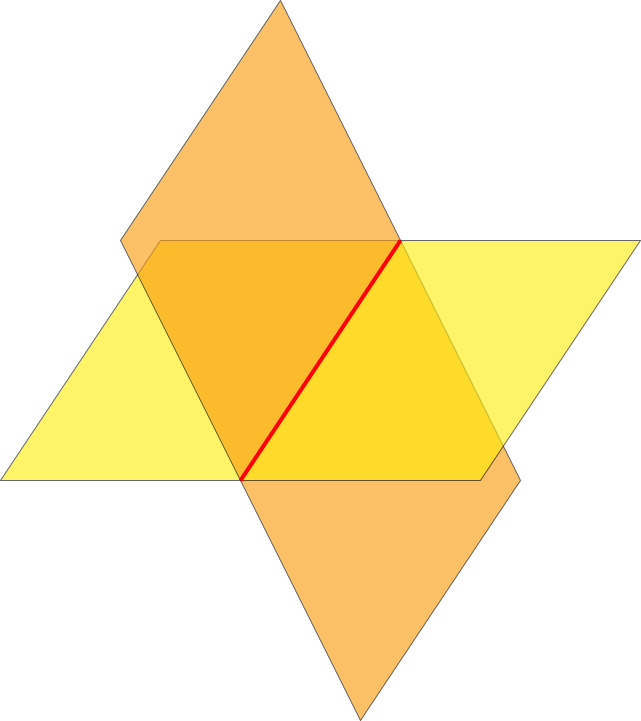}
\qquad
\includegraphics[width=1.5in]{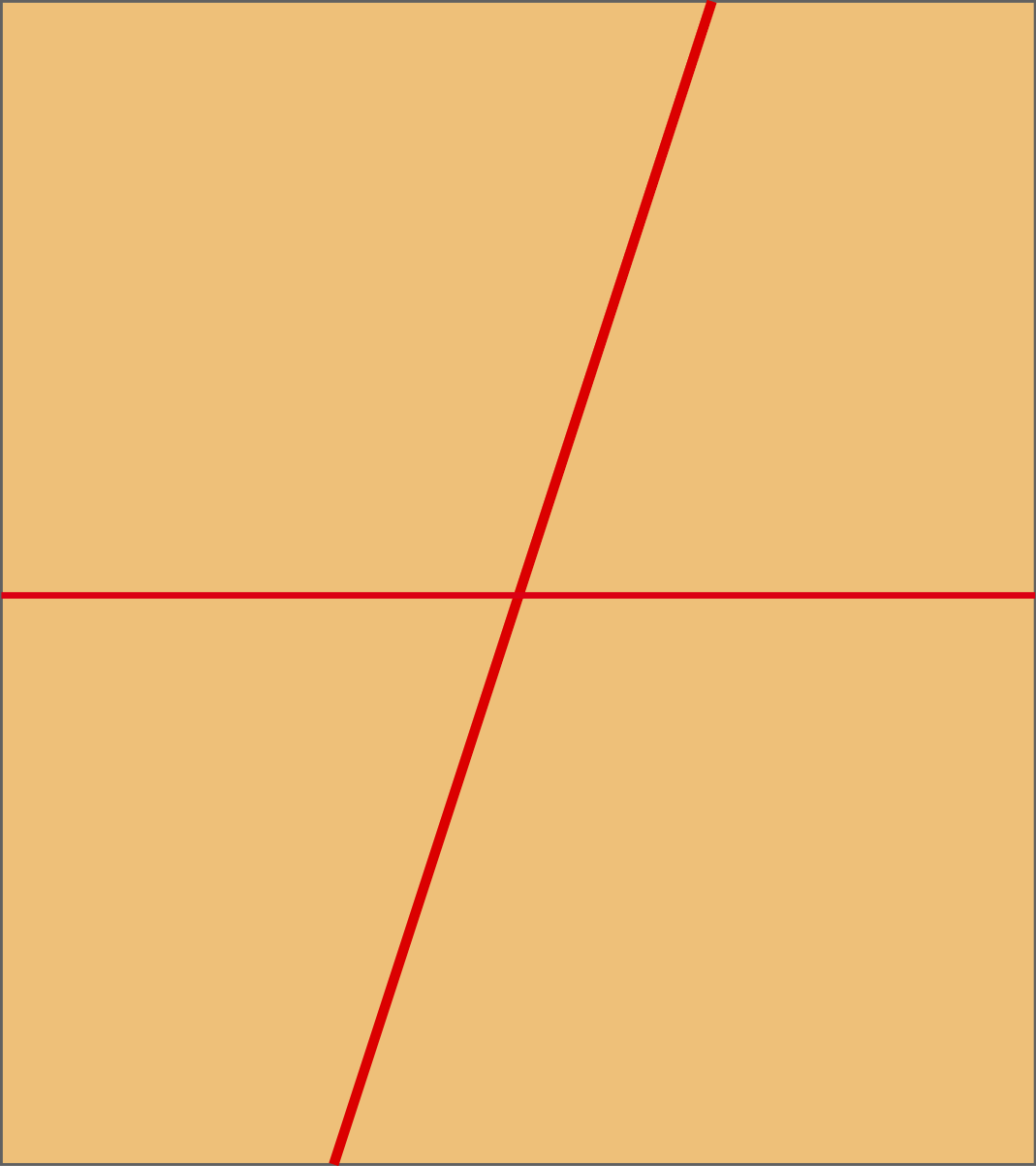}
\caption{The $\wedge$ as intersection and the $\vee$ as linear sum of subspaces in quantum logic }
\label{ }
\end{center}
\end{figure}

In quantum mechanics all physical observables can be constructed out of projectors. 
For general, non necessarily commuting projectors $\mathbf{A}$ and $\mathbf{B}$ on subspaces $A$ and $B$ one still associate propositions $\mathbf{a}$ and $\mathbf{b}$. The negation $\neg\mathbf{a}$, the ``and''  (or ``meet'') $\mathbf{a}\wedge\mathbf{b}$ and the ``or'' (or ``join'') $\mathbf{a}\vee\mathbf{b}$ are still defined by the geometrical operations $\perp$, $\cap$ and $+$ on subspaces given by
\ref{AandorB}. The ``implies'' $\implies$ is also defined by the $\subset$ as in \ref{AimpliesB}

However the fact that in a Hilbert space projectors do not necessarily commute implies that the standard distributivity law of propositions
\index{Distributivity}
\begin{equation}
\label{ }
A\wedge(B\vee C)=(A \wedge B)\vee(A\wedge C)\qquad\vee=\text{or}\qquad\wedge=\text{and}
\end{equation}
does not hold. It is replaced by the weaker condition ($A$, $B$, $C$ are the linear subspaces associated to the projectors $\mathbf{A}$, $\mathbf{B}$, $\mathbf{C}$)
\begin{equation}
\label{ }
A\cap(B + C)\supset ((A \cap B)+(A\cap C))
\end{equation}
which corresponds in terms of propositions (projectors) to
\begin{equation}
\label{ }
(\mathbf{a} \wedge \mathbf{b})\vee(\mathbf{a}\wedge \mathbf{c})\implies \mathbf{a}\wedge(\mathbf{b}\vee \mathbf{c})
\end{equation}
or equivalently
\begin{equation}
\label{ }
\mathbf{a}\vee(\mathbf{b}\wedge \mathbf{c})
\implies 
(\mathbf{a} \vee \mathbf{b})\wedge(\mathbf{a}\vee \mathbf{c})
\end{equation}

A simple example is depicted on fig.~\ref{nondistplane}. The vector space $V$ in the plane (dim=2) and the subspaces $A$, $B$ and $C$ are three different coplanar lines (dim=1). $B+C=V$, hence $A\cap(B+C)=A\cap V=A$, while $A\cap B=A\cap C =\{0\}$; hence $A\cap B+A\cap C =\{0\}$.

\begin{figure}[h]
\begin{center}
\includegraphics[width=2in]{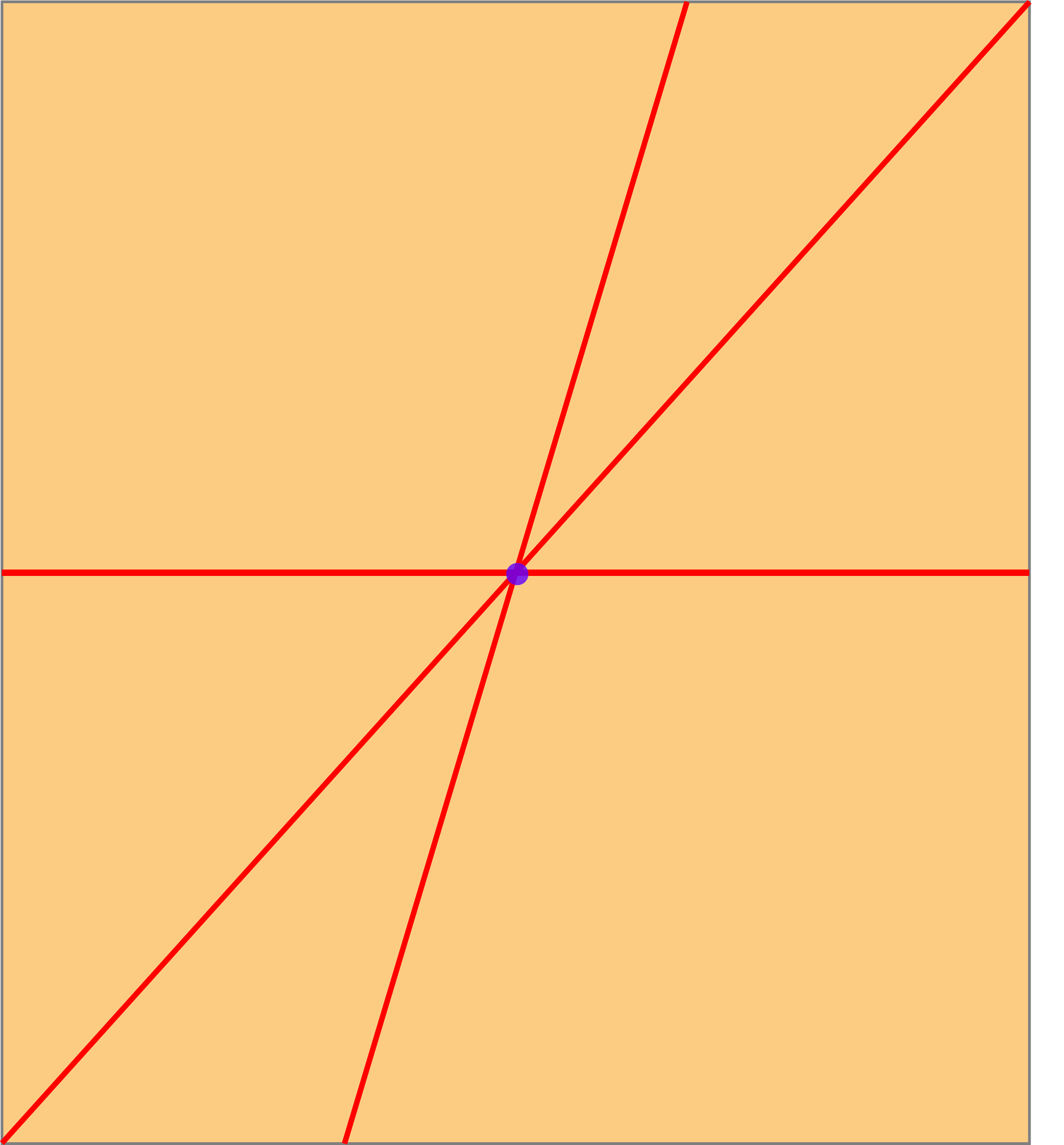}
\caption{A simple example of non-distributivity}
\label{nondistplane}
\end{center}
\end{figure}

Therefore the set of projectors on a Hilbert space do not generate a Boolean algebra. 
The purpose of the quantum logic approach is to try to understand what are the minimal set of consistency requirements on such propositions/measurements, based on logical consistency (assuming that internal consistency has something to do with the physical world), and on physical requirements (in particular causality, reversibility and locality) and what are the consequences for the formulation of physical laws.
I discuss the conservative approach where one does not try to use a non-classical logic (whatever it means) but discuss in a classical logic framework the statements which can be made on quantum systems.

There are many variants of the formalism: some insist on the concept and the properties of the propositions (the test), some others on those of the states (the probabilities). They are often equivalent.
Here I present a version based primarily on the propositions.

\section{A presentation of the principles}

\subsection{Projective measurements as propositions}

\index{Ideal measurement}
As explained above, in the standard formulation of quantum mechanics, projectors are associated to ``ideal'' projective measurements  (``projective measurement ``of the first kind'', or ``non-demolition'' projective measurements). The fundamental property of such measurements is that if the system is already in an eigenstate of the projector, for instance $\mathbf{P}|\psi\rangle=|\psi\rangle$, then after  measurement  the state of the system is unchanged. This means that successive measurements of $P$ give always the same result ($1$ or $\TRUE$).
Without going into  a discussion of measurements in quantum physics, let me stress that this is of course an idealisation of actual measurements. In general physical measurements are not ideal measurements, they may change the state of the system, while gaining some information on the system we in general loose some other information, they may and in general do destroy part or the whole of the system studied. Such general processes may be described by the formalism of POVM's (Projective Operator Valued Measures).

In the following presentation, I assume that such ideal repeatable measurements are (in principle ) possible for all the observable properties of a quantum system. The formalism here tries to guess what is a natural and minimal set of physically reasonable and logically consistent axioms for such measurements.

\subsection{Causality, POSET's and the lattice of propositions}
One starts from a set of propositions or tests $\mathcal{L}$ (associated to ideal measurements of the first kind on a physical system)
and from a set of states $\mathcal{E}$ (in a similar sense as in the algebraic formulation, to be made more precise along the discussion).
On a given state $\varphi$ the test (measurement) of the proposition $a$ can give TRUE (i.e. YES or $1$) or FALSE (NO or $0$).
It gives TRUE with some probability. In this case one has extracted information on the system, which is now (considered to be) in a state $\varphi_a$.

I note $\varphi(a)$
the probability that $a$ is found $\TRUE$, assuming that the system was in state $\varphi$ before the test.
I shall not discuss at that stage what I mean exactly by probability (see the previous discussions).
\\

\subsubsection{Causal order relation:}
The first ingredient is to assume that there an order relation $a\preceq b$ between propositions. Here it will be defined by the causal relation
\begin{equation}
\label{preceqdef}
a\preceq b\quad\iff\quad\text{for any state $\phi$, if $a$ is found true, then $b$ will be found true}
\end{equation}
Note that this definition is  causal (or dynamical) from the start, as to be expected in quantum physics.
It is equivalent to 
\begin{equation}
\label{preceqstate}
a\preceq b\quad\iff\quad\forall\ \phi\ ,\ \ \phi_a(b)=1
\end{equation}

One assumes that this causal relation has the usual properties of a partial order relation.
This amounts to enforce relations between states and propostions.
First one must have:
\begin{equation}
\label{aperceqa}
a\preceq a
\end{equation}
This means that if $a$ has been found true, the system is now in a state such that $a$ will always be found true.
Second one assumes also that
\begin{equation}
\label{preceqeq}
a\preceq b\ \text{and}\ b\preceq c\ \implies \ a\preceq c
\end{equation}
This is true in particular when, if the system is in a state $\psi$ such that $b$ is always true, then after measuring $b$, the system is still is the same state $\psi$. In other word, $\psi(b)=1\implies \psi_b=\psi$. This is the concept of repeatability discussed above.

These two properties makes $\preceq$ a \emph{preorder relation}.
\index{Preorder relation}

One also assumes that
\begin{equation}
\label{precprec}
a\preceq b\ \text{and}\ b\preceq a\quad \implies \quad a=b
\end{equation}
This means that tests which give the same results on any states are indistinguishable.
This also means that one can identify a proposition $a$ with the set of states such that $a$ is always found to be true (i.e. $\psi(a)=1$).
\ref{precprec} makes $\preceq$ a partial order relation and $\mathcal{L}$ a \emph{partially ordered set} or POSET.
\index{POSET}

\subsubsection{AND (meet $\wedge$):}
The second ingredient is the notion of logical cunjunction AND.
One assumes that for any pair of test $a$ and $b$, there is a \emph{unique greater proposition} $a\wedge b$ such that
\begin{equation}
\label{andcunj}
a\wedge b\preceq a \ \text{and}\ \ a\wedge b\preceq b
\end{equation}
in other word, there is a unique $a\wedge b$ such that
$$c\preceq a\ \text{and}\  c\preceq b \ \ \implies\ \ c\preceq a\wedge b$$
NB: this is a non trivial assumption, not a simple consequence of the previous ones. 
It can be justified  using the notion of filters (see Jauch) or that of questions associated to propositions (see Piron).
Here to make things simpler I just present it as an assumption. 
On the other hand it is very difficult to build anything without this assumption.
Note that \ref{andcunj} implies \ref{precprec}.

This definition extends to any set $\mathcal{A}$ of propositions 
\begin{equation}
\label{ }
\bigwedge{\mathcal A}=\mathop{\bigwedge}\{{a\in\mathcal{A}}\}=\text{greatest}\ c:\ c\preceq a,\ \forall a\in\mathcal{A}
\end{equation}
I do not discuss if the set $\mathcal{A}$ is finite or countable.

\subsubsection{Logical OR (join $\vee$):}

From this we can infer the existence of a logical OR (by using Birkhoff theorem)
\begin{equation}
\label{ }
a\vee b=\bigwedge\{c\,:\ a\preceq c\ \text{and}\ b\preceq c\}
\end{equation}
which extents to sets of propositions
\begin{equation}
\label{ }
\bigvee\mathcal{A}=\bigwedge\{b\,:\ a\preceq b\ ,\ \forall a\in\mathcal{A}\}
\end{equation}

\subsubsection{Trivial $\mathbf{1}$ and vacuous $\emptyset$ propositions:}

It is natural to assume that there is a proposition $\mathbf{1}$ that is always true 
\begin{equation}
\label{ }
\text{for any state $\phi$,  $\mathbf{1}$ is always found to be true, i.e. $\phi(\mathbf{1})=1$}
\end{equation}
and another proposition $\emptyset$ that is never true
\begin{equation}
\label{ }
\text{for any state $\phi$,  $\emptyset$ is never found to be true, i.e. $\phi(\emptyset)=0$}
\end{equation}
Naturally one has
\begin{equation}
\label{ }
\mathbf{1}=\bigvee\mathcal{L}\ \text{and}\ \ \emptyset=\bigwedge\mathcal{L}
\end{equation}
With these assumptions and definitions the set of propositions $\mathcal{L}$  has now the structure of a  \textbf{complete lattice}.
\index{Complete lattice}

\subsection{Reversibility and  orthocomplementation}

\subsubsection{Negations $\mathbf{a}'$ and $'\mathbf{a}$}
I have not yet discussed what to do if a proposition is found to be false.
To do so one must introduce the seemingly simple notion of negation or complement.
In classical logic this is easy. The subtle point is that for quantum systems, where causality matters, there are two inequivalent ways to introduce the negation.
These two definitions becomes equivalent only if one assumes that propositions on quantum systems share a property of causal reversibility. 
In this case, one recovers the standard negation of propositions in classical logic, and ultimately this will lead to the notion of orthogonality and of scalar product of standard quantum mechanics. 
Thus here again, as in the previous section, reversibility appears to be one of the essential feature of the principles of quantum physics.

\paragraph{Negation - définition 1:}

To any proposition $a$ one can associate its negation (or complement proposition) $a'$ defined as
\begin{equation}
\label{ }
\text{for any state $\phi$, if $a$ is found to be true, then $a'$ will be found to be false}
\end{equation}
$a'$ can be defined equivalently as
\begin{equation}
\label{ }
a'=\bigvee\{b\,\text{such that}\ \text{on any state $\phi$, if $a$ is found true, then $b$ will be found false}\}
\end{equation}
%
%
\paragraph{Negation - définition 2:}
It is important at that stage to realize that, because of the causality ordering in the definition, there is an alternate definition for the complement,  that I denote $'\!a$, given by
\begin{equation}
\label{ }
\text{for any state $\phi$, if $'\!a$ is found to be true, then $a$ will be found to be false}
\end{equation}
or equivalently
\begin{equation}
\label{ }
'\!a=\bigvee\{b\,;\ \text{such that on any state $\phi$ , if $b$ is found  true, then $a$ will be found false}\}
\end{equation}
These two definitions are \emph{not equivalen}t, and they do not necessarily fulfill the properties of the negation in classical propositional logic
\footnote{The point discussed here is a priori not connected to the classical versus intuitionist logics debate. Remember that we are not discussing a logical system.}
.
\begin{equation*}
\label{ }
\neg{(\neg a)}\mathop{=}a\ \ \text{and}\ \ \ \neg(a\wedge b)\mathop{=}  \neg a\vee \neg b
\end{equation*}
These problems come from the fact that the definition for the causal order $a\preceq b$ does not implies that $b'\preceq a'$, as in classical logic. Indeed  the definition \ref{preceqdef} for $a\preceq b$ implies that for every state
\begin{equation}
\label{ }
\text{if $b$ is found false, then $a$ was found false}
\end{equation}
while $b'\preceq a'$ would  mean
\begin{equation}
\label{ }
\text{if $b$ is found false, then $a$ will be found false}
\end{equation}
or equivalently
\begin{equation}
\label{ }
\text{if $a$ is found true, then $b$ was found true}
\end{equation}

\subsubsection{Causal reversibility and negation}

In order to build a formalism consistent with what we know of quantum physics, we need to enforce the condition that the causal order structure on propositions is in fact independent of the choice of a causal arrow \og if $\cdots$, then $\cdots$ will $\cdots$ \fg{} versus \og if $\cdots$, then $\cdots$ was $\cdots$ \fg{} . This is nothing but the requirement of causal reversibility and it is enforced by the following simple but very important condition.
\index{Negation}
\index{Reversibility}

\paragraph{Causal reversibility:}
One assumes that the negation $a'$ is such that
\begin{equation}
\label{negrev}
a\preceq b\quad\iff\quad b'\preceq a'
\end{equation}

With this assumption, it is easy to show that the usual properties of  negation are satisfied. 
The two alternate definitions of negation are now equivalent
\begin{equation}
\label{ }
a'={'}a=\neg a
\end{equation}
and may be denoted by the standard logical symbol $\neg$.
We then have
\begin{equation}
\label{aprimeprime}
{(a')}'=a
\end{equation}
and
\begin{equation}
\label{awedgebprime}
(a\wedge b)'=a'\vee b'\
\end{equation}
as well as
\begin{equation}
\label{zerooneprime}
\emptyset=\mathbf{1}'\ ,\ \ \emptyset'=\mathbf{1}
\end{equation}
and
\begin{equation}
\label{aaprime}
a\wedge a'=\boldsymbol{\emptyset}
\ ,\ \ 
a\vee a'=\mathbf{1}
\end{equation}

A lattice $\mathcal{L}$ with a complement with the properties \ref{negrev}-\ref{aaprime} is called an \emph{orthocomplemented complete lattice} (in short OC lattice).
For such a lattice, the couple $(a,a')$ describes what is called a perfect measurement.
\index{Orthocomplementation}

\paragraph{NB:} Note that in Boolean logic, the implication $\to$ can be defined from the negation $\lnot$. Indeed  $a\to b$ means $\neg a\lor b$. Here it is the negation $\neg$ which is defined out of the implication $\preceq$.

\subsubsection{Orthogonality}
With reversibility and complement, the set of propositions starts to have properties similar to the set of projections on  linear subspaces of a Hilbert space%
\footnote{One should be careful for infinite dimensional Hilbert spaces and general operator algebras. Projectors correspond in general to orthogonal projections on closed subspaces.}. 
The complement $a'$ of a proposition $a$ is similar to the orthogonal subspace $P^\perp$ of a subspace $P$.
This analogy can be extended to the general concept of orthogonality.

\paragraph{Orthogonal propositions:}
\begin{equation}
\label{ }
\text{Two proposition $a$ and $b$ are orthogonal,  if $b\preceq a'$ (or equivalently $a\preceq b'$).}
\end{equation}
This is noted
\begin{equation}
\label{ }
a\perp b
\end{equation}

\paragraph{Compatible propositions:}\ \\
OC lattices contain also the concept of classical propositions. 
A subset of an OC lattice $\mathcal{L}$ is a sublattice $\mathcal{L}'$ if it is  stable under the operations $\wedge$, $\vee$ and $'$ (hence it is itself   an OC lattice). 
To any subset $\mathcal{S}\subset\mathcal{L}$ on can associate the sublattice $\mathcal{L}_\mathcal{S}$ generated by $\mathcal{S}$, defined as the smallest sublattice $\mathcal{L}'$ of $\mathcal{L}$ which contains $\mathcal{S}$.

A (sub)lattice is said to be \emph{Boolean} if it satisfy the distributive law of classical logic $a\wedge(b\vee c)=(a\wedge b)\vee(a\wedge  c)$.

A subset $\mathcal{S}$ of an OC lattice is said to be a subset of \emph{compatible propositions} if the generated lattice $\mathcal{L}_\mathcal{S}$ is Boolean.

Compatible propositions are the analog of commuting projectors, i.e. compatible or commuting observables in standard quantum mechanics.
For a set of compatible propositions, one expects that the expectations of the outcomes YES or NO will satisfy the rules of ordinary logic.

\paragraph{Orthogonal projection:}
\index{Orthogonal projector}
The notion of orthogonal projection onto a subspace can be also formulated in this framework as
\begin{equation}
\label{ }
\text{projection of $a$ onto $b$ $=\Phi_b(a)=b\wedge (a\vee b')$}
\end{equation}
This projection operation is often called the Sasaki projection. Its dual $(\Phi_b(a'))'=b'\vee(a\wedge b)$ is called the Sasaki hook $(b\mathop{\to}\limits^S a)$.
It has the property that even if $a\npreceq b$, if for a state $\psi$ the Sasaki hook  $(a\mathop{\to}\limits^S b)$ is always true, then for this state $\psi$, if $a$ is found true then $b$ will always  be  found true.
\index{Sasaki projection}
\index{Sasaki hook}


\subsection{Subsystems of propositions and orthomodularity}

\subsubsection{What must replace distributivity?}

The concept of orthocomplemented lattice of propositions is not sufficient to reconstruct a consistent quantum formalism.
There are mathematical reasons and physical reasons.

One reason is that if the distributive law $A\land((B\lor C)=(A\land B)\lor(A\land C)$ is known not to apply, assuming no restricted distributivity condition is not enough and leads to too many possible structures.
In particular in general a lattice with an orthocomplementation $\neg$ may be endowed with several inequivalent ones! 
This is  problematic for the physical interpretation of the complement as $a\to\TRUE\ \iff\ a'\to\FALSE$.
\index{Distributivity}

Another problem is that in physics one is led to consider conditional states and conditional  propositions. 
In classical physics this would correspond to the restriction to some  subset  $\Omega'$ of the whole phase space $\Omega$ of a physical system, or to the projection $\Omega\to\Omega'$.
Such projections or restrictions are necessary if there are some constraints on the states of the system, if one has access only to some subset of all the physical observables of the system, or if one is interested only in the study of a subsystem of a larger system.
In particular such a separation of the degrees of freedom is  very important when discussing locality: we are interested in the properties of the system we can associate to (the observables measured in) a given interval of space and time, as already discussed for algebraic QFT.
It is also very important when discussing effective low energy theories: we want to separate (project out) the (un-observable) high energy degrees of freedom from the (observable) low energy degrees of freedom.
And of course this is crucial to discuss open quantum systems, quantum measurement processes, decoherence processes, and the emergence of classical degrees of freedom and classical behaviors in quantum systems.

\subsubsection{Sublattices and weak-modularity}
\index{Weak modularity}
In general a subsystem is defined from the observables (propositions) on the system which satisfy some constraints.
One can reduce the discussion to one constraint $a$.
If $\mathcal{L}$ is an orthocomplemented lattice and $a$ a proposition of  $\mathcal{L}$, let us considers the subset $\mathcal{L}_{<a}$ of all propositions which imply $a$
\begin{equation}
\label{Linfadef}
\mathcal{L}_{<a}=\{b\in\mathcal{L}: \ b\preceq a\}
\end{equation}
One may also consider the subset of propositions $\mathcal{L}_{>a}$ of propositions implied by $a$
\begin{equation}
\label{Lsupadef}
\mathcal{L}_{>a}=\{b\in\mathcal{L}: \ a\preceq b\}=\left(\mathcal{L}_{<a'}\right)'
\end{equation}
The question is: is this set of propositions $\mathcal{L}_{<a}$  still an orthocomplemented lattice? One takes as order relation $\preceq$, $\vee$ and $\wedge$ in $\mathcal{L}_{<a}$ the same than in $\mathcal{L}$ and as trivial and empty propositions $\mathbf{1}_{<a}=a$, $\emptyset_{<a}=\emptyset$. 
Now, given a proposition $b\in\mathcal{L}_{<a}$, one must define what is its complement  $b'_{\scriptscriptstyle{<a}}$ in $\mathcal{L}_{<a}$. A natural choice is 
\begin{equation}
\label{ }
b'_{\scriptscriptstyle{<a}}=b'\wedge a
\end{equation}
but in general with such a choice  $\mathcal{L}_{<a}$ is not an orthocomplemented lattice, since it is easy to find for general AC lattices counterexamples such that one may have 
$b\vee b'_{\scriptscriptstyle{<a}}\neq a$. 

\paragraph{Weak-modularity:}
In order for  $\mathcal{L}_{<a}$ to be an  orthocomplemented lattice  (for any $a\in\mathcal{L}$), the orthocomplemented lattice $\mathcal{L}$ must satisfy the weak-modularity condition
\begin{equation}
\label{ }
b\preceq a\ \  \implies\ \ (a\wedge b')\vee b=a
\end{equation}
This condition is also sufficient.

\subsubsection{Orthomodular lattices}
\index{Orthomodular lattice}
\paragraph{Orthomodularity:}
An OC lattice which satisfies the weak-modularity condition is said to be an \emph{orthomodular lattice} (or OM lattice)\footnote{In French: treillis orthomodulaire, in German: Orthomodulare Verband.}. 
Clearly if $\mathcal{L}$ is OM, for any $a\in\mathcal{L}$, $\mathcal{L}_{<a}$ is also OM, as well as $\mathcal{L}_{>a}$.

\paragraph{Equivalent definitions:}
Weak-modularity has several equivalent definitions. 
Here are two interesting ones:
\begin{itemize}
  \item $a\preceq b\ \implies\ a\ \text{and}\ b\ \text{are compatible}.$
  \item the orthocomplementation  $a\to a'$ is unique in $\mathcal{L}$. 
\end{itemize}

\paragraph{Irreducibility:}
For such lattices one can also define the concept of irreducibility.
We have seen that two elements $a$ and $b$ of $\mathcal{L}$ are compatible (or commute) if they generate a Boolean lattice. The center  $\mathcal{C}$ of a lattice $\mathcal{L}$ is  the set of $a\in\mathcal{L}$ which commute with all the elements of the lattice $\mathcal{L}$. It is obviously a Boolean lattice.
A lattice is irrreducible if its center $\mathcal{C}$ is reduced to the trivial lattice $\mathcal{C}=\{\emptyset,\mathbf{1}\}$.
\index{Irreducible lattice}

\subsubsection{Weak-modularity versus modularity}
\paragraph{NB:} The (somewhat awkward) denomination \og weak-modularity\fg{} is historical. 
Following Birkhoff and von Neumann the stronger  \og modularity\fg{} condition for lattices was first considered.
\index{Modularity}
Modularity is defined as
\begin{equation}
\label{ }
a\preceq b \ \implies\ (a\lor c)\land b=a\lor(c\land b)
\end{equation}
Modularity is equivalent to weak modularity for finite depth lattices (as a particular case the set of projectors on a finite dimensional Hilbert space for a modular lattice).
But modularity turned out to be inadequate for infinite depth lattices (corresponding to the general theory of projectors in infinite dimensional Hilbert spaces).
The theory of modular lattice has links with some  W$^*$-algebras and the theory of ``continuous geometries'' (see e.g. \cite{Von-Neumann:1960fk}).

\subsection{Pure states and AC properties}
Orthomodular (OM) lattices are a good starting point to consider the  constraints that we expect for the set of ideal measurements on a physical system, and therefore to study how one can represent its states. 
In fact one still needs two more assumptions, which seem technical, but which are also very important (and quite natural from the point of view of quantum information theory). They rely on the concept of atoms, or minimal proposition, which are the analog for propositions of the concept of minimal projectors on of pure states in the algebraic formalism.

\subsubsection{Atoms}
An element $a$ of an OM lattice is said to be  an atom if
\index{Atom}
\begin{equation}
\label{ }
b\preceq a\quad\text{and}\quad b\neq a\quad\implies\quad b=\emptyset
\end{equation}
This means that $a$ is a minimal non empty proposition; it is not possible to find another proposition compatible with $a$ which allows to obtain more information on the system than the information obtained if $a$ is found to be TRUE. 

Atoms are the analog of projectors on pure states in the standard quantum formalism (pure propositions). Indeed, if the system in in some state $\psi$, before the measurement of $a$, if $a$ is an atom and is found to be true, the system will be in a pure state $\psi_a$ after the measure.

\subsubsection{Atomic lattices}
\index{Atomic lattice}
A  lattice is said to be atomic if any non trivial proposition $b\neq\emptyset$ in $\mathcal{L}$ is such that there is at least one atom $a$ such that $a\preceq b$ (i.e. any proposition \og contains\fg{} at least one minimal non empty proposition). For an atomic OM lattice one can show that any proposition $b$ is then the union of its atoms (atomisticity).

\subsubsection{Covering property}
\index{Covering property}
Finally one needs also the covering property. The formulation useful in the quantum framework is to state that if $a$ is a proposition and $b$ an atom not in the complement $a'$ of $a$, then the Sasaki projection of $b$ onto $a$, $\Phi_a(b)=a\wedge(b \vee a')$, is still an atom. 

The original definition of the covering property for atomic lattices by Birkhoff is: for any $b\in\mathcal{L}$ and any atom $a\in\mathcal{L}$ such that $a\wedge b=\emptyset$, $a\vee b$ covers $b$, i.e. there is no $c$ between $b$ and $a\vee b$ such that $b\prec c\prec a\vee b$.

This covering property is very important. It means that when reducing a system to a subsystem by some constraint (projection onto $a$), one cannot get a non-minimal proposition out of a minimal one. 
This would mean that one could get more information out of a subsystem than from the greater system.
In  other word, if a system is in a pure state, performing a perfect measurement can only map it onto another pure state. Perfect measurements cannot decrease the information on the system.

\textit{The covering property is in fact also related to the superposition principle. Indeed, it implies that (for irreducible lattices) for any two difference atoms $a$ and $b$, there must be a third atom $c$ different from $a$ and $b$ such that $c\prec a\vee b$. Thus, in the weakest possible sense (remember we have no addition) $c$ is a superposition of $a$ and $b$.
CHECK}

\medskip
An atomistic lattive with the covering property is said to be an AC  lattice.
As mentionned before these properties can be formulated in term of the properties of the set of states on the lattice rather than in term of the propositions.
I shall not discuss this here.

\section{The geometry of orthomodular AC lattices}

I have given one (possible and personal) presentation of the principles at the basis of the quantum logic formalism. It took some time since I tried  to explain both the mathematical formalism and the underlying physical ideas.
I now explain the main mathematical result: the definition of the set of propositions (ideal measurements) on a quantum system as an orthomodular AC lattice can be equivalently represented as the set of orthogonal projections on some \og generalized Hilbert space \fg{}.

\subsection{Prelude: the fundamental theorem of projective geometry}
\index{Projective geometry}
The idea is to extend a classical and beautiful theorem of geometry, the Veblen-Young theorem. 
Any abstract projective geometry can be realized as the geometry of the affine subspaces of some  left-module (the analog of vector space) on a division ring $K$ (a division ring is a non-commutative field). This result is known as the \og coordinatization of projective geometry\fg{}.
Classical references on geometry are the books by E. Artin Geometric Algebra \cite{Artin:1957hs}, R. Baer \textsl{Linear algebra and projective geometry} \cite{Baer:2005kc} or Conway .
More precisely, a geometry on a linear space is simply defined by a set of points $X$, and a set of lines $\mathcal{L}$ of $X$ (simply a set of subsets of $X$). 
\index{Veblen's axiom}
\index{Veblen-Young theorem}
\index{Division ring}

\paragraph{Theorem:}
If the geometry satisfies the following axioms:
\begin{enumerate}
  \item Any line contains at least 3 points,
  \item Two points lie in a unique line,
  \item A line meeting two sides of triangle, not at a vertex of the triangle, meets the third side also (Veblen's axiom),
  \item There are at least 4 points non coplanar (a plane is defined in the usual way from lines),
  \end{enumerate}
 then the corresponding geometry is the geometry of the affine subspaces of a left module $M$ on a division ring $K$ (a division ring is a in general non-commutative field).
 
 \paragraph{Discussion:}
 The theorem here is part of the Veblen-Young theorem, that encompasses the cease when the 4th axiom is not satisfied.
The first two axioms define a line geometry structure such that lines are uniquely defined by the pairs of points, but with some superposition principle.
 The third axiom is represented on \ref{Veblen}.
 \begin{figure}[h]
\begin{center}
\includegraphics[width=4in]{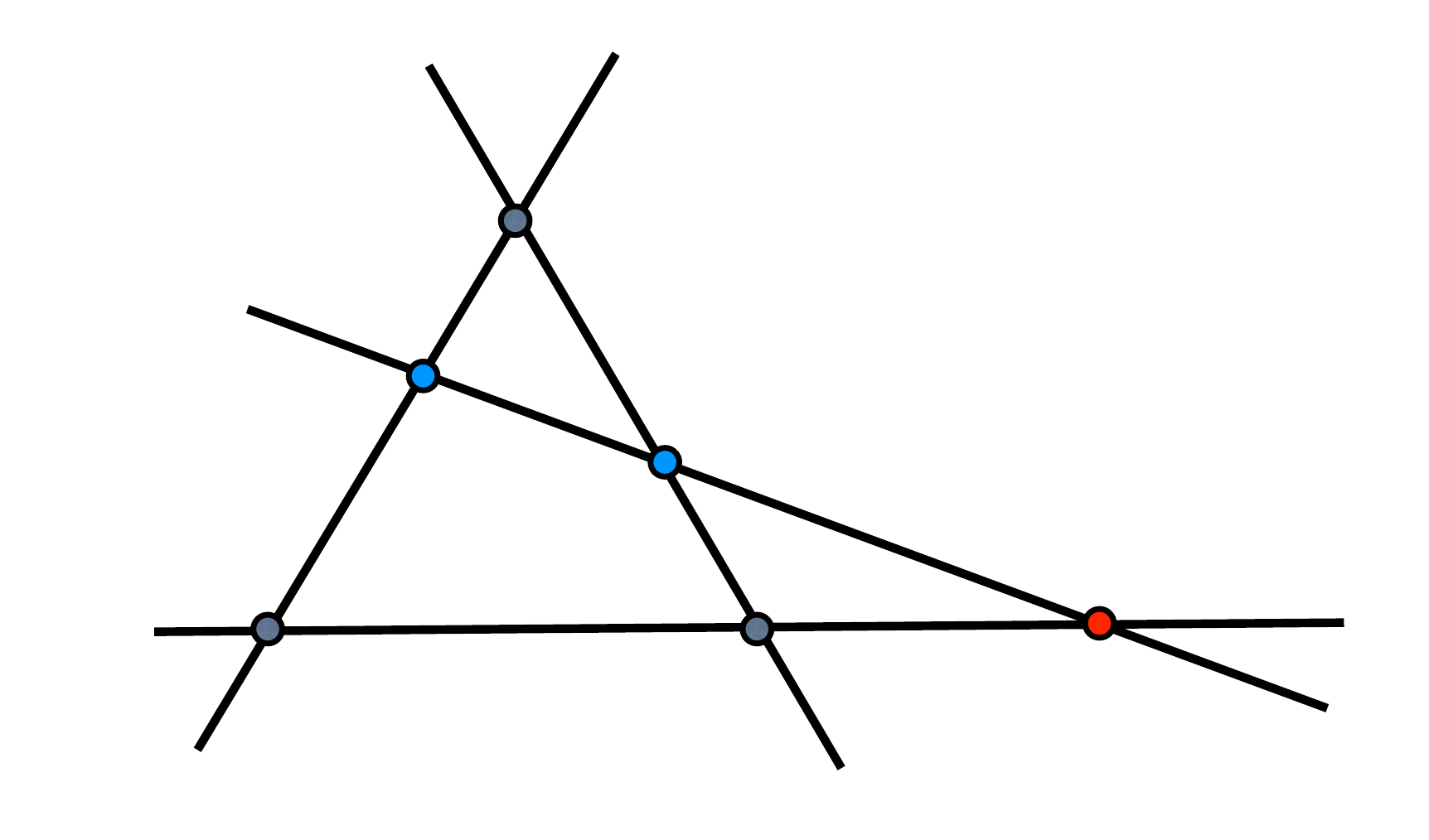}
\caption{Veblen's axiom}
\label{Veblen}
\end{center}
\end{figure} 

 The fourth one is necessary to exclude some special non-Desarguesian geometries.

Let me note that the division ring $K$ (an associative algebra with an addition $+$, a multiplication $\times$ and an inverse $x\to x^{-1}$) is constructed out of the symmetries of the geometry, i.e. of the automorphisms, or applications $X\to X$, $\mathcal{L}\to\mathcal{L}$, etc. which preserve the geometry. Without giving any details, let me illustrate the case of the standard real projective plane (where $K=\mathbb{R}$). The field structure on $\mathbb{R}$ is obtained by identifying $\mathbb{R}$ with a projective line $\ell$ with three points $0$, $1$ and $\infty$. The ``coordinate'' $x\in\mathbb{R}$ of a point $X\in\ell$ is identified with the cross-ratio $x=(X,1;0,\infty)$.
On Fig.~\ref{projplustimes} are depicted the  geometrical construction of the addition $X+Y$ and of the multiplication $X\times Y$ of two points $X$ and $Y$ on a line $\ell$.
\begin{figure}[h]
\begin{center}
\includegraphics[width=2.99in]{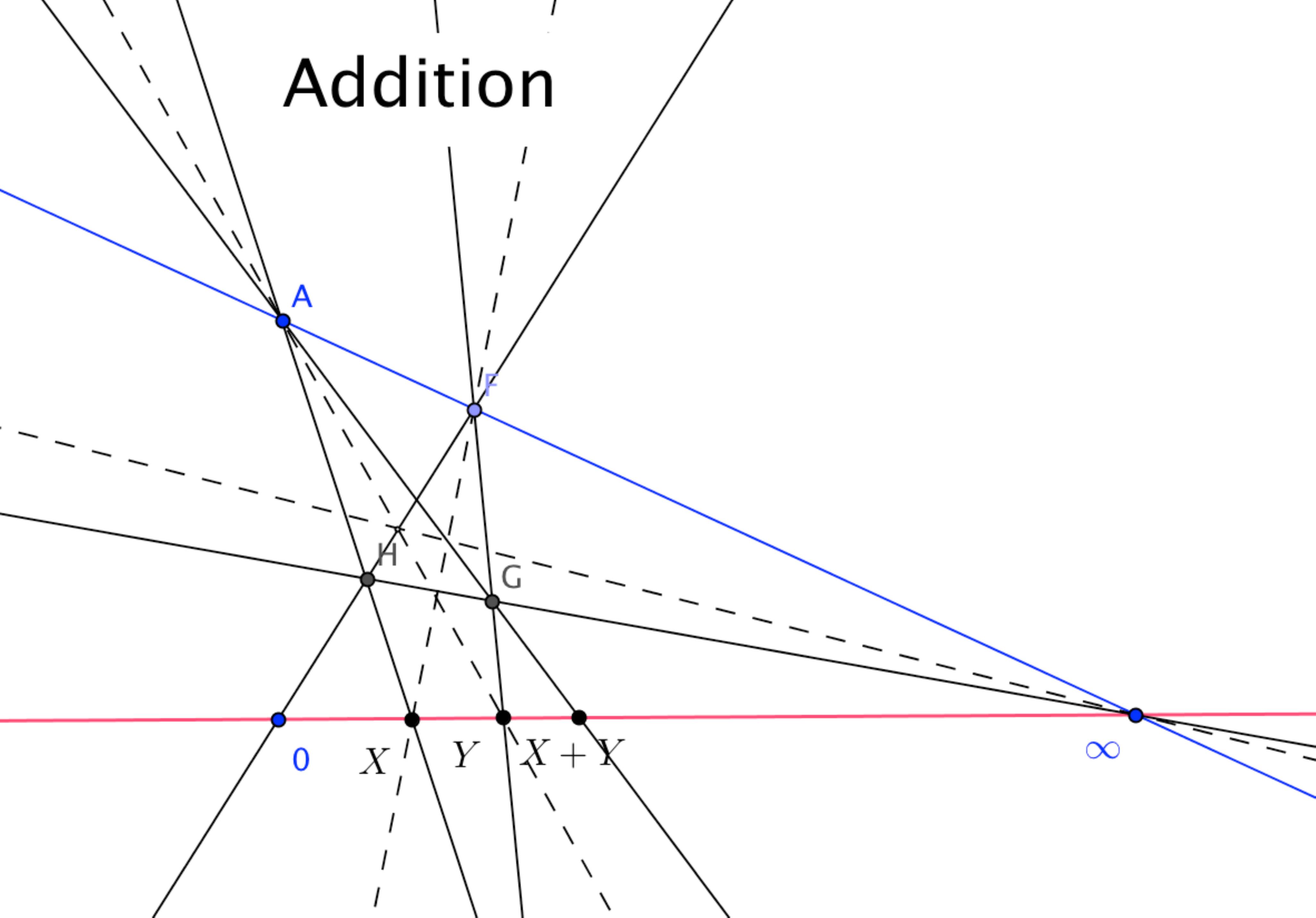}
\quad
\includegraphics[width=2.99in]{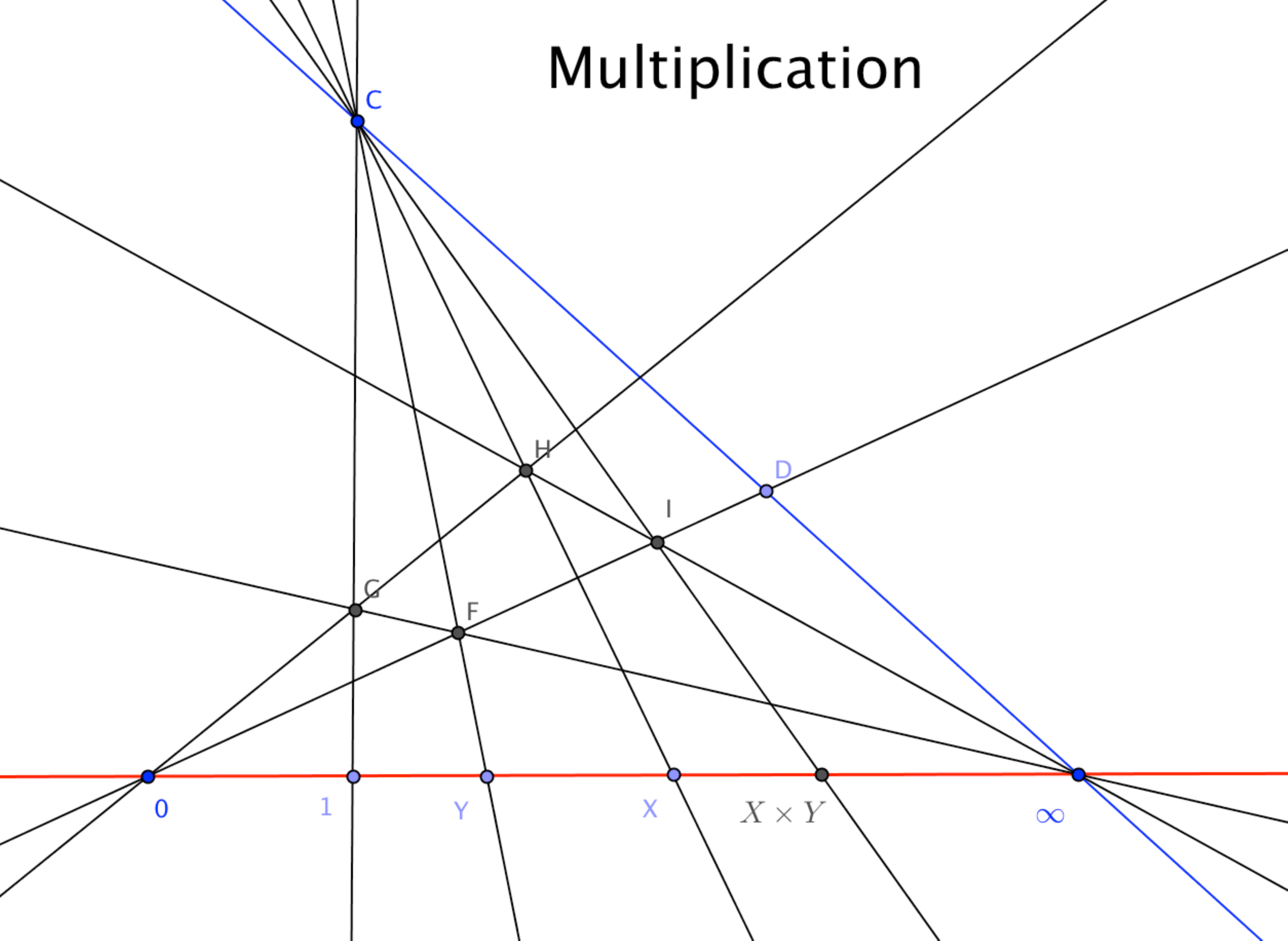}
\caption{Construction of $+$ and $\times$ in the  projective real plane}
\label{projplustimes}
\end{center}
\end{figure}

\subsection{The projective geometry of orthomodular AC lattices}

\subsubsection{The coordinatization theorem}
\index{Coordinatization theorem}
\index{Orthomodular lattice}
\index{Atomic lattice}
\index{Covering property}
\index{Division ring}
\index{Scalar product}

Similar ``coordinatization''  theorems hold for the orthomodular AC lattices that have been introduced in the previous section. The last axioms AC (atomicity and covering) play a similar role as the axioms of abstract projective geometry, allowing to define ``points'' (the atoms), lines, etc , with properties similar to the first 3 axioms of linear spaces. The difference with projective geometry is the existence of the orthocomplementation (the negation $\neg$) which allows to define an abstract notion of orthogonality $\perp$, and the specific property of weak-modularity (which will allows to define in a consistent way what are projections on closed subspaces).

Let me first state the main theorem

\paragraph{Theorem:}
Let $\mathcal{L}$ be a complete irreducible orthocomplemented AC lattice with length $>3$ (i.e. at least three 4 different levels of proposition $\emptyset\prec a\prec b\prec c\prec d\prec 1)$.
Then the ``abstract'' lattice $\mathcal{L}$ can be represented as the lattice $\mathcal{L}(V)$ of the closed subspaces of a left-module%
\footnote{A module is the analog of a vector space, but on a ring instead of a (commutative) field}
  $V$ on a division ring%
  \footnote{A division ring is the analog of a field, but without commutativity}
   $K$ with a Hermitian form $f$.
The ring $K$, the module $V$ and the form $f$ have the following properties:
\begin{itemize}
  \item The division ring $K$ has an involution $^*$  such that $(xy)^*=y^* x^*$
  \item The vector space $V$ has a non degenerate Hermitian (i.e. sesquilinear) form $f:\ V\times V \to K$
  \begin{equation}
\label{ }
\mathbf{a},\mathbf{b}\in V \ \ ,\ \ f(a,b)=\langle \mathbf{a}|\mathbf{b}\rangle\in K
  \quad,\quad \langle\mathbf{a}|\mathbf{b}\rangle=\langle\mathbf{b}|\mathbf{a}\rangle^*
\end{equation}
    \item The Hermitian form $f$ defines an orthogonal projection and associates to each linear subspace $M$ of $V$ its orthogonal $M^\perp$. 
    \begin{equation}
\label{ }
M^\perp=\{\mathbf{b}\in v:\ \langle\mathbf{b}|\mathbf{a}\rangle=0\ \ \forall\mathbf{a}\in M\}
\end{equation}
\item   The closed subspaces of $V$ are the subspaces $M$ such that $(M^\perp)^\perp=M$.
  \item The Hermitian form is orthomodular, i.e. for any closed subspace, $M^\perp+M=V$.
\item The OM structure $(\preceq\, ,\, \wedge\, ,\, ')$ on the lattice $\mathcal{L}$ is isomorphic to the standard lattice structure $(\subseteq\ ,\,\cap \,,\,\perp)$ (subspace of, intersection of, orthogonal complement of) over the space $\mathcal{L}(V)$ of closed linear subspaces of $V$.
\item Moreover, $V$ and $K$ are such that there is some element $a$ of $V$ with \og norm\fg{}  unity  $f(a,a)=1$ (where $1$ is the unit element of $K$).
\end{itemize} 

\medskip
I do not give the proof. I refer to the physics literature: ( \cite{BeltCassi81}  chapter 21, \cite{Piron64,Piron76}, and to the original mathematical literature  
\cite{BirkVNeumann36}
\cite{maeda1971theory}
\cite{Varadarajan1985}.

\medskip

Thus this theorem states that an OM AC lattice can be represented as the lattice of orthogonal projections over the closed linear subspaces of some ``generalized Hilbert space'' with a quadratic form defined over some non-commutative field $K$. This is very suggestive of the fact that Hilbert spaces are not abstract and complicated mathematical objects (as still sometimes stated), but are the natural objects to describe and manipulate ideal measurements in quantum physics.
In particular  the underlying ring $K$ and the algebraic structure of the space $V$ come out naturally from the symmetries of the lattice of propositions $\mathcal{L}$.

\subsubsection{Discussion: which division ring $K$?}
\label{sssRing}
The important theorem discussed before is very suggestive, but is not sufficient to ``derive'' standard quantum mechanics. 
The main question is which division algebra $K$ and which involution $^*$ and Hermitian form $f$ are physically allowed? Can one construct physical theories based on other rings than the usual $K=\mathbb{C}$ (or $\mathbb{R}$ or $\mathbb{H}$)?

The world of division rings is very large! The simplest one are finite division rings, where the first Wedderburn theorem implies that $K$ is a (product of) Galois fields $\mathbb{F}_p=\mathbb{Z}/\mathbb{Z}_p$ ($p$ prime). Beyond $\mathbb{C}$, $\mathbb{R}$ and $\mathbb{H}$, more complicated  ones are rings of rational functions $F(X)$, up to very large ones (like surreal numbers...), but still commutatives, to non-commutatives rings.

However, the requirements that $K$ has an involution, and that $V$ has a non degenerate hermitian form, so that $\mathcal{L}(V)$ is a OM lattice, put already very stringent constraints on $K$.
For instance, it is well known that finite fields like the $\mathbb{F}_p$ ($p$ prime) do not work. 
Indeed, it is easy to see that the lattice $\mathcal{L}(V)$ of the linear subspaces of the finite dimensional subspaces of the $n$-dimensional vector space $V=({\mathbb{F}_p})^n$ is not orthomodular and cannot be equipped with a non-degenerate quadratic form! Check with $p=n=3$!
But still many more exotic division rings $K$ than the standard $\mathbb{R}$, $\mathbb{C}$ (and $\mathbb{H}$) are possible at that stage
.
\index{Galois field}

\subsection{Towards Hilbert spaces}
There are several arguments that point towards the standard solution: $V$ is a Hilbert space over $\mathbb{R}$, $\mathbb{C}$ (or $\mathbb{H}$).
However none is completely mathematically convincing, if most would satisfy a physicist.
Remember that real numbers are expected to occur in physical theory for two reasons. Firstly we are trying to compute probabilities $p$, which are real numbers. Secondly quantum physics must be compatible with the relativistic concept of space-time, where space (and time) is described by continuous real variables. Of course this is correct as long as one does not try to quantize gravity. 

We have not discussed yet precisely the structure of the states $\psi$, and which constraints they may enforce on the algebraic structure of propositions.
Remember that it is the set of states $\mathcal{E}$ which allows to discuss the partial order relation $\preceq$ on the set of propositions $\mathcal{L}$. Moreover states $\psi$ assign probabilities $\psi(a)\in[0,1]$ to propositions $a$, with the constraints that if $a\perp b$, $\psi(a\lor b)=\psi(a)+\psi(b)$.
Moreover the  propositions $a\in\mathcal{L}$ (projective measurements) define via the Sasaki orthogonal projections $\pi_a$ a set of transformations $\mathcal{L}\to\mathcal{L}$, which form  a so called Baer $^*$-semi group.
On the same time, propositions $a\in\mathcal{L}$ define mappings $\psi\to\psi_a$ on the states. 
Since as in the algebraic formalism, convex linear combinations of states are states, $\mathcal{E}$ generate a linear vector space $E$, and form a convex subset $\mathcal{E}\subset E$. 
Thus there is more algebraic structure to discuss than what I explained up to now. I refer to \cite{BeltCassi81}, chapters 16-19, for more details.
I shall come back to states when discussing Gleason's theorem in the next section.

Assuming some ``natural'' continuity or completeness conditions for the states leads to theorems stating that the division ring $K$ must contain the field of real numbers $\mathbb{R}$, hence is $\mathbb{R}$, $\mathbb{C}$ or $\mathbb{H}$, and that the involution $^*$ is continuous, hence corresponds to the standard involution $x^*=x$,  $x^*=\bar x$ or $x^*=x^\star$ respectively.
See \cite{BeltCassi81}, chapter 21.3.

 Another argument comes from an important theorem in the theory of orthomodular lattices, which holds for lattices of projections in infinite dimensional modules.

\paragraph{Sol{\`e}r's Theorem:} (Solèr 1995) Let $\mathcal{L}=\mathcal{L}_K(V)$ be an irreducible OM AC lattice of compact linear suspaces in a left-module $V$ over a divison ring $K$, as discussed above. 
If there is an infinite family $\{v_i\}$ of orthonormal vectors in $V$ such that $\langle v_i|v_j\rangle=\delta_{ij} f$ with some $f\in K$ 
 then the division ring $K$ can only be $\mathbb{R}$, $\mathbb{C}$ or $\mathbb{H}$.
 \index{Soler's theorem}

\medskip
The proof of this highly non trivial theorem is given in \cite{Soler95}. It is discussed in more details in \cite{Holland95}.

\medskip

The assumptions of the theorem state that there an infinite set of mutually compatible atoms $\{a_i\}_{i\in I}$ in $\mathcal{L}$ (commuting, or causally independent elementary propositions $a_i$), and in addition that there is some particular symmetry between the generators $v_i\in V$ of the linear spaces (the lines or rays) of these propositions.

The first assumption is quite natural if we take into account space-time and locality in quantum physics.
Let me consider the case where the physical space in which the system is defined to be infinite (flat) space or some regular lattice, so that it can be separated into causally independent pieces $\mathcal{O}_\alpha$ (labelled by $\alpha\in\Lambda$ some infinite lattice). See for instant Fig.~\ref{StrCauDia}. 
It is sufficient to have one single proposition $a_\alpha$ relative to each $\mathcal{O}_\alpha$ only (for instance ``there is one particle in $\mathcal{O}_\alpha$'') to build an infinite family of mutually orthogonal propositions $b_\alpha=a_\alpha\wedge(\bigwedge_{\beta\neq\alpha}\neg a_\beta)$ in $\mathcal{L}$. 
Out of the $b_\alpha$, thanks to the atomic property (A), we can extract an infinite family of orthogonal atoms $c_\alpha$.
\begin{figure}[h]
\begin{center}
\includegraphics[width=4in]{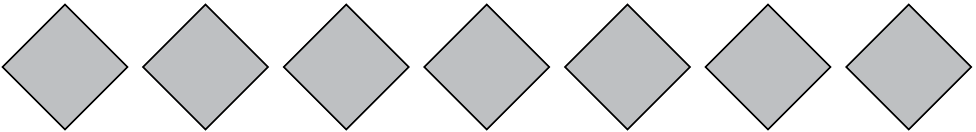}
\caption{A string of causal diamonds (in space-time)}
\label{StrCauDia}
\end{center}
\end{figure}

However this does not ensure the second assumption: the fact that the corresponding $v_i\in V$ are orthonormals. The group of space translations $\mathfrak{T}$ must act as a group of automorphism on the lattice of linear subspaces $\mathcal{L}=\mathcal{L}_K(V)$ (a group of automorphisms on a OC lattice $\mathcal{L}=\mathcal{L}_K(V)$ is a group of transformations which preserves the OC lattice structure  $(\preceq\,,\, \wedge\,,\, ')$ or equivalently $(\subseteq\, ,\ \cap\, ,\, \perp)$). There must correspond an action (a representation) of the translation group $\mathfrak{T}$ on the vector space $V$, and on the underlying field $K$. If the action is trivial the conditions of Soler's theorem are fulfilled, but this is not ensured a priori. See for instance \cite{Casinelli-Lahti-2012} for a recent discussion of symmetries in orthomodular geometries. However I am not aware of a counterexample where a non standard orhomodular geometry (i.e. different from that of a Hilbert space on $\mathbb{C}$ (or $\mathbb{R}$) carries a representation of a ``physical'' symmetry group such as the Poincaré or the Galilean group of space-time transformations (representations of these groups should involves the field of real numbers $\mathbb{R}$ in some form).
\index{Poincaré invariance}
\index{Hilbert space}

From now on we assume that a quantum system may indeed be described by projectors in a real or complex Hilbert space.

One last remark. The coordinatization theorem depends crucially on the fact that the OM lattice $\mathcal{L}$ is atomic, hence contain minimal propositions (atoms). They are the analog of minimal projectors in the theory of operator algebras. Hence the formalism discussed here is expected to be valid mathematically to describe only type I  von Neumann algebras.
I shall not elaborate further.

\section{Gleason's theorem and the Born rule}

\subsection{States and probabilities}
\index{State}
\index{Probabilities}
In the presentation of the formalism we have not put emphasis on the concept of states, although states are central in the definition of the causality order relation $\preceq$ and  of the orthocomplementation $\ '$.
We recall that to each state $\psi$ and to each proposition $a$ is associated the probability $\psi(a)$ for $a$ to be found true on the state $\psi$.
In other word, states are probability measures on the set of propositions, compatible with the causal structure.
As already mentionned, the lattice structure of propositions can be formulated from the properties of the states on $\mathcal{L}$.

At that stage we have almost derived the standard mathematical formulation of quantum mechanics. 
Proposition (yes-no observables) are represented by othonormal projectors on a Hilbert space $\mathcal{H}$. Projectors on pure states corresponds to projectors on one dimensional subspaces, or rays of $\mathcal{H}$ so the concept of pure states is associated to the vectors of $\mathcal{H}$.

Nevertheless it remains to understand which are the consistent physical states, and what are the rules which determine the probabilities for a proposition $a$ to be true in a state $\psi$, in particular in a pure state. 
We remind that the states are in fact characterized by these probability distributions $a\to\psi(a)$ on $\mathcal{L}$.
Thus states must form a convex set of functions $\mathcal{L}\to [0,1]$ and by consistency with the OM structure of $\mathcal{L}$ they must satisfy four conditions. These conditions define ``quantum probabilities''

\paragraph{Quantum probabilities:}
\begin{align}
\label{psicond1}
(1)\qquad    & \psi(a)\in [0,1]  \\
\label{psicond1b}
(2)\qquad    &  \psi(\emptyset)=0\ , \quad\psi(\mathbf{1})=1\\
\label{psicond1c}
(3)\qquad    & a\neq b\ \implies\ \exists\, \psi\ \text{such that}\ \psi(a)\neq\psi(b)\\
\label{psicond1d}
(4)\qquad    & a\perp b\ \implies\ \text{}\quad \psi(a\vee b)=\psi(a)+\psi(b)
\end{align}
Conditions (1) and (2) are the usual normalization conditions for probabilities.
Condition (3)  means that observables are distinguishable by their probabilities.
Condition (4)  is simply the fact that if $a$ and $b$ are orthogonal, they generate a Boolean algebra, and the associated probabilities must satisfy the usual sum rule.
These conditions imply in particular that for any state $\psi$,  $\psi(\neg a)=1-\psi(a)$, and that if $a\preceq b$, then $\psi(a)\le\psi(b)$, as we expect.

It remains to understand if and why all states $\psi$  can be represented by density matrices $\rho_\psi$, and the probabilities for propositions $\mathbf{a}$ given by $\psi(\mathbf{a})=\tr(\rho_\psi P_\mathbf{a})$, where $P_\mathbf{a}$ is the projector onto the linear -subspace associated to the proposition $\mathbf{a}$.
This is a consequence of a very important theorem in operator algebras, Gleason's theorem \cite{Gleason57}.

\subsection{Gleason's theorem}
\label{ssGleason}
\index{Gleason's theorem}
It is easy to see that to obtain quantum probabilities that satisfy the conditions \ref{psicond1}- \ref{psicond1d}, it is sufficient to consider atomic propositions, i.e. projections onto 1 dimensional subspaces (rays) generated by vectors $\vec e=|e\rangle$ (pure states) of the Hilbert space $\mathcal{H}$. Indeed, using \ref{psicond1d}, the probabilities for general projectors can be reconstructed (by the usual sum rule) from the probabilities for projections on rays. Denoting since there is no ambiguity for a state $\psi$ the probability for the atomic proposition $\mathbf{e}$ represented by the projection $P_{\vec e}$ onto a vector $\vec e=|\vec e\rangle\in\mathcal{H}$ as
\begin{equation}
\label{}
\psi(\mathbf{e})=\psi(P_{\mathbf{ e}})=\psi(\vec e)
\qquad
P_{\vec e}=|\vec e\rangle\langle\vec e|\ =\ \text{projector onto\ }\vec e
\end{equation}
The rules  \ref{psicond1}- \ref{psicond1d} reduces to the conditions.

\paragraph{Quantum probabilities for projections on pure states:}
For any states $\psi$, the function $\psi(\vec e)$ considered as a function on the ``unit sphere'' of the rays over the Hilbert Space $\mathcal{H}$ (the projective space) $\mathcal{S}=\mathcal{H}^*/K^*$  must satisfy
\begin{align}
\label{frame1a}
  (1)\qquad   &   \psi(\vec e)=\psi(\lambda\vec e)\quad\text{for any\ }\lambda\in K\ \text{such that\ } |\lambda |=1\\
 \label{frame1b} (2) \qquad &   0\le \psi(\vec e)\le 1\\
 \label{frame1c}(3)\qquad   &  \text{For any complete orthonormal basis of}\  \mathcal{H},\   \{\vec e_i\},\ \text{one has}\ \sum_i\psi(\vec e_i)=1
\end{align}

Gleason's theorem states the fundamental result that any such function is in one to one correspondence with a density matrix.

\paragraph{Gleason's theorem:} \ \\
If the Hilbert space $\mathcal{H}$ over $K=\mathbb{R}$ or $\mathbb{C}$ is such that
\begin{equation}
\label{ }
\dim(\mathcal{H})\ge 3
\end{equation}
then any function $\psi$ over the unit rays of $\mathcal{H}$ that satisfies the three conditions \ref{frame1a}--\ref{frame1c} is of the form
\begin{equation}
\label{frame2qform}
\psi(\vec e)=(\vec e \cdot \rho_\psi \cdot\vec e) =\langle \vec e|\rho_\psi |\vec e\rangle
\end{equation}
where $\rho_\psi$ is a positive quadratic form (a density matrix) over $\mathcal{H}$ with the expected properties for a density matrix
\begin{equation}
\label{ }
\rho_\psi=\rho_\psi^\dagger\quad,\qquad \rho_\psi\ge 0\quad,\qquad \tr(\rho_\psi)=1
\end{equation}
Reciprocally, any such quadratic form defines a function $\psi$ with the three properties \ref{frame1a}--\ref{frame1c}.

\medskip\medskip
Gleason's theorem is fundamental. As we shall discuss more a bit later, it implies the Born rule. It is also very important when discussing (and excluding a very general and most natural class of) hidden variables theories. So let us discuss it a bit more, without going into the details of the proof.

\subsection{Principle of the proof} 

The theorem is remarkable since there are non conditions on the regularity or measurability of the function $\psi$. In the original derivation by Gleason \cite{Gleason57} he considers real ``frame functions'' $f$ of weight $W$ over $\mathcal{H}^*=\mathcal{H}\backslash \{0\}$ such that
\begin{align}
\label{frame2a}
  (1)\qquad   &   f(\vec e)=f(\lambda\vec e)\quad\text{for any\ }\lambda\neq 0 \in K\\
 \label{frame2b} (2) \qquad &   f\quad\text{is bounded}\\
 \label{frame2c}(3)\qquad   &  \text{For any complete orthonormal basis of}\  \mathcal{H},\   \{\vec e_i\},\ \sum_i f(\vec e_i)=W=\text{constant}
\end{align}
and proves that such a function must be of the form \ref{frame2qform}
\begin{equation}
\label{ }
f(\vec e)=(\vec e{\cdot}Q{\cdot}\vec e)\quad,\qquad Q\quad \text{quadradic form such that\quad } \tr(Q)=W
\end{equation}
It is easy to see that this is equivalent to the theorem as stated above, since one can add constants and rescale the functions $f$ to go from \ref{frame2a}--\ref{frame2c} to \ref{frame1a}--\ref{frame1c}. The original proof goes into three steps

\begin{enumerate}
  \item Real Hilbert space, $\dim(\mathcal{H})=3$
  and $f$ a continuous frame function $\implies$ the theorem\\
  This is the easiest part, involving some group theory.
   Any frame function $f$ is a real function on the unit two dimensional sphere $\mathcal{S}_2$ and if continuous it is square summable and can be decomposed into  spherical harmonics 
  \begin{equation}
\label{fYlm}
f(\vec n)=\sum_{l,m} f_{l,m} Y^{m}_{l}(\theta,\varphi)
\end{equation}
The theorem amounts to show that if $f$ is a frame function of weight $W=0$, then only the $l=2$ components of this decomposition \ref{fYlm} are non zero. Some representation theory (for the SO(3) rotation group) is enough.
Any orthonormal (oriented) basis $(\vec e_1,\vec e_2,\vec e_3)$ of $\mathbb{R}^3$ is obtained by applying a rotation $R$ to the basis $(\vec e_x,\vec e_y,\vec e_z)$. Thus one can write
\begin{equation}
\label{fbasis}
f(\vec n_1)+f(\vec n_2)+f(\vec n_3)=\sum_l \sum_{m,m'}  f_{l,m}\  D^{(l)}_{m,m'}(R)\  V^{(l)}_{m'}
\end{equation}
with the $D^{(l)}_{m,m'}(R)$ the Wigner $D$ matrix for the rotation $R$, and the $V^{(l)}_{m'}$ the components of the vectors $\vec V^{(l)}$ in the spin $l$ representation of $SO(3)$, with components
\begin{equation}
\label{ }
\vec V^{(l)}=\{V^{(l)}_m\}\quad,\qquad V^{(l)}_m= Y_{l,m}(0,0)+Y_{l,m}(\pi/2,0)+Y_{l,m}(\pi/2,\pi/2)
\end{equation}
If $f$ is a frame function of weight $W=0$, the l.h.s. of \ref{fbasis} is zero for any $R\in\mathrm{SO(3)}$. This implies that for a given $l$, the coefficients $f_{lm}$ must vanish if the vector $\vec V^{(l)}\neq 0$, but are free if $\vec V^{(l)}=0$.
An explicit calculation shows that indeed
\begin{equation}
\label{ }
\vec V^{(l)}\ \begin{cases}
      \neq 0& \text{if}\ l\neq 2, \\
      =0 & \text{if}\ l=2.
\end{cases}
\end{equation}
This establishes the theorem in case (1).

  \item Real Hilbert space, $\dim(\mathcal{H})=3$
  and $f$ any  frame function $\implies$ $f$ continuous.\\
  This is the most non-trivial part: assuming that the function is bounded, the constraint \ref{frame2c} is enough to imply that the function is continuous! It involves a clever use of spherical geometry and of the frame identity $\sum\limits_{i=1,2,3}f(\vec e_i)=W$. 
The basic idea is to start from the fact that since $f$ is bounded, it has a lower bound $f_{min}$ which can be set to 0. Then for any $\epsilon>0$, take a vector $\vec n_0$ on the sphere such that $|f(\vec n_0)-f_{min}|<\epsilon$. It is possible to show that there is a neibourhood $\mathcal{O}$ of $\vec n_0$ such that $|f(\vec n_1)-f(\vec n_2)|< C\, \epsilon$ for any $\vec n_1$ and  $\vec n_2\in\mathcal{O}$. $C$ is a universal constant. It follows that the function $f$ is continuous at its minimum! Then it is possible, using rotations to show that the function $f$ is continuous at any points on the sphere.

  \item Generalize to $\dim(\mathcal{H})>3$ and to complex Hilbert spaces.\\
This last part is more standard and more algebraic. Any frame function $f(\vec n)$ defined on unit vectors $\vec n$ such that $\| \vec n\|=1$ may be extended to a quadratic function over vectors $f(\vec v)=\|\vec v\|^2 f(\vec v/\|\vec v\|)$.

For a real Hilbert space with dimension $d>3$, the points (1) and (2) implies that the restriction of a frame function $f(\vec n)$ to any 3  dimensional subspace is a quadratic form $f(\vec v)=(\vec v{\cdot}Q{\cdot}\vec v)$.
A simple and classical theorem by Jordan and von Neumann shows that this is enough to define a global real quadratic form $Q$ on the whole Hilbert space $\mathcal{H}$ through the identity
$2(\vec x{\cdot}Q{\cdot}\vec y)=f(\vec x+\vec y)-f(\vec x-\vec y)$.

For complex Hilbert spaces, the derivation is a bit more subtle. One can first apply the already obtain results to the restriction of frame functions over real submanifolds of $\mathcal{H}$ (real submanifolds are real subspaces of $\mathcal{H}$ such that ($\vec x{\cdot}\vec y)$ is always real). One then extends the  obtained real quadratic form over the real submanifolds to a complex quadratic form on $\mathcal{H}$.

\end{enumerate}

\subsection{The Born rule}
\index{Born rule}
\index{Density matrix}
\index{Pure state}
The Born rule is a simple consequence of Gleason theorem.
Indeed, any state (in the general sense of statistical state) corresponds to a positive quadratic form (a density matrix) $\rho$ and given a minimal atomic proposition, which corresponds to a projector $P=|\vec a\rangle\langle\vec a|$ onto the ray corresponding to a single vector (pure state) $|\vec a\rangle$, the probability $p$ for $P$ of being true is
\begin{equation}
\label{ }
p=\langle P\rangle=\tr(\rho P)=\langle \vec a|\rho|\vec a\rangle
\end{equation}
The space of states $\mathcal{E}$ is thus the space of (symmetric) positive density matrices with unit trace 
\begin{equation}
\label{ }
\text{space of states}\ =\ \mathcal{E}=\{\rho:\ \rho=\rho^\dagger,\ \rho\ge 0\ , \tr(\rho)=1\}
\end{equation}
It is a convex set. Its extremal points, which cannot be written as a linear combination of twi different states, are the pure states of the system, and are the density matrices of rank one, i. e. the density matrices which are themselves projectors onto a vector $|\psi\rangle$ of the Hilbert space.
\begin{equation}
\label{ }
\rho\ =\ \text{pure state}\quad\implies\quad \rho=|\psi\rangle\langle\psi |
\quad,\quad \|\psi\|=1
\end{equation}
One thus derives the well known fact that the pure states are in one to one correspondence with the vectors (well... the rays) of the Hilbert space $\mathcal{H}$ that was first introduced from the bassc observables of the theory, the elementary atomic propositions (the projectors $P$).
Similarily, one recovers the simplest version of the Born rule: the probability to measure a pure state $|\varphi\rangle$ into another pure state $|\psi\rangle$ (to ``project''  $|\varphi\rangle$ onto $|\psi\rangle$) is the square of the norm of the scalar product
\begin{equation}
\label{ }
p(\varphi\to\psi)\ =\ |\langle\varphi|\psi\rangle|^2
\end{equation}

\subsection{Physical observables}
\index{Physical observable}
One can easily reconstruct the set of all physical observables, and the whole algebra of observables $\mathcal{A}$ of the system.
I present the line of the argument, without any attempt of mathematical rigor.

Any ideal physical mesurement of some observable $O$ consists in fact in taking a family of mutually  orthogonal propositions $a_i$, i.e. of commuting symmetric projectors $P_i$ on $\mathcal{H}$ such that
\begin{equation}
\label{ }
P_i^2=P_i\ ,\quad P_i=P_i^*\ ,\quad P_i P_j=P_j P_i=0\ \ \text{if}\ i\neq j
\end{equation}
performing all the tests (the order is unimportant since the projectors commute) and 
assigning a \emph{real number} $o_i$ to the result of the measurement (the value of the observable $O$) if $a_i$ is found true (this occurs for at most one $a_i$) and zero otherwise.
In fact one should take an appropriate limit when the number of $a_i$ goes to infinity, but I shall not discuss these important points of mathematical consistency.
If you think about it, this is true for any imaginable measurement (position, speed, spin, energy, etc.).
The resulting physical observable $O$ is thus associated to the symmetric operator
\begin{equation}
\label{ }
O=\sum_i o_i P_i
\end{equation}
This amounts to the spectral decomposition of symmetric operators in the theory of algebras of operators.
\index{Spectral decomposition}

Consider a system in a general state given by the density matrix $\rho$.
From the general rules of quantum probabilities, the probability to find the value $o_i$ for the measurement of the observable $O$ is simply the sum of the probabilities to find the system in a eigenstate of $O$ of eigenvalue $o_i$, that is
\begin{equation}
\label{ }
p(O\to o_i)= \tr(P_i\rho)
\end{equation}
and for a pure state $|\varphi\rangle$ it is simply 
\begin{equation}
\label{ }
\langle \varphi|P_i|\varphi\rangle= |\langle\varphi|\varphi_i\rangle|^2
\quad,\qquad |\varphi_i\rangle ={1\over \|P_i|\varphi\rangle\|}\, P_i | \varphi \rangle
\end{equation}
Again the Born rule!
\index{Born rule}
The expectation value for the result of the measurement of $O$ in a pure state $\psi_a$ is obviously 
\begin{equation}
\label{ }
\mathbb{E}[O;\psi]\ =\ \langle O\rangle_\psi= \sum_i o_i \ p(O\to o_i)\ =\ \sum_i o_i \langle\psi | P_i|\psi\rangle\ =\  \langle\psi | O |\psi\rangle
\end{equation}
This is the standard expression for expectation values of physical observables as diagonal matrix elements of the corresponding operators. Finally for general (mixed) states one has obviousy
\begin{equation}
\label{ }
\mathbb{E}[O;\rho]\ =\ \langle O\rangle_\rho= \sum_i o_i \, p(O\to o_i)\ =\ \sum_i o_i \,\tr(P_i \rho)\ =\  \tr( O \rho)
\end{equation}

We have seen that the pure states generate by convex combinations the convex set $\mathcal{E}$ of all (mixed) states $\psi$ of the system. 
Similarily the symmetric operators $O=O^\dagger$ generates (by operator multiplication and linear combinations) a C$^*$-algebra $\mathcal{A}$ of bounded operators $\mathcal{B}(\mathcal{H})$ on the Hilbert space $\mathcal{H}$.
States are normalized positive linear forms on $\mathcal{A}$ and we are back to the standard algebraic formulation of quantum physics.
The physical observables generates an algebra of operators, hence an abstract algebra of observables, as assumed in the algebraic formalism.
We refer to the section about the algebraic formalism for the arguments for preferring complex Hilbert spaces to real or quaternionic ones.

\section{Discussion}
This was a sketchy and partial introduction to the quantum logic approach for the formulation of the principle of quantum mechanics. 
I hope to have shown its relation with the algebraic formulation. It relies  on the concepts of states and of observables as the algebraic formulation. However the observables are limited to the physical subset of yes/no proposition, corresponding to ideal projective measurements, without assuming a priori some algebraic structure between non-compatible  propositions (non-commuting observables in the algebraic framework). I explained how the minimal set of axioms on these propositions and their actions on states, used in the quantum logic approach, is related to the physical concepts of causality, reversibility and separability/locality.
The canonical algebraic structure of quantum mechanics comes out from the symmetries of the ``logical structure'' of the lattice of propositions.
The propositions corresponding to ideal projective measurements are realized on orthogonal projections on a (possibly generalized) Hilbert space.
Probabilities/states are given by quadratic forms, and the Born rule follows from the logical structure of quantum probabilities through Gleason's theorem.

\index{von Neumann algebra}
\index{Bipartite system}
\index{Tensor product}
This kind of approach is of course not completely foolproof. We have seen that the issue if the possible division algebras $K$ is not completely settled. The strong assumptions of atomicity and covering are essential, but somehow restrictive compared to the algebraic approach (type II and III von Neumann algebras). It is sometime stated that it cannot treat properly the case of a system composed of two subsystems since there is no concept of ```tensorial product'' of two OM-AC lattices as there is for Hilbert spaces and operator algebras. Note however that one should in general always think about multipartite systems as parts of a bigger system, not the opposite! Even in the algebraic formulation it is not known in the infinite dimensional case if two commutting subalgebras $\mathcal{A}_1$ and $\mathcal{A}_2$ of a bigger algebra $\mathcal{A}$ always correspond  to the decomposition of the Hilbert space $\mathcal{H}$ into a tensor product of two subspaces $\mathcal{H}_1$ and $\mathcal{H}_2$.

\chapter{Additional discussions}
\label{c:misc}
\section{Quantum information approaches}
\label{s:QuanInfo}
\index{Quantum information}
Quantum information science has undergone enormous developments in the last 30 years. I do not treat this wide and fascinating field here, but shall only discuss briefly some relations with the question of formalism. Indeed information theory leads to new ways to consider and use  quantum theory. This renewal is sometimes considered as a real change of paradigm.

The interest in the relations between Information Theory and Quantum Physics started really in the 70's from several questions and results:
\begin{itemize}
  \item The relations and conflicts between General Relativity and Quantum Physics: the theoretical discovery of the Bekenstein-Hawking quantum entropy for black holes, the black hole evaporation (information) paradox, the more general Unruh effect and quantum  thermodynamical aspects of gravity and of events horizons (with many recent developments in quantum gravity and string theories, such as ``Holographic gravity'', ``Entropic Gravity'', etc.).
  \item The general ongoing discussions on the various interpretations of the quantum formalism, the meaning of quantum measurement processes, and whether a quantum state represent the ``reality'', or some ``element of reality'' on a quantum system, or simply the observer's information on the quantum system.
  \item Of course the theoretical and experimental developments of quantum computing. See for instance \cite{Nielsen:2010fk}. It started from the realization that quantum entanglement and quantum correlations can be used as a resource for performing calculations and the transmission of information in a more efficient way than when using classical correlations with classical channels.
   \item This led for instance to the famous ``It from Bit'' idea (or aphorism) of J. A. Wheeler (see e.g. in \cite{Zurek:1990fk})
and others (see for instance the book by Deutsch \cite{Deutsch:1997fk}, or talks by Fuchs \cite{Fuchs-2001,Fuchs-2002}). Roughly speaking this amounts to reverse the famous statement of Laudauer ``Information is Physics'' into ``Physics is Information'', and to state that Information is the good starting point to understand the nature of the physical world and of the physical laws. 
\end{itemize}

This point of view has been developed and advocated by several authors in the area of quantum gravity and quantum cosmology. 
Here I shall just mention some old or recent attempts to use this point of view to discuss the formalism of ``standard'' quantum physics, not taking into account the issues of quantum gravity.

In the quantum information inspired approaches a basic concept is that of 
``device'', or ``operation'', which represents the most general manipulation on a quantum system. In a very oversimplified presentation\footnote{slightly more general than in some presentations}, such a device  is a ``black box'' with both a quantum input system A and  quantum output system B, and with a set $I$ of classical settings $i\in I$ and a set $O$ of classical responses $o\in O$. 
\index{Quantum device}
%
The input and output systems $A$ and $B$ may be different, and may be multipartite systems, e.g. may consist in collections of independent subsystems $A=\mathop\cup\limits_\alpha A_\alpha$, $B=\mathop\cup\limits_\alpha B_\beta$.

This general concept of device encompass the standard concepts of \emph{state} and of \emph{effect}.
\index{State (preparation)}\index{Effect}
A \emph{state} corresponds to the preparation of a quantum system $S$ in a definite state; there is no input $A=\emptyset$, the setting $i$ specify the state, there is no response, and  $B=S$ is the system.
An \emph{effect}, corresponds to a destructive measurement on a quantum system $S$;  the input $A=S$ is the system, there are no output $B=\emptyset$, no settings $i$, and the response set $O$ is the set of possible output measurements $o$.
This concept of device contains also the general concept of a \emph{quantum channel};  then $A=B$, there are no settings or responses.
\begin{figure}[h]
\begin{center}
\includegraphics[width=2in]{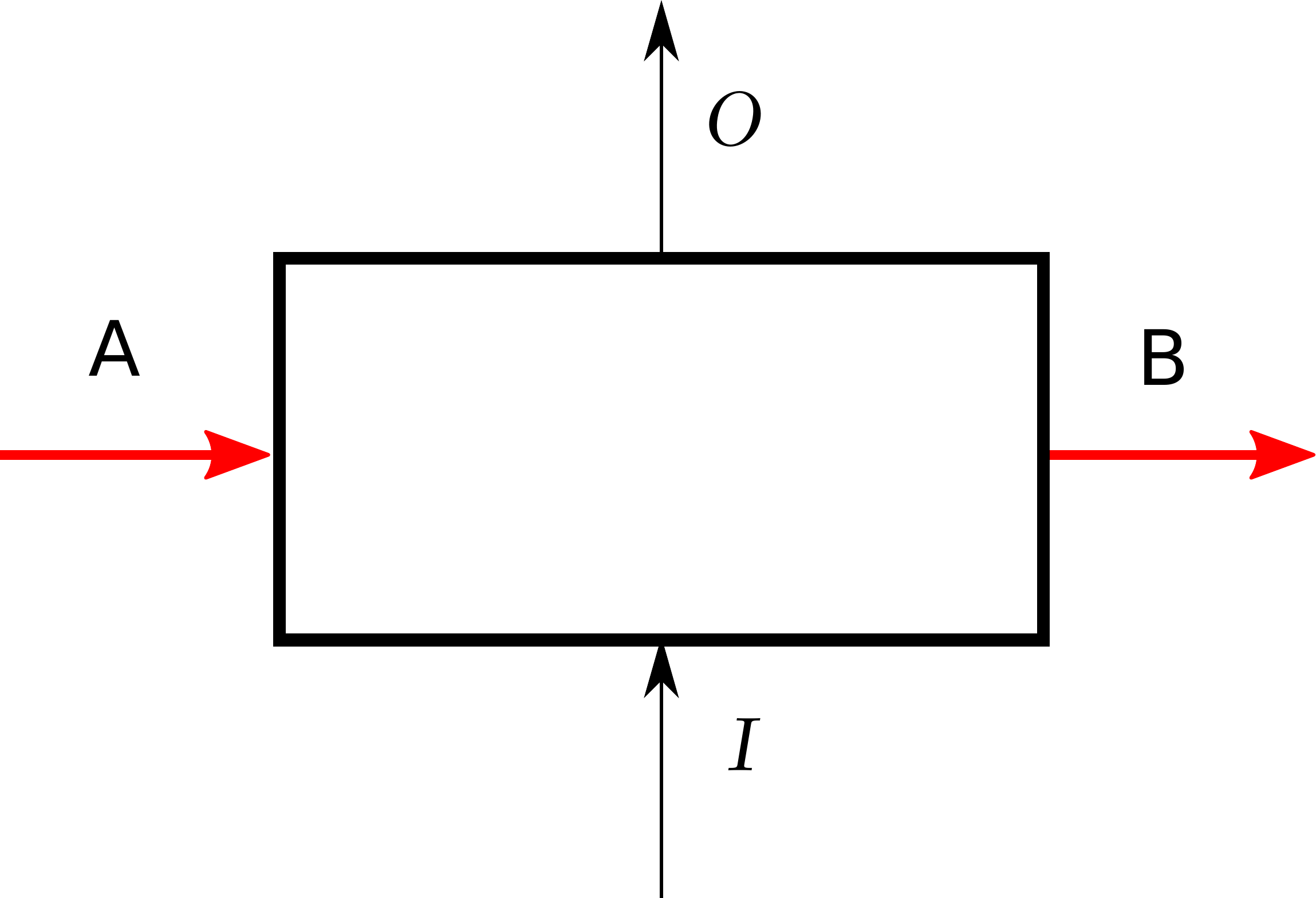}
\qquad
\raisebox{-.9ex}{\includegraphics[width=1.2in]{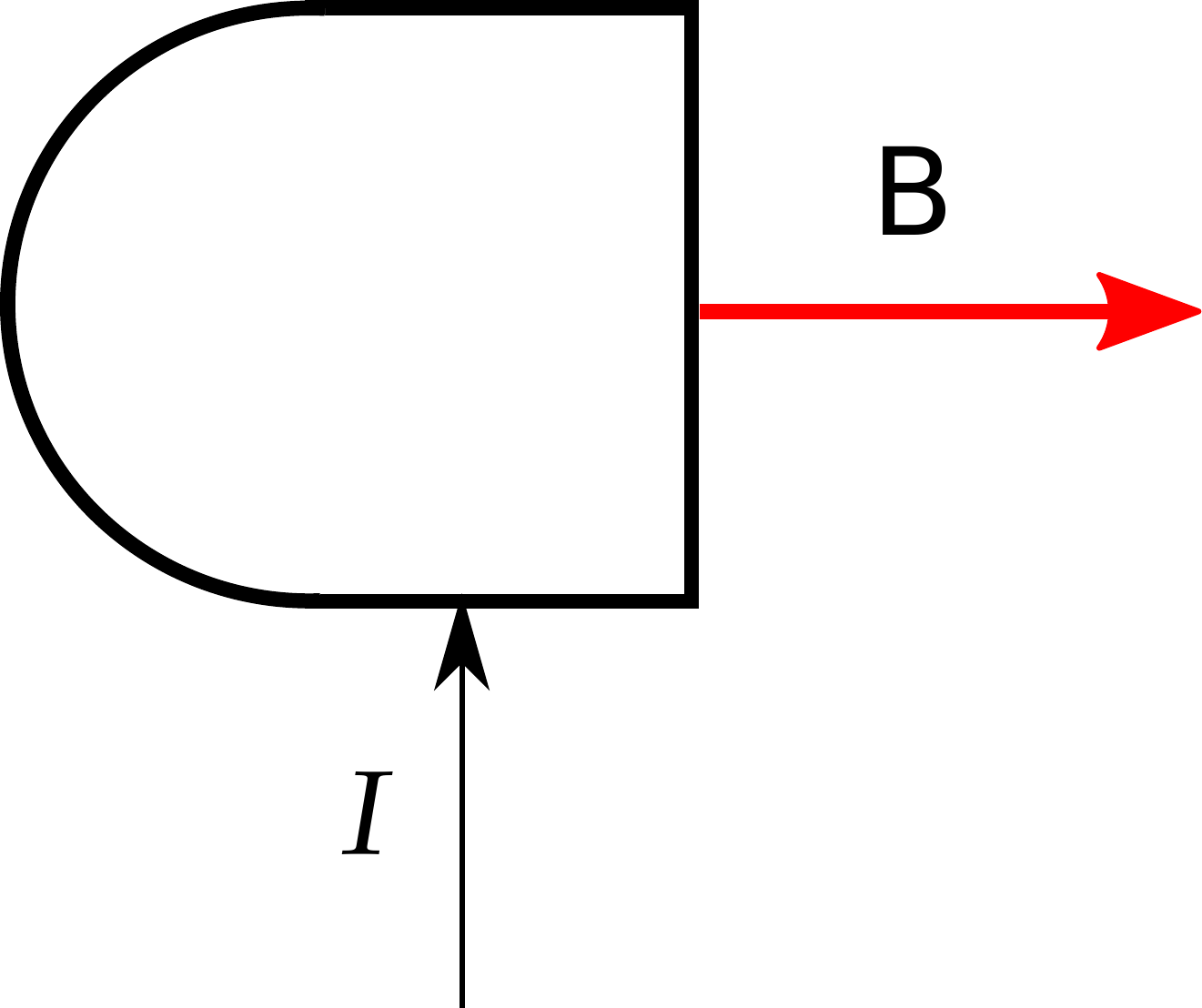}}
\qquad
\raisebox{4.ex}{\includegraphics[width=1.2in]{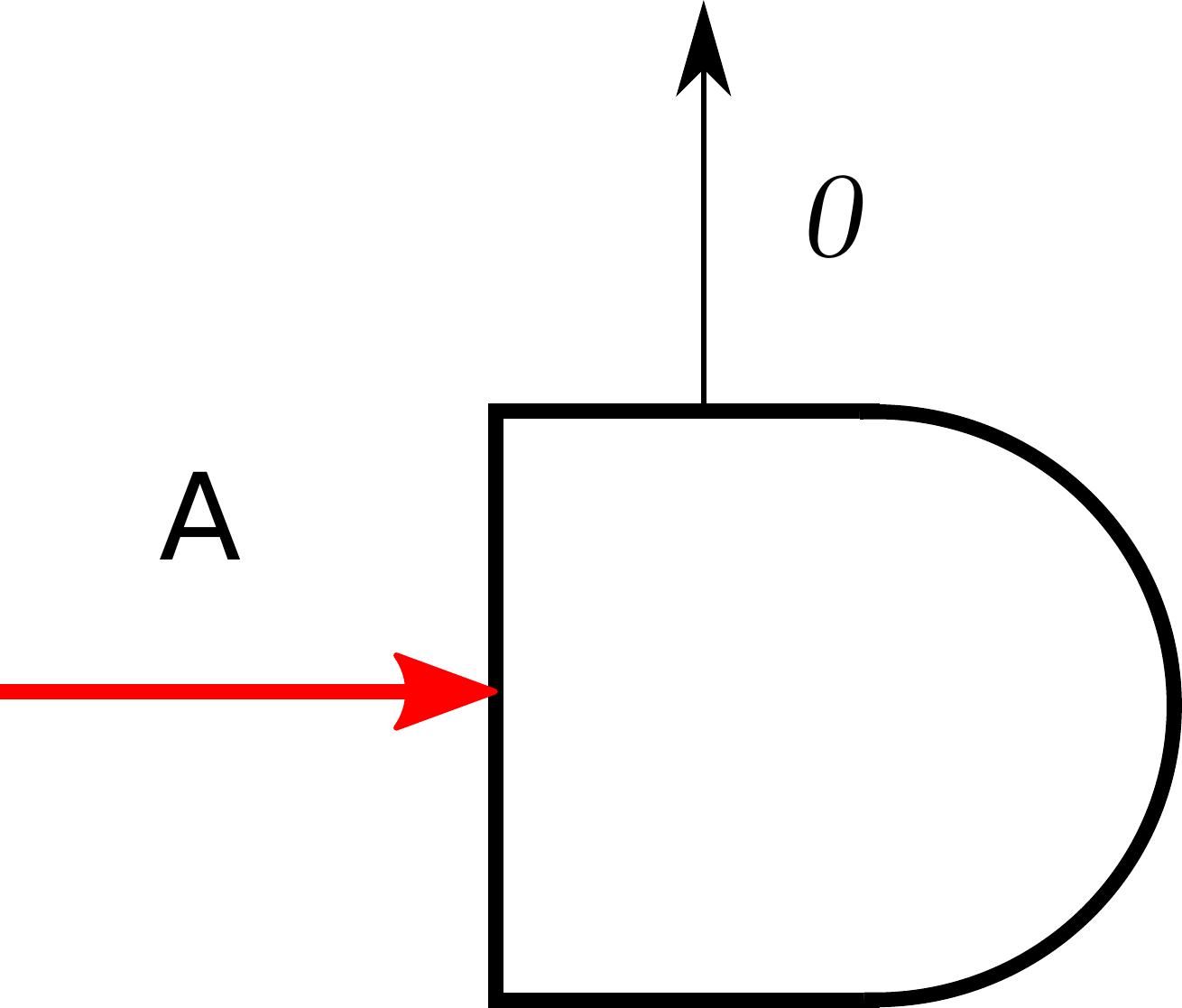}}
\caption{A general device, a state and an effect }
\label{fdevice}
\end{center}
\end{figure}
\index{Probabilities}
Probabilities $p(i | o)$  are associated to the combination of a state and an effect, this correspond to the standard concept of probability of observing some outcome $o$ when making a measurement on a quantum state (labeled by $i$).
\begin{figure}[h]
\begin{center}
\includegraphics[width=2in]{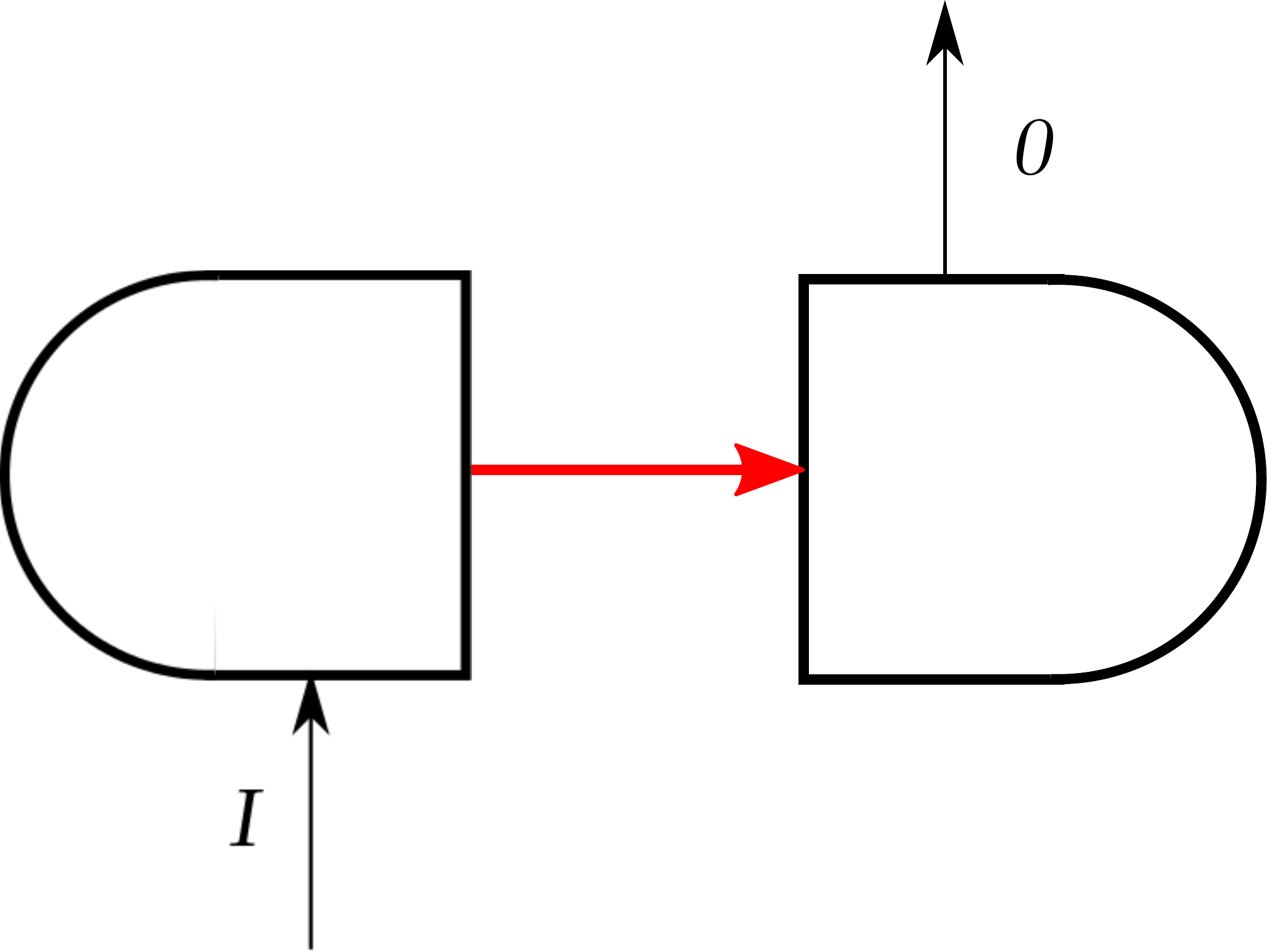}
\caption{ Probabilities are associated to a couple state-effect}
\label{ }
\end{center}
\end{figure}

General information processing quantum devices are constructed by building causal circuits out of these devices used as building blocks, thus constructing complicated apparatus out of simple ones.
An information theoretic formalism is obtained by choosing axioms on the properties of such devices (states and effects) and  operational rules to combine these devices and circuits and the associated probabilities, thus obtaining for  instance what is called in \cite{PhysRevA.84.012311} an operational probabilistic theory.
This kind of approach is usually considered for finite dimensional theories (which in the quantum case correspond to finite dimensional Hilbert spaces), both for mathematical convenience, and since this is the kind of system usually considered in quantum information science.

This approach  leads to a pictorial formulation of quantum information processing. It shares similarities with the ``quantum pictorialism" logic formalism, more based on category theory, and presented for instance in
\cite{Coecke-2010}.

It can also be viewed as an operational and informational extension of the convex set approach (developed notably by G. Ludwig, see
\cite{Ludwig:1985uq} and  \cite{Auletta},\cite{BeltCassi81} for details). This last approach puts more emphasis on the concept of states than on observables in QM.

I shall not discuss in any details these approaches. 
Let me just highlight amongst  the most recent attempts those of Hardy \cite{Hardy-2001,Hardy-2011} and those of Chiribella, D'Ariano \& Perinotti  
\cite{PhysRevA.81.062348, PhysRevA.84.012311} (
see \cite{Physics.4.55} for a short presentation of this last formulation).
See also \cite{MasMue2011}.
In \cite{PhysRevA.84.012311} the standard complex Hilbert space formalism of QM is derived from 6 informational principles: 
Causality,
Perfect Distinguishability,
Ideal Compression,
Local Distinguishability,
Pure conditioning,
Purification principle.

The first 2 principles are not very different from the principles of other formulations (causality is defined in a standard sense, and distinguishability is related to the concept of differentiating states by measurements).
The third one is related to existence of reversible maximally efficient compression schemes for states. 
The four and the fifth are about the properties of bipartite states and for instance the possibility to performing local tomography and the effect of separate atomic measurements on such states.
The last one, about ``purification'' distinguishes quantum mechanics from classical mechanics, and states that any mixed state of some system $\mathcal{S}$ may be obtained from a pure state of a composite larger system $\mathcal{S}+\mathcal{S}'$.
See \cite{Physics.4.55} for a discussion of the relation of this last purification principle with the discussions of the ``cut'' between the system measured and the measurement device done for instance by Heisenberg in \cite{pittphilsci8590}, but see of course the previous discussion by von Neumann in \cite{vonNeumann32G}.

\section{Quantum correlations}
\label{sQuCor}
The world of quantum correlations is richer, more subtle and more interesting than the world of classical ones. Most of the puzzling features and seeming paradoxes of quantum physics come from these correlations, and in particular from the phenomenon of entanglement. 
Entanglement is probably \emph{the} distinctive feature of quantum mechanics, and is a consequence of the superposition principle when considering quantum states for composite systems. Here I discuss briefly some basic aspects.
Entanglement describes the particular quantum correlations between two quantum systems which (for instance after some interactions) are in a non separable pure state,  so that each of them considered separately, is not in a pure state any more.
Without going into history, let me remind that if the terminology ``entanglement'' (``Verschränkung'') was introduced in the quantum context  by E. Schrödinger in 1935 (when discussing the famous EPR paper).
However the mathematical concept 
 is older and goes back to the modern formulation of quantum mechanics. 
 For instance, some peculiar features of entanglement and its consequences have been  discussed already around the 30' in relation with the theory of quantum measurement by Heisenberg, von Neumann, Mott, etc. Examples of interesting entangled many particles states are provided by the Stater determinant for many fermion states, by the famous Bethe ansatz for the ground state of the spin 1/2 chain, etc.
 
\subsection{Entropic inequalities}
\paragraph{von Neumann entropy:} The difference between classical and quantum correlations is already visible when considering the properties of the von Neumann entropy of states of composite systems.
Remember that the von Neumann entropy of a mixed state of a system $A$, given by a density matrix $\rho_A$, is given by
\index{von Neumann entropy}
\begin{equation}
\label{ }
S(\rho_A)=-\tr(\rho_A\log\rho_A)
\end{equation}
In quantum statistical physics, the $\log$ is usually the natural logarithm
\begin{equation}
\label{ }
\log=\log_{\emath}=\ln
\end{equation}
while in quantum information, the $\log$ is taken to be the binary logarithm
\begin{equation}
\label{ }
\log=\log_{2}
\end{equation}
The entropy measures the amount of ``lack of information" that we have on the state of the system.  But in quantum physics, at variance with classical physics, one must be very careful about the meaning of ``lack of information'', since one cannot speak about the precise state of a system before making measurements. 
So the entropy could (and should) rather be viewed as a measure of the number of independent measurements we can make on the system before having extracted all the information, i.e. the amount of information we can extract of the system. It can be shown also that the entropy give the maximum information capacity of a quantum channel that we can build out of the system.
See \cite{Nielsen:2010fk} for a good introduction to quantum information and in particular on entropy viewed from the information theory point of view.

When no ambiguity exists on the state $\rho_A$ of the system $A$, I shall use the notations
\begin{equation}
\label{ }
S_A=S(A)=S(\rho_A)
\end{equation}

The von Neumann entropy shares many  properties of the classical entropy. It has the same  convexity properties
\begin{equation}
\label{ }
S[\lambda\rho+(1-\lambda)\rho']\ge \lambda S[\rho]+(1-\lambda)S[\rho']\quad,\qquad  0\le\lambda\le 1
\end{equation}
It is minimal $S=0$ for systems in a  pure state and maximal for systems in a equipartition state $S=\log(N)$ if $\rho={1\over N} \mathbf{1}_N$.
It is extensive for systems in separate states.

\paragraph{Relative entropy:} 
\index{Relative entropy}
The relative entropy (of a state $\rho$ w.r.t. another state $\sigma$ for the same system) is defined as in classical statistics (Kullback-Leibler entropy) as 
\begin{equation}
\label{ }
S(\rho\|\sigma)=\tr(\rho\log\rho)-\tr(\rho\log\sigma)
\end{equation}
with the same convexity properties.

The differences with the classical entropy arise for composite systems. For such a system $AB$, composed of two subsystems $A$ and $B$, a general mixed state is given by a density matrix $\rho_{AB}$ on $\mathcal{H}_{AB}=\mathcal{H}_{A}\otimes\mathcal{H}_{B}$.
The reduced density matrices for $A$ and $B$ are 
\begin{equation}
\label{ }
\rho_A=\tr_B(\rho_{AB})
\quad,\qquad
\rho_B=\tr_A(\rho_{AB})
\end{equation}
This corresponds to the notion of marginal distribution w.r.t. $A$ and $B$ of the general probability distribution of states for $AB$ in classical statistics.
Now if one considers
\begin{equation}
\label{ }
S({AB})=-\tr(\rho_{AB}\log\rho_{AB})\quad,\qquad
S({A})=-\tr(\rho_{A}\log\rho_{A})\quad,\qquad
S({B})=-\tr(\rho_{B}\log\rho_{B})
\end{equation}
one has the following definitions.
\paragraph{Conditional entropy:} 
\index{Conditional entropy} 
The \emph{conditional entropy} S(A|B) (the entropy of $A$ conditional to $B$ in the composite system $AB$) is
\begin{equation}
\label{ }
S(A|B)=S(AB) - S(B)
\end{equation}
The conditional entropy $S(A|B)$ corresponds to the remaining uncertainty (lack of information) on $A$ if $B$ is known.
\paragraph{Mutual information:} 
\index{Mutual information}
The \emph{mutual information} (shared by $A$ and $B$ in the composite system $AB$)
\begin{equation}
\label{MutInf}
S(A:B)=S(A)+S(B)-S(AB)
\end{equation}
\paragraph{Subadditivity:} 
\index{Subadditivity}
The entropy satisfies the general inequalities (triangular inequalities)
\begin{equation}
\label{subaddS}
|S(A)-S(B)|\le S({AB})\le S(A)+S(B)
\end{equation}
The rightmost inequality $S({AB})\le S(A)+S(B)$ is already valid for classical systems, but the leftmost  is quantum.
Indeed for classical systems the classical entropy $H_\mathrm{cl}$ satisfy only the much stronger lower bound
\begin{equation}
\label{CaddS}
\max(H_\mathrm{cl}(A),H_\mathrm{cl}(B)) \le H_\mathrm{cl}(AB)
\end{equation}

Subadditivity implies that if $AB$ is in a pure entangled state, $S(A)=S(B)$.
It also implies that the mutual information in a bipartite system is always positive
\begin{equation}
\label{MutInfPos}
S(A:B)\ge 0
\end{equation}

In the classical case the conditional entropy is always positive $H_\mathrm{cl}(A|B)\ge 0$.
In the quantum case the conditional entropy may be negative $S(A|B)< 0$ if the entanglement between $A$ and $B$ is large enough.
This is a crucial feature of quantum mechanics. If $S(A|B)<0$ it means that $A$ and $B$ share information resources (through entanglement) which get lost if one gets information on $B$ only (through a measurement on $B$ for instance). 

\paragraph{Strong subadditivity:}
\index{Strong subadditivity}
Let us consider a tripartite systems $ABC$. The entropy satisfies another very interesting inequality
\begin{equation}
\label{StrSubAdd1}
S(A)+S(B)\le S({AC})+S({BC})
\end{equation}
It is equivalent to (this is the usual form)
\begin{equation}
\label{StrSubAdd2}
S({ABC})+S({C})\le S({AC})+S({BC})
\end{equation}
Note that \ref{StrSubAdd1} is also true for the classical entropy, but then for simple reasons. In the quantum case it is a non trivial inequality.

The strong subadditivity inequality implies the triangle inequality for tripartite systems
\begin{equation}
\label{ }
S(AC)\le S(AB)+S(BC)
\end{equation}
so the entropic inequalities can be represented graphically as in fig.~\ref{fSSubT}
\begin{figure}[h]
\begin{center}
\includegraphics[width=2in]{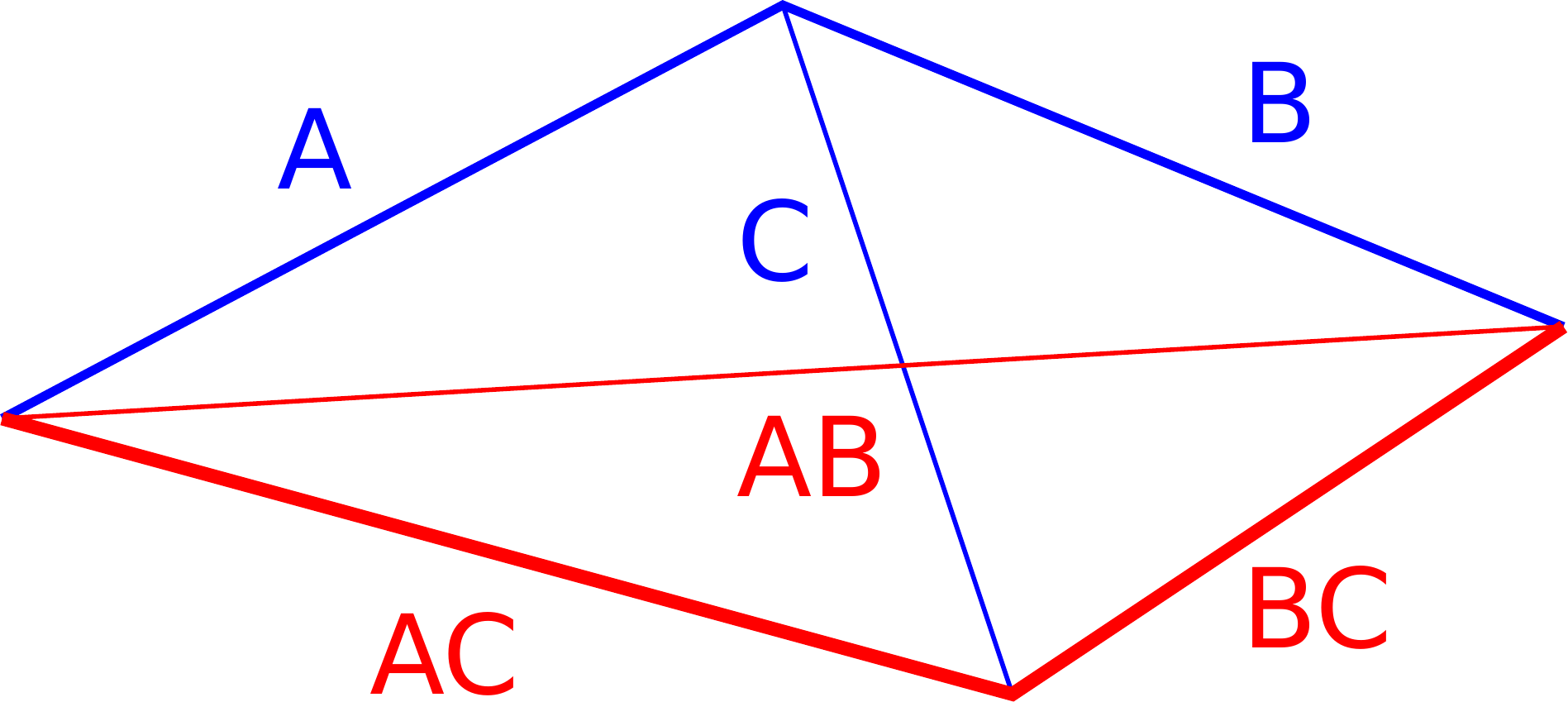}
\caption{Entropic inequalities: the length of the line ``$X$''  is the von Neumann entropy $S(X)$. The tetrahedron has to be ``oblate", the sum {\color{red}{AC+BC}} (fat red lines) is always $\ge$ the sum {\color{blue}{A+B}} (fat blue lines). }
\label{fSSubT}
\end{center}
\end{figure}

The strong subadditivity inequality has important consequences for conditional entropy and mutual information (see \cite{Nielsen:2010fk}). 
Consider a tripartite composite system $ABC$.
It implies for instance 
\begin{equation}
\label{ }
S(C|A)+S(C|B) \ge 0
\end{equation}
and
\begin{equation}
\label{ }
S(A|BC)\le S(A|B) 
\end{equation}
which means that conditioning A to a part of the external subsystem (here C inside BC) increase the information we have on the system (here A).
One has also for the mutual information
\begin{equation}
\label{ }
S(A:B)\le S(A:BC) 
\end{equation}
This means that discarding a part of a multipartite quantum system (here $C$) increases the mutual information (here between $A$ and the rest of the system).
This last inequality is very important. It implies for instance that if one has a composite system $AB$, performing some quantum operation on $B$ without touching to $A$ cannot increase the mutual information between $A$ and the rest of the system. 

Let us mention other subadditivity inequalities for tri- or quadri-partite systems.
\begin{align}
\label{ }
S(AB|CD)&\le S(A|C) + S(B|D)
\\
\label{ }
S(AB|C)&\le S(A|C) + S(B|C)
\\
\label{ }
S(A|BC)&\le S(A|B) + S(A|C)
\end{align}

%
%
%
%
%
%
%

\subsection{Bipartite correlations:}
The specific properties of quantum correlations between two causally separated systems are known to disagree with
what one would expect from a ``classical picture'' of quantum theory, where the quantum probabilistics features come just from some lack of knowledge of underlying ``elements of reality''.
I shall come back later on the very serious problems with the ``hidden variables'' formulations of quantum mechanics.
But let us discuss already some of the properties of these quantum correlations in the simple case of a bipartite system. 

I shall discuss briefly one famous and important result: the  Tsirelson bound.
The general context is that of the discussion of non-locality issues and of Bell's \cite{Bell:1964kc} and CHSH inequalities \cite{PhysRevLett.23.880} in bipartite systems. 
However, since these last inequalities are more of relevance when discussing hidden variables models, I postpone their discussion to
the next section \ref{sHidVar}.

This presentation is standard and simply taken from \cite{Laloe-book}.

\subsubsection{The Tsirelson Bound}
\label{ssTsiBn}
\index{Tsirelson bound}

\paragraph{The two spin system:} Consider a simple bipartite system consisting of two spins 1/2, or q-bits 1 and 2.
If two observers (Alice $\mathcal{A}$ and Bob $\mathcal{B}$) make independent measurements of respectively the value of the spin 1 along some direction $\vec n_1$ (a unit vector in 3D space)
and of the spin 2 along $\vec n_2$, at each measurement they get results (with a correct normalization) $+1$ or $-1$.
Now let us compare the results of four experiments, depending whether $\mathcal{A}$ choose to measure the spin 1 along a first direction $\vec a$ or a second direction $\vec a'$, and wether $\mathcal{B}$ chose (independently) to measure the spin 2 along a first direction $\vec b$ or a second direction $\vec b'$. Let us call the corresponding observables $\mathbf{A}$, $\mathbf{A'}$, $\mathbf{B}$, $\mathbf{B'}$, and by extension the results of the corresponding measurements in a single experiment $A$ and $A'$ for the first  spin, $B$ and $B'$ for the second spin.

\begin{equation}
\label{ }
\text{spin 1 along }\ \vec a \quad \to\quad A= \pm 1
\quad;\quad
\text{spin 1 along }\ \vec a' \quad \to\quad A'= \pm 1
\end{equation}

\begin{equation}
\label{ }
\text{spin 2 along }\ \vec b \quad \to\quad B= \pm 1
\quad;\quad
\text{spin 2 along }\ \vec b' \quad \to\quad B'= \pm 1
\end{equation}

Now consider the following combination $M$ of products of observables, hence of products of results of experiments
\begin{equation}
\label{McorBHSH}
M=AB-A B' + A' B + A' B'
\end{equation}
and consider the expectation value $\langle M\rangle_\psi$ of $M$ for a given quantum state $|\psi\rangle$ of the two spins system.
In practice this means that we prepare the spins in state  $|\psi\rangle$, chose randomly (with equal probabilities) one of the four observables, and to test locality $\mathcal{A}$ and $\mathcal{B}$ may be causally deconnected, and choose independently (with equal probabilities) one of their own two observables, i.e. spin directions. Then they make their measurements. The experiment is repeated a large number of time  and the right average combination $M$ of the results of the measurements is calculated afterwards.

A simple explicit calculation shows the following inequality, known as the Tsirelson bound \cite{springerlink:10.1007/BF00417500}

\paragraph{Tsirelson bound:}
For any state and any choice orientations $\vec a$, $\vec a'$, $\vec b$ and $\vec b'$, one has
\begin{equation}
\label{TsirBound}
|\langle M\rangle |\le 2\sqrt{2}
\end{equation}
while, as discussed later, ``classically'', i.e. for theories where the correlations are described by contextually-local hidden variables attached to each subsystem, one has the famous Bell-CHSH bound \index{CHSH inequality}
\begin{equation}
\label{ }
\langle|M|\rangle_\mathrm{``classical''}\le 2.
\end{equation}
The Tsirelson bound is saturated if the state $|\psi\rangle$ for the two spin is the singlet
\begin{equation}
\label{ }
|\psi\rangle=|\text{singlet}\rangle={1\over \sqrt{2}}\left(  |\uparrow\rangle\otimes | \downarrow\rangle - |\downarrow\rangle\otimes | \uparrow\rangle \right)
\end{equation}
and the directions for  $\vec a$, $\vec a'$, $\vec b$ and $\vec b'$ are coplanar, and such that $\vec a\perp\vec a'$, $\vec b\perp\vec b'$, and the angle between $\vec a$ and $\vec b$ is $\pi/4$, as depicted on \ref{figBCHSH}.
\begin{figure}[h]
\begin{center}
\includegraphics[width=1.8in]{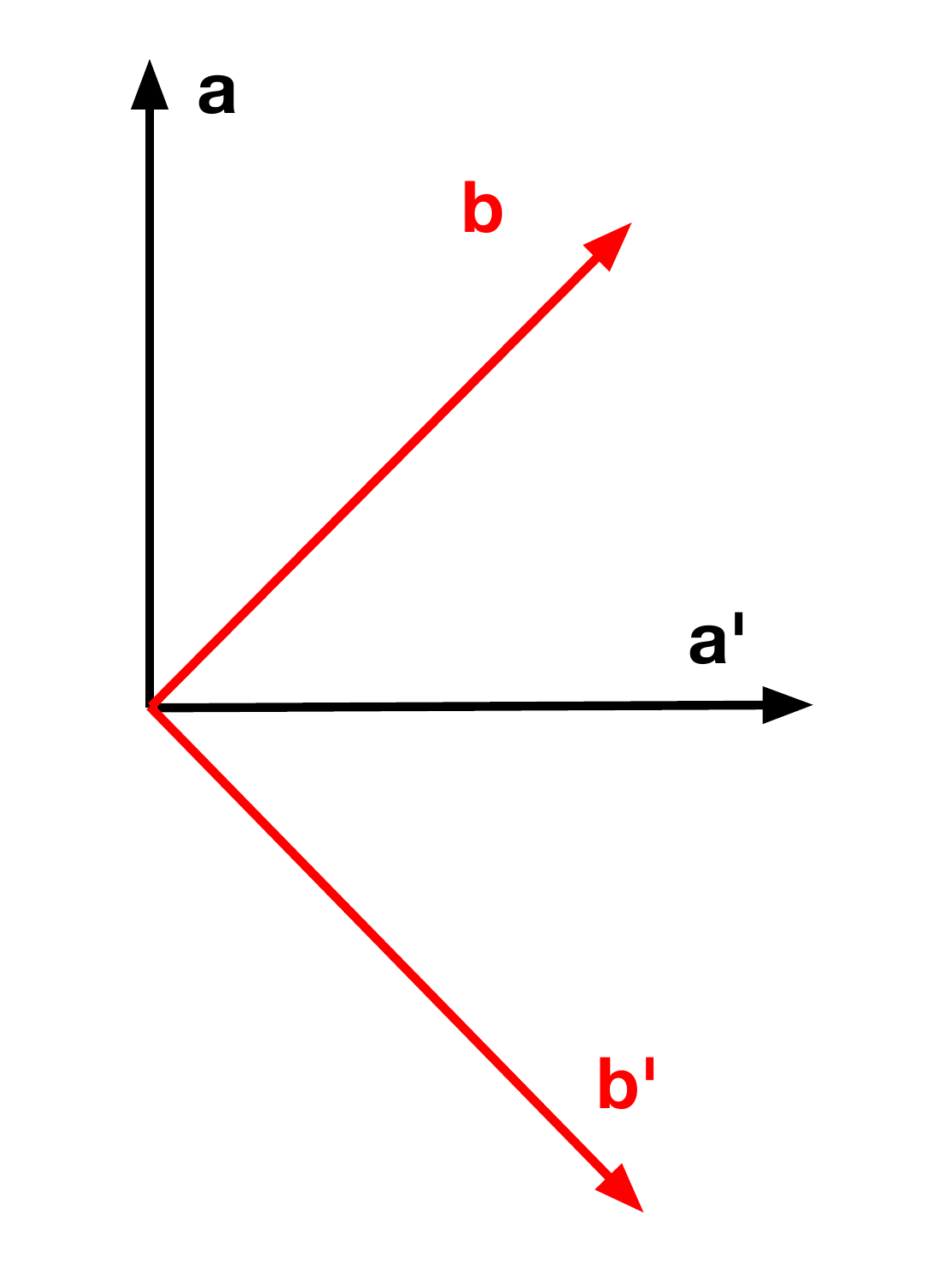}
\caption{Spin directions for saturating the Tsirelson bound and maximal violation of the Bell-CHSH inequality}
\label{figBCHSH}
\end{center}
\end{figure}

\subsubsection{Popescu-Rohrlich boxes}
\index{Popescu-Rohrlich boxes}
\paragraph{Beyond the Tsirelson bound ?}
Interesting questions arise when one consider what could happen if there are ``super-strong correlations'' between the two spins (or in general between two subsystems) that violate the Tsirelson bound.
Indeed, the only mathematical bound on $M$ for general correlations is obviously $|\langle M\rangle|\le 4$.
Such hypothetical systems are considers in the theory of quantum information and are denoted Popescu-Rohrlich boxes
\cite{springerlink:10.1007/BF02058098}
. 
With the notations of the previously considered 2 spin system, BR-boxes consist in a collection of probabilities $P(A,B|a,b)$
for the outputs $A$ and $B$ of the two subsystems, the input or settings $a$ and $b$ being fixed. 
The $(a,b)$ correspond to the settings $I$ and the $(A,B)$ to the outputs $O$ of fig.~\ref{fdevice} of the quantum information section.
In our case we can take for the first spin
\begin{equation}
\label{ }
a=1\ \to\ \text{chose orientation}\  \vec a\quad,\qquad
a=-1\ \to\ \text{chose orientation}\  \vec a'
\end{equation}
and for the second spin
\begin{equation}
\label{ }
b=1\ \to\ \text{chose orientation}\  \vec b\quad,\qquad
b=-1\ \to\ \text{chose orientation}\  \vec b'
\end{equation}
The possible outputs being always $A=\pm 1$ and $B=\pm 1$.

\begin{figure}[h]
\begin{center}
\includegraphics[width=2in]{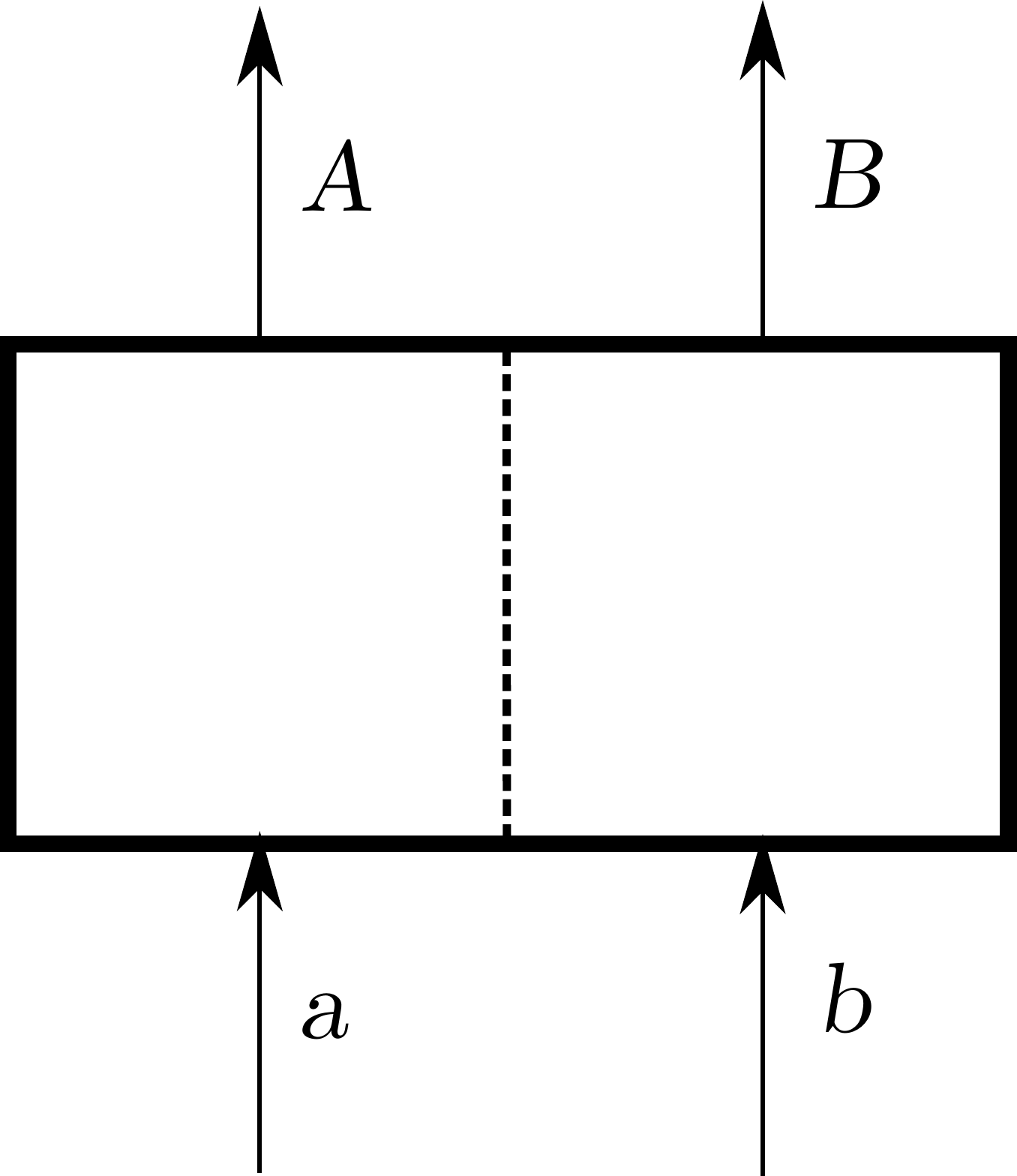}
\caption{a Popescu-Rohrlich box}
\label{fPRbox}
\end{center}
\end{figure}

The fact that the $P(A,B|a,b)$ are probabilities means that
\begin{equation}
\label{ }
0\le P(A,B|a,b)\le 1\quad,\qquad \sum_{A,B}P(A,B|a,b)=1\quad\text{for}\quad a,b\ \text{fixed}
\end{equation}
\paragraph{Non signalling:} 
If the settings $a$ and $b$ and the outputs $A$ and $B$ are relative to two causally separated parts of the system, corresponding to manipulations by two independent agents (Alice and Bob), enforcing causality means that Bob cannot guess which setting ($a$ or $a'$) Alice has chosen from his choice of setting ($b$ and $b'$) and his output ($B$ or $B$), without knowing Alices' output $A$. The same holds for Alice with respect to Bob.
This requirement is enforced by the non-signaling conditions
\begin{align}
\label{}
    \sum_A P(A,B|a,b) = &  \sum_A P(A,B|a',b) \\
    \sum_B P(A,B|a,b) = &  \sum_B P(A,B|a,b')  
\end{align}

A remarkable fact is that there are choices of probabilities which respect the non-signaling condition (hence causality) but violate the Tsirelson  bound and even saturate the absolute bound $|\langle M\rangle|=4$ .
Such hypothetical devices  would allow to use ``super-strong correlations'' (also dubbed ``super-quantum correlations'') to manipulate and transmit information in a more efficient way that quantum systems (in the standard way of quantum information protocols, by sharing some initially prepared bipartite quantum system and exchanges of classical information) 
\cite{PhysRevLett.96.250401}
\cite{vanDam2005}
\cite{10.1038/nature08400}
.  However, besides these very intriguing features of ``trivial communication complexity", such devices are problematic. 
For instance it seems that no interesting dynamics can be defined on such systems  \cite{PhysRevLett.104.080402}.

\section{The problems with hidden variables}
\label{sHidVar}
\subsection{Hidden variables and ``elements of reality''}
\index{Hidden variable}
In this section I discuss briefly some features of quantum correlations which are important when discussing the possibility that the quantum probabilities may still have, to some extent, a ``classical interpretation'' by reflecting our ignorance of inaccessible ``sub-quantum'' degrees of freedom or ``elements of reality'' of quantum systems, which could behave in a more classical and deterministic way. In particular a question is: which general  constraints on such degrees of freedom are enforced by quantum mechanics?

This is the general idea of the ``hidden variables'' program and of the search of explicit hidden variable models. These ideas go back to the birth of quantum mechanics, and were for instance proposed by L. de Broglie in his first ``pilot wave model'', but they were abandoned by most physicist after 1927 Solvay Congress and the advances of the 1930' , before experiencing some revival  and setbacks in the 1960', from the works by Bohm and de Broglie, and the discussions about locality and Bell-like inequalities.

The basic idea is that when considering a quantum system $\mathcal{S}$, its state could be described by some (partially or totally) hidden variables $\mathfrak{v}$ in some space $\mathfrak{V}$, with some unknown statistics and dynamics. 
Each $\mathfrak{v}$ may represent a (possibly infinite) collection of more fundamental variables.
But they are such that the outcome of a measurement operation of a physically accessible  observables $A$ is  determined by the hidden variable $\mathfrak{v}$. \begin{equation}
\label{m2oVD}
\text{mesurement of}\quad A\ \to \ \text{outcome}\quad a=f(A,\frak{v})\quad\text{(a real number)}
\end{equation}
Quantum undeterminism should come from our lack of knowledge on the exact state of the hidden variables.
In other word, the pure quantum states $|\psi\rangle$ of the system should correspond to some classical probability distribution $p_\psi(\mathfrak{v})$ on $\mathcal{V}$.
Of course a measurement operation could back react on the hidden variables $\mathfrak{v}$.

This is probably an oversimplified presentation of the idea, since there are several versions and models. But for instance in the hidden variable model of de Broglie and Bohm (for a single particle obeying the Schrödinger equation), the hidden variable $\mathfrak{v}=(\psi,x)$, where $\psi=\{\psi(y)\}$ is the whole ``pilot" wave function, and $x$ the position of the particle.

In its simplest version, one could try to consider hidden variables (element of reality) that are in one to one correspondence with the possible outcomes $a$ of all the observables $A$ of the system, and in particular which obeys the addition law
\begin{equation}
\label{CHVvN}
C=A+B \quad\implies\quad c=a+b\quad\text{i.e}\qquad f(C,\frak{v})=f(A,\frak{v})+f(B,\frak{v})
\end{equation}
This possibility is already discussed by J. von Neumann in his 1932 book
\cite{vonNeumann32G,vonNeumann32}, 
\index{von Neumann J.}
where it is shown to be clearly inconsistent. Indeed if $A$ and $B$ do not commute, the possible outcomes of $C$ (the eigenvalues of the operator $C$) are not in general sums of outcomes of $A$ and $B$ (sums of eigenvalues of $A$ and $B$), since $A$ and $B$ do not have common eigenvectors.
See \cite{Bub:2010kx} for a detailed discussion of the argument and of its historical significance.


\subsection{Context free hidden variables ?}
\index{Noncontextuality}
Hidden variable models have been rediscussed a few decades later, from a more realistic point of view, in particular by J. Bell.
In a modern language, the models considered are  ``context free'' or ``non contextual''  hidden variables models. 
The idea is that one should consider only the correlations between results of  measurements  on a given system for \emph{sets of commuting observables}. Indeed only such measurements can be performed independently and in any possible order (on a single realization of the system), and without changing the statistics of the outcomes. Any such given set  of observables can be thought as a set of classical observables, but of course this classical picture is not consistent from one set to another.

Thus the idea is still that a hidden variable $\mathfrak{v}$ assigns to any observable $A$ an outcome $a=f(A,\mathfrak{v})$ as in \ref{m2oVD}.
This assumption is often called ``value definiteness'' (VD).
\index{Value definiteness} 

However the very strong constraint \ref{CHVvN} should be replaced by the more realistic constraint for the set of outcomes $\{f(A,\mathfrak{v})$

\begin{align}
\label{CHVNCon}
\text{if $A$ and $B$ commute, then}\quad\begin{cases}
      & f(A+B,\mathfrak{v})=f(A,\mathfrak{v})+f(B,\mathfrak{v})\\
      &\hskip 5em \text{and} \\
      & f(AB,\mathfrak{v})=f(A,\mathfrak{v})f(B,\mathfrak{v})
\end{cases}
\end{align}
Moreover, these conditions are extended to  any  family $\mathcal{F}=\{A_i,\,i=1,2,\cdots\}$ of commuting operators.

Here I consider purely deterministic HV. This means that the assignement  $A\to a=f(A,\frak{v})$ is unique, and thus in QM $a$ is one of the eigenvalues of the operator $A$. 

The term ``context free'' means that the outcome $a$ for the measurement of the first  observable $A$  is supposed to be independent of the choice of the second observable  $B$. 
In other word, the outcome of a measurement depends on the hidden variable, but not of the ``context'' of the measurement, that is of the other  quantities measured at the same time.

We shall discuss the possibility that $a$ is a random variable (with a law fixed by $\mathfrak{v}$) later.

\subsection{Gleason's theorem and contextuality}
\label{ssGTCont}
\index{Gleason's theorem}
\index{Contextuality}
These kind of models seem much more realistic. 
However, they are immediatly excluded by  Gleason's theorem \cite{Gleason57}, as already argued by J. Bell in \cite{RevModPhys.38.447}.

Indeed, if to any $\mathfrak{v}$ is associated a function $f_\mathfrak{v}$, defined as
\begin{equation}
\label{ }
f_\mathfrak{v}\quad;\qquad A\to f_\mathfrak{v}(A)=f(A,\mathfrak{v})
\end{equation}
which satisfy  the consistency conditions \ref{CHVNCon}, this is true in particular for any family of commuting projectors $\{P_i\}$, whose outcome in $0$ or $1$
\begin{equation}
\label{ }
P\ \text{projector such that}\ \ P=P^\dagger=P^2\quad\implies\quad f_\mathfrak{v}(P)=0\ \text{or}\ 1
\end{equation}
In particular, this is true for the family of projectors $\{P_i\}$ onto the vector of any orthonormal basis $\{\vec e_i\}$ of the Hilbert space $\mathcal{H}$ of the system. This means simply that defining the function $f$ on the unit vectors $\vec e$ by 
\begin{equation}
\label{eNCdef}
f(\vec e)=f_\mathfrak{v}(P_{\vec e})\quad,\qquad P_{\vec e}=|\vec e\rangle\langle \vec e|
\end{equation}
(remember that $\mathfrak{v}$ is considered fixed), this function must satisfy for any orthonormal basis 
\begin{equation}
\label{eNCframe}
\{\vec e_i\}\quad\text{orthonormal basis}\qquad\implies\qquad \sum_i f(\vec e_i)=1
\end{equation}
while we have for any unit vector
\begin{equation}
\label{eNCproj}
f(\vec e)\ =\ 0\ \text{or}\ 1
\end{equation}
This contradicts strongly Gleason's theorem (see \ref{ssGleason}), as soon as the Hilbert space of the system $\mathcal{H}$ has dimension $\text{dim}(\mathcal{H}) \ge 3$! 
Indeed, \ref{eNCframe} means that the function $f$ is a frame function (in the sense of Gleason), hence is continuous, while $\ref{eNCproj}$ (following from the fact that $f$ is function on the projectors) means that $f$ cannot be a continuous function. So
\begin{equation}
\label{ }
\text{dim}(\mathcal{H}) \ge 3\quad\implies\quad\text{no context-free HV can describe  all the quantum correlations}
\end{equation}
Gleason's theorem is a very serious blow to the HV idea. However, some remaining possibilities can be considered, for instance:
\begin{enumerate}
 \item There are still context-free HV, but they describe only some specific subset of the quantum correlations, not all of them.
  \item There are HV, but they are fully contextual.
\end{enumerate} 
We now discuss two famous cases where the first  option has been explored, but appears to be still problematic.
The second one raises also big questions, that will be shortly discussed in \ref{ssHVdisc}.
 
\subsection{The Kochen-Specker theorem}
\index{Kochen-Specker theorem}
The first option is related to the idea that some subset of the correlations of a quantum system have a special status, being related to some special explicit ``elements of reality'' (the ``be-ables'' in the terminology of J. Bell), by contrast to the ordinary observables which are just ``observ-ables''. Thus a question is whether for a given quantum system there are  finite families of non commuting observables which can be associated to context-free HV.

In fact the problems with non-contextual HV have been shown to arise already for very small such subsets of observables, first by S. Kochen and E. Specker \cite{KochenSpecker67}. These issues started to be discussed by J. Bell in
\cite{RevModPhys.38.447}. 
This is the content of the  Kochen-Specker theorem. This theorem provides in fact examples of finite families of unit vectors $\mathcal{E}=\{\vec e_i\}$ in a Hilbert Space $\mathcal{H}$ (over $\mathbb{R}$ or $\mathbb{C}$) of finite dimension ($\mathrm{dim}(\mathcal{H})=n$), such that it is impossible to find any  frame function such that
\begin{equation}
\label{ }
f(\vec e_i)=0\ \text{or}\ 1
\quad\text{and}\quad
(\vec e_{i_1},\cdots, \vec  e_{i_n})\quad\text{orthonormal basis}\quad\implies\quad \sum_{a=1}^n f(\vec e_{i_a})=1
\end{equation}
The original example of \cite{KochenSpecker67} involved a set with 117 projectors in a 3 dimensional Hilbert space and is a very nice example of non-trivial geometry calculation.  Simpler examples in dimension $n=3$ and $n=4$ with less projectors have been provided  by several authors (Mermin, Babello, Peres, Penrose).

I do not discuss more these examples and their significance. But this shows that the non--contextual character of quantum correlations is a fundamental feature of quantum mechanics.

\subsection{Bell / CHSH inequalities and non-locality}
Another important situation where non-contextuality is explored, in relation with locality, is found in the famous 1964 paper by J. Bell 
\cite{Bell:1964kc}. 
\index{Bell inequality}
\index{CHSH inequality}
Consider a bipartite system $\mathcal{S}$ consisting of two causally separated subsystems $\mathcal{S}_1$ and $\mathcal{S}_2$, for instance a pair of time-like separated photons in a  Bell-like experiment. One is interested in the correlations between the measurements that are performed independently on $\mathcal{S}_1$ and $\mathcal{S}_2$. 
Any pair of corresponding observables $A$ and $B$ (or more exactly $A\otimes 1$ and $1\otimes B$) commute, and thus one expect that the result of a measurement on $\mathcal{S}_1$, if it depends on some HV, should not depend on the measurement made on $\mathcal{S}_2$. In other word, the result of a measurement on $\mathcal{S}_1$ should not depend on the context of $\mathcal{S}_2$. The reciprocal statement being true as well. 

Thus, following Bell, let us assume that some HV's underlie the bipartite system $\mathcal{S}$, and that it is local in the sense that it is $\mathcal{S}_1$-versus-$\mathcal{S}_2$ context free. But it may not -- and in fact  it cannot -- be context-free with respect to $\mathcal{S}_1$ or $\mathcal{S}_2$ only. This means that a given HV $\mathfrak{v}$ should determine separately the relation \emph{observable $A\to$ outcome $a$}  for  $\mathcal{S}_1$  and  \emph{observable $B\to$ outcome $b$} for $\mathcal{S}_2$. In other word, such a hidden variable assigns a pair of probability distributions for all the observables relative to $\mathcal{S}_1$ and $\mathcal{S}_2$
\begin{equation}
\label{locHV}
\mathfrak{v}\quad
\mapsto\quad (p_1(a|A),p_2(b|B))
\end{equation}
The function $p_1(a|A)$ give the probability for the outcome $a$ when measuring $A$ on $\mathcal{S}_1$, the function $p_1(a|A)$  the probability for the outcome $b$ when measuring $B$ on $\mathcal{S}_2$.

One may assume that these probabilities can be decomposed into subprobabilities associated to local hidden variables $\mathfrak{w}_1$ and $\mathfrak{w}_2$  for the two subsystems $\mathcal{S}_1$ and $\mathcal{S}_2$. In this case $\mathfrak{v}$ is itself a pair of probability distribution $(q_1,q_2)$ over the $\mathfrak{w}_1$'s and $\mathfrak{w}_2$'s respectively.
\begin{equation}
\label{ }
\mathfrak{v}=(q_1,q_2)
\quad,\qquad q_1:\ \mathfrak{w}_1\to q_1(\mathfrak{w}_1)\ ,\ \ q_2:\ \mathfrak{w}_2\to q_1(\mathfrak{w}_2)
\end{equation} 
while it is the HV $\mathfrak{w}_1$ (respectively $\mathfrak{w}_2$) that determines the outcome $A\to a$ (respectively $B\to b$). 
These HV,s have to be contextual if one wants the relations $A\to a$ and $B\to b$  to be consistent with quantum mechanics for the two subsystems.

But one may also take the probability distributions $p_1(a|A)$ and $p_2(b|B)$ to be fully quantum mechanical, thus corresponding, using Gleason's theorem, to some density matrices $\rho_1$ and $\rho_2$ 
\begin{equation}
\label{ }
\mathfrak{v}=(\rho_1,\rho_2)
\end{equation} 
such that 
$p_1(a|A)=\tr(\delta(a-A)\rho_1)$ and $p_2(b|B)=\tr(\delta(b-B)\rho_2)$.

In any case,  hidden variables of the form \ref{locHV} are denoted ``local hidden variables''.
One might perhaps rather call them 
``locally-contextual-only hidden variables" 
but let us keep the standard denomination.

A quantum state $\psi$ of $\mathcal{S}$ corresponds to some probability distribution $q(\mathfrak{v})$ over the HV's $\mathfrak{v}$.
$q(\mathfrak{v})$ represent our ignorance about the ``elements of reality''  of the system.
If this description is correct, the probability for the pair of outcomes $(A,B)\to (a,b)$ in the state $\psi$ is given by the famous representation
\begin{equation}
\label{probABloc}
p(a,b|A,B)=\sum_{\mathfrak{v}} q(\mathfrak{v})\,p_1(a|A)\,p_2(b|B)
\end{equation}
It is this peculiar form which implies the famous Bell and BHSH inequalities on the correlations between observables on the two causally independent subsystems. Let us repeat the argument for the CHSH inequality.
If we consider for observables for $\mathcal{S}_1$ (respectively   $\mathcal{S}_2$) two (not necessarily commuting) projectors $P_1$ and $P'_1$ (respectively $Q_2$ and $Q'_2$), with outcome $0$ or $1$, and redefine them as
\begin{equation}
\label{ }
A=2 P_1-1\,\quad A'=2 P'_1-1\,\quad B=2 Q_1-1\,\quad B'=2 Q'_1-1\,\quad
\end{equation}
so that the outcomes are $-1$ or $1$,
if one perform a series of experiments on an ensemble of independently prepared instances of the bipartite system $\mathcal{S}$, choosing randomly with equal probabilities to measure $(A,B)$, $(A',B)$, $(A,B')$ or $(A',B')$, and combine the results to compute the average 
\begin{equation}
\label{ }
\langle M\rangle = \langle AB\rangle -\langle A B' \rangle +\langle  A' B\rangle  +\langle  A' B'\rangle 
\end{equation}
the same argument than in \ref{ssTsiBn}, using the general inequality
\begin{equation}
\label{ }
a,a',b,b'\in[-1,1] \quad\implies\quad a(b-b')+a'(b+b')\in[-2,2]
\end{equation}
implies the CSHS inequality
\begin{equation}
\label{CSHSineq}
-2\le \langle M\rangle\le  2
\end{equation}
\index{Bell inequality}
\index{CHSH inequality}
This inequality is known to be violated for some quantum states (entangled states) and some choice of observables. Indeed $\langle M\rangle$ may saturate the Tsirelon's bound $|\langle M\rangle |\le 2\sqrt{2}$.
\index{Tsirelson bound}
The reason is simple. Assuming that all quantum states give probabilities of the form  \ref{probABloc} and that the probabilities $p_1(a|A)$ and $p_2(b|B)$ obey the quantum rules and are representable by density matrices means that any quantum state (mixed or pure) $\psi $ can be represented by a density matrix of the form
\begin{equation}
\label{ }
\rho=\sum_{\mathfrak{v}} q(\mathfrak{v})\  \rho_1(\mathfrak{v}) \otimes \rho_2(\mathfrak{v})
\end{equation}
Such states are called separable states.
\index{Separable state}
\index{Non-locality}
But not all states are  separable. 
 For a bipartite system, 
this is the case indeed for pure  entangled states.

I do not discuss the many and very interesting generalizations and variants of Bell inequalities (for instance the spectacular GHZ example for tripartite systems) and the possible consequences and tests of non-contextuality.

I do not review either all the experimental tests of violations of Bell-like inequalities in various contexts, starting from the first experiments by Clauser, and those by Aspect et al., up to the most recent ones. They are in full agreement with the predictions of standard Quantum Mechanics and more precisely of Quantum Electro Dynamics. See for instance \cite{Laloe-book} or a recent and very complete review. 

\subsection{Discussion}
\label{ssHVdisc}
The significance and consequences of Bell and CHSH inequalitys and of the Kochen-Specker theorem have been enormously discussed, and some debates are still going on. To review and summarize these discussions is not the purpose of these notes. 
Let me just try to make some simple remarks.

The assumption of context-free value definiteness is clearly not tenable, from Gleason's theorem. This means that one must be very careful when discussing quantum physics about correlations between results of measurements. To quote a famous statement by Peres: ``Unperformed experiments have no results''  \cite{Peres78}.

Trying to assign some special ontological status to a (finite and in practice small) number of observables to avoid the consequence of the Kochen-Specker theorem may be envisioned, but raises other problems. 
For instance, if one wants to keep the main axioms of QM, and non-contextuality, by using a finite number of observables, one would expect the quantum logic formalism would lead to QM on a finite division ring (a Galois field), but it is known that this is not possible (see the discussion in \ref{sssRing}).
Note however that relaxing some basic physical assumptions like reversibility and unitarity has been considered for instance in \cite{tHooft2007}.

It is also clear that non-local quantum correlations are present in non-separable quantum states, highlighted by the violations of Bell's and CHSH-like inequalities (and their numerous and interesting variants). They represent some of the most non-classical and counter-intuitive features of quantum physics.
In connexion with the discussions of the ``EPR-paradox'', this non-local aspect of quantum physics has been often -- and is still sometimes --presented as  a contradiction between the principles of quantum mechanics and those of special relativity.
This is of course not the case.
These issues must be discussed in the framework of relativistic quantum field theory, where the basic objects are quantum fields, not (first quantized) particles (or classical fields).
See the section \ref{ssAQFTdsh}.
In this formalism a quantum state of a field is (some kind of) wave function over fields configurations over the entire space, and is intrinsically a non-local and non-separable object. 
The physical requirements of causality, and locality, implying no faster-than-light signaling (or any kind of real ``spooky-at-a-distance action''), are requirements on the observables, i.e. on the self-adjoint operators of the theory.

\index{Contextuality}
Finally, the option (2) at the end of \ref{ssGTCont} -- \textit{There are hidden variables but that they are fully contextual} -- is also very problematic and raises more questions than solutions (in my opinion). 
For instance, I would expect that even assuming non-contextual value definiteness, the sum and product relations \ref{CHVNCon} should still holds for commuting observables with fully non-degenerate spectrum. Then a problem of definiteness arises when considering a projector as a limit of such observables (in some sense the result of a measurement should depend not only on all the measurements you can perform, but on those you will not perform).
Another problem is that contextuality leads to consider that there are non-local hidden correlation between the system and the measurement apparatus before any measurement, which in some sense pushes the problem one rug further without really solving it.
Nevertheless, contextuality has been considered by several authors in connexion with some interpretations of quantum mechanics like the so called ``modal interpretations''. I am however unable to discuss this further.

\medskip
To summarize the discussions of these last two sections \ref{sQuCor} and \ref{sHidVar}: 
Contrary to classical physics, there is an irreducible quantum uncertainty in the description of any quantum system. Not all its physical observable can be characterized at the same time. This is of course the uncertainty principle.
Contrary to a simple reasoning, this does not mean that a quantum system is always more uncertain or ``fuzzy''  than a classical system. Indeed, the quantum correlations are stronger than the classical correlations, as exemplified by the quantum entropic inequalities \ref{subaddS} and 
\ref{StrSubAdd1}, and the Tsirelson bound \ref{TsirBound} compared to their classical analog, the entropic bound \ref{CaddS} and the B-CHSH inequality \ref{CSHSineq}.
This can be represented by the little drawing of Fig.~\ref{QYantra}.
This is why the results by J. Bell and the subsequent ones turned out to have a long term impact. They contributed to the realization of what is not quantum mechanics, and to the rise of quantum information: using quantum correlations and entanglement, it is possible to  transmit and manipulate information, perform calculations, etc.  in ways which are impossible by classical means, and which are much more efficient.

\begin{figure}[h]
\begin{center}
\includegraphics[width=2in]{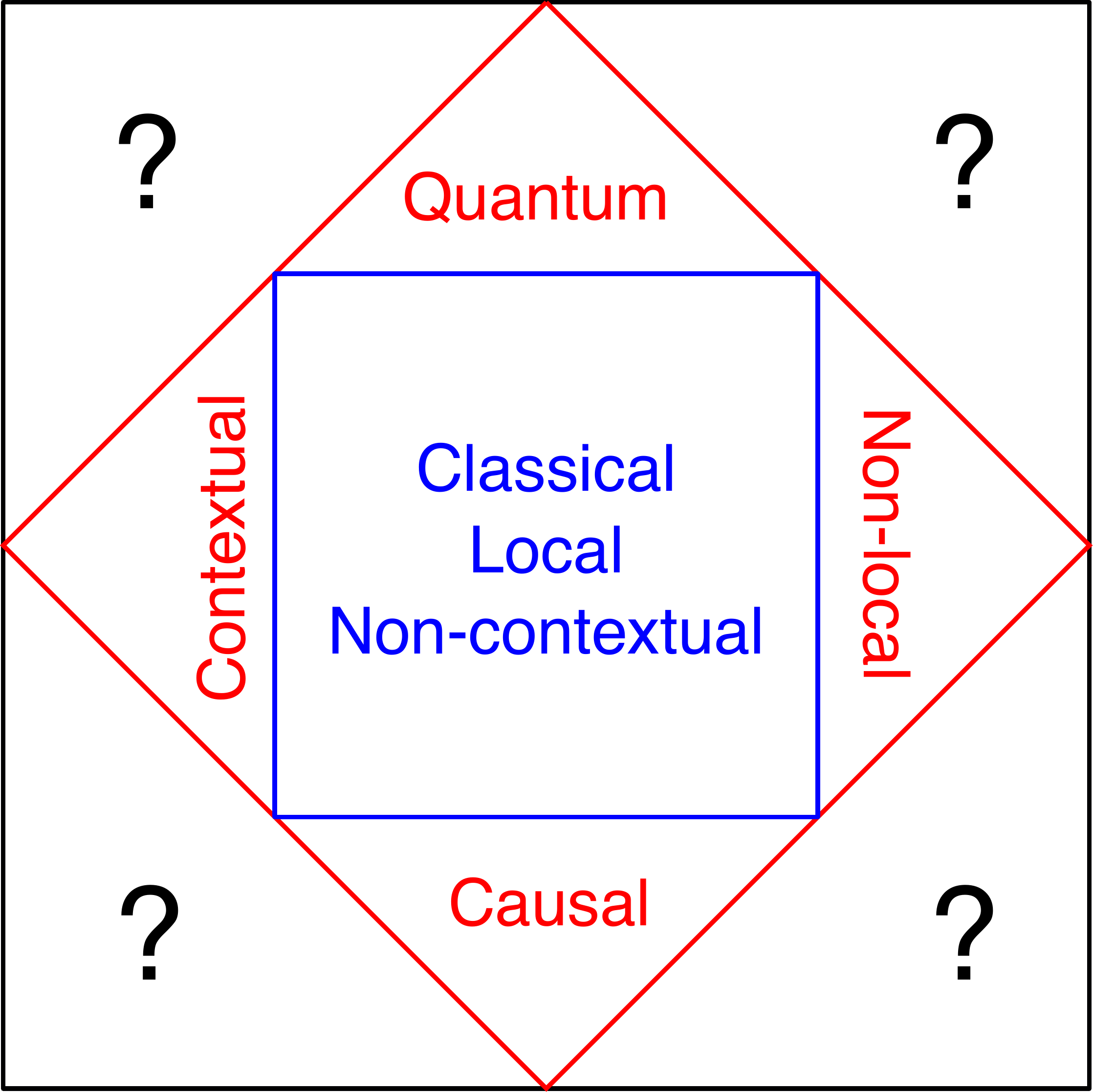}
\caption{Schematic of the worlds of classical correlations, quantum correlations  and ``super-strong'' unphysical correlations}
\label{QYantra}
\end{center}
\end{figure}

\section{Measurements}
\label{sMeasure}
\subsection{What are the questions?}
\index{Measurement}
Up to now I have not  discussed much the question of quantum measurements.
I simply took the standard point of view that  (at least in principle) ideal projective measurements are feasible and one should look at the properties of the outcomes. 
The question is of course highly more complex. In this section I just recall some basic points about quantum measurements.

The meaning of the measurement operations is at the core of quantum physics. It was considered as such from the very beginning.
See for instance the proceedings of the famous Solvay 1927 Congress \cite{Solvay27}, and the 1983 review by Wheeler and Zurek \cite{WheelerZurek83}. Many great minds have thought about the so called ``measurement problem''  and the domain has been revived in the last decades by the experimental progresses, which allows now to manipulate simple quantum system and implement effectively ideal measurements.

On one hand, quantum measurements represent one of the most puzzling features of quantum physics. They are non-deterministic processes (quantum mechanics predicts only probabilities for outcomes of measurements). They are irreversible processes (the phenomenon of the ``wave-function collapse''). They reveal the irreducible uncertainty of quantum physics (the uncertainty relations).
This makes quantum measurements very different from ``ideal classical measurements''. 

On the other hand, quantum theory is the first physical theory that addresses seriously the problem of the interactions between the observed system (the object) and the measurement apparatus (the observer).
Indeed in classical physics the observer is considered as a spectator, able to register the state of the real world (hence to have its own state modified by the observation), but without perturbing the observed system in any way. 
Quantum physics shows that this assumption is not tenable. 
Moreover, it seems to provide a logically satisfying answer\footnote{If not satisfying every minds, every  times...} to the basic question: what are the minimal constraints put on the results of physical measurements by the basic physical principles\footnote{Well...  as long as gravity is not taken into account!}.

It is often stated that the main problem about quantum measurement is the problem of the uniqueness of the outcome. 
For instance, why do we observe a spin 1/2 (i.e. a q-bit) in the state $|{\uparrow}\rangle$ or in the state $|{\downarrow}\rangle$ when we start in a superposition $|\psi\rangle=\alpha|{\uparrow}\rangle+\beta |{\downarrow}\rangle$? However by definition a measurement is a process which gives one single classical outcome (out of several possible).
Thus in my opinion the real questions, related to the question of the ``projection postulate'', are: (1) Why do repeated ideal measurements should give always the same answer? 
(2) Why is it not  possible to ``measure'' the full quantum state $|\psi\rangle$ of a q-bit by a \emph{single} measurement operation, but only its projection onto some reference frame axis?

Again, the discussion that follows is very sketchy and superficial. A good recent reference, both on the history of the ``quantum measurement problem'', a detailed study of explicit dynamical models for quantum measurements, and a complete bibliography, is the research and review 
article \cite{AlBaNi2012}. 
\subsection{The von Neumann paradigm}
The general framework to discuss quantum measurements in the context of quantum theory is provided by J. von Neumann
\index{von Neumann J.}  in his 1932 book \cite{vonNeumann32G,vonNeumann32}.
\index{Non-destructive measurement}
\index{Ideal measurement}
Let me present it on the simple example of the q-bit.

But before, let me insist already on the fact that this discussion will not provide a derivation of the principle of quantum mechanics (existence of projective measurements, probabilistic features and Born rule), but rather a self-consistency argument of compatibility between the axioms of QM about measurements and what QM predicts about measurement devices.

An ideal measurement involves the interaction between the quantum system $\mathcal{S}$ (here a q-bit) and a measurement apparatus $\mathcal{M}$ which is a \emph{macroscopic object}.
The idea is that $\mathcal{M}$ must be treated as a quantum object, like $\mathcal{S}$.
An ideal non destructive measurement on $\mathcal{S}$ that does not change the orthogonal states  $|{\uparrow}\rangle$ and  $|{\downarrow}\rangle$ of $\mathcal{S}$ (thus corresponding to a measurement of the spin along the $z$ axis, $S_z$), correspond to introducing for a finite (short) time an interaction between $\mathcal{S}$ and $\mathcal{M}$, and to start from a well chosen initial state $|I\rangle$ for $\mathcal{M}$. The interaction and the dynamics of $\mathcal{M}$ must be such that, if one starts from an initial separable state where $\mathcal{S}$ is in a superposition state
\begin{equation}
\label{UMeasEv}
|\psi\rangle=\alpha\, |{\uparrow}\rangle+\beta\, |{\downarrow}\rangle
\end{equation}
after the measurement (interaction) the whole system (object+apparatus) is in an entangled state
\begin{equation}
\label{ }
|\psi\rangle\otimes |I\rangle\qquad\to\qquad\alpha\,|{\uparrow}\rangle\otimes |F_+\rangle + \beta\, |{\downarrow}\rangle\otimes|F_-\rangle
\end{equation}
The crucial point is that the final states $|F_+\rangle$ and $|F_-\rangle$ for $\mathcal{M}$ must be \emph{orthogonal} 
\footnote{as already pointed out in \cite{vonNeumann32G}   }

\begin{equation}
\label{ }
\langle F_+ | F_-\rangle=0
\end{equation}
Of course this particular evolution \ref{UMeasEv} is unitary for any choice of $|\psi\rangle$, since it transforms a pure state into a pure state.
\begin{equation}
\label{vNDecProj}
|\psi\rangle\otimes |I\rangle
\qquad\to\qquad
\alpha\,|{\uparrow}\rangle\otimes |F_+\rangle + \beta\, |{\downarrow}\rangle\otimes|F_-\rangle
\end{equation}

One can argue that this is sufficient to show that the process has all the characteristic expected from an ideal measurement, within the quantum formalism itself. Indeed, using the Born rule, this is consistent with the fact that the state $\alpha|{\uparrow}\rangle$ is observed with probability $p_+=|\alpha|^2$ and the state $\alpha|{\downarrow}\rangle$ with probability $p_-=|\beta|^2$. 
Indeed the reduced density matriices both for the system $\mathcal{S}$ and for the system $\mathcal{M}$ (projected onto the two pointer states) is that of a completely mixed state
\begin{equation}
\label{ }
\rho_\mathcal{S}=
\begin{pmatrix}
   p_+   &    0\\
   0   &  p_-
\end{pmatrix}
\end{equation}

For instance, as discussed in   \cite{vonNeumann32G}\cite{vonNeumann32}, if one is in the situation where the observer $\mathcal{O}$, really observe the measurement apparatus $\mathcal{M}$, not the system $\mathcal{S}$ directly, the argument can be repeated as
\begin{equation}
\label{ }
|\psi\rangle\otimes |I\rangle\otimes |O\rangle
\qquad\to\qquad
\alpha\, | {\uparrow}\rangle \otimes |F_+\rangle \otimes |O_+\rangle + \beta\,  |{\downarrow}\rangle\otimes|F_-\rangle \otimes |O_-\rangle
\end{equation}
and it does not matter if one puts the fiducial separation between object and observer between $\mathcal{S}$ and $\mathcal{M}+\mathcal{O}$ or between $\mathcal{S}+\mathcal{M}$ and $\mathcal{O}$. This argument being repeated ad infinitum.

A related argument is that once a measurement has been performed, if we repeat it using for instance another copy $\mathcal{M}'$ of the measurement apparatus, after the second measurement we obtain
\begin{equation}
\label{ }
|\psi\rangle\otimes |I\rangle\otimes |  I' \rangle\qquad\to\qquad\alpha\, |{\uparrow}\rangle\otimes |F_+\rangle \otimes |F'_+\rangle+ \beta \, |{\downarrow}\rangle\otimes|F_-\rangle\otimes|F'_-\rangle
\end{equation}
so that we never observe both  $|{\uparrow}\rangle$ and  $|{\downarrow}\rangle$ in a successive series of measurements (hence the measurement is really a projective measurements). 
The arguments holds also if the outcome of the first measurement is stored on some classical memory device $\mathcal{D}$ and the measurement apparatus reinitialized to $|I\rangle$.
This kind of argument can be found already in 
\cite{Mott29}.

The discussion here is clearly outrageously oversimplified and very sketchy. For a  precise discussion, one must distinguish among the degrees of freedom of the measurement apparatus $\mathcal{M}$ the (often still macroscopic) variables which really register the state of the observed system, the so called\emph{ pointer states}, from the other (numerous) microscopic degrees of freedom of $\mathcal{M}$, which are present anyway since $\mathcal{M}$ is a macroscopic object, and which are required both for ensuring decoherence (see next section) and to induce dissipation, so that the pointer states become stable and store in a efficient way the information about the result of the measurement. One must also take into account the coupling of the system $\mathcal{S}$ and of the measurement apparatus $\mathcal{M}$ to the environment $\mathcal{E}$.

\subsection{Decoherence and ergodicity (mixing)}
As already emphasized, the crucial point is that starting from the same initial state $|I\rangle$, the possible final pointer states for the measurement apparatus, $|F_+\rangle$ and $|F_-\rangle$, are orthogonal. 
This is now a well defined dynamical problem, which can be studied using the theory of quantum dynamics for closed and open systems. The fact that $\mathcal{M}$ is macroscopic, i.e. that its Hilbert space of states in very big, is essential, and the crucial concept is  \emph{decoherence} (in a general sense).

\index{Decoherence}
The precise concept and denomination of quantum decoherence was introduced in the 70's (especially by Zeh) and developed and popularized in the 80's (see the reviews
\cite{JoosZehKiefer2003}, \cite{Zurek2003}). But the basic idea seems much older and for our purpose one can probably go back to the end of the 20' and to von Neumann's 
\index{von Neumann J.}
 quantum ergodic theorem \cite{vonNeumann29} (see \cite{springerlink:10.1140/epjh/e2010-00008-5} for the english translation and \cite{GoldLebMasTumZan2010} for historical and physical perspective).
\index{Quantum ergodic theorem}

One starts from the simple geometrical remark \cite{vonNeumann29} that if $| e_1 \rangle$ and $| e_2 \rangle$ are two \emph{random 
unit vectors }in a $N$ dimensional Hilbert space $\mathcal{H}$ (real or complex), their average ``overlap'' (squared scalar product) is of order
\begin{equation}
\label{ }
\overline{|\langle e_1 | e_2 \rangle  |^2}\ \simeq {1\over N}
\quad,\qquad N=\dim (\mathcal{H})
\end{equation}
hence it is very small, and for all practical purpose equal to $0$, if $N$ is very large. Remember that for a quantum system made out of $M$ similar subsystems, $N\propto (N_0)^M$, $N_0$ being the number of accessible quantum states for each subsystem.

A simple idealized model to obtain a dynamics of the form \ref{vNDecProj} for $\mathcal{S}+\mathcal{M}$ is to assume that both $\mathcal{S}$ and $\mathcal{M}$ have no intrinsic dynamics and that the evolution during the interaction/measurement time interval is given by a interaction Hamiltonian (acting on the Hilbert space $\mathcal{H}=\mathcal{H}_\mathcal{S}\otimes\mathcal{H}_\mathcal{M}$ of $\mathcal{S}+\mathcal{M}$)  of  the form
 \begin{equation}
\label{Hdec}
H_\mathrm{int}=  |  {\uparrow}\rangle \langle {\uparrow} |  \otimes  H_+ +  |  {\downarrow}\rangle \langle  {\downarrow} |  \otimes  H_-
\end{equation}
where $H_+$ and $H_-$ are two \emph{different} Hamiltonians (operators) acting on $\mathcal{H}_\mathcal{M}$.
It is clear that if the interaction between $\mathcal{S}$ and $\mathcal{M}$ takes place during a finite time $t$, and is then switched off, the final state of the system is an entangled one of the form \ref{vNDecProj}, with
\begin{equation}
\label{F+F-Exp}
|F_+ \rangle = \emath^{{t\over \imath\hbar} H_+} |I\rangle
\quad,\qquad
|F_- \rangle = \emath^{{t\over \imath\hbar} H_-} | I \rangle
\end{equation}
so that
\begin{equation}
\label{F+F-Overlap}
\langle F_+|F_-\rangle = \langle I | \emath^{-{t\over \imath\hbar} H_+} \cdot\emath^{{t\over \imath\hbar} H_-} |I\rangle
\end{equation}
It is quite easy to see that if $H_+$ and $H_-$ are not (too much) correlated (in a sense that I do not make more precise), the final states $|F_+\rangle$ and $|F_-\rangle$ are quite uncorrelated with respect to each others and with the initial state $|I\rangle$  after a very short time, and may be considered as random states in $\mathcal{H}_\mathcal{M}$, so that 
\begin{equation}
\label{F+F-OSm}
|\langle F_+|F_-\rangle|^2\simeq {1\over\dim(\mathcal{H}_\mathcal{M})}\ll\!\! \ll 1
\end{equation}
so that for all practical purpose, we may assume that
\begin{equation}
\label{F+F-zero}
\langle F_+ | F_-\rangle =0
\end{equation}
This is the basis of the general phenomenon of \emph{decoherence}. 
The interaction between the observed system and the measurement apparatus has induced a decoherence between the states $|\uparrow\rangle$ and $| \downarrow \rangle$ of $\mathcal{S}$, but also a decoherence between the pointer states $| F_+\rangle$ and $| F_-\rangle$ of $\mathcal{M}$.

Moreover, the larger $\dim(\mathcal{H}_\mathcal{M})$, the smaller the ``decoherence time'' beyond which $\langle F_+ | F_-\rangle \simeq 0$ is (and it is often in practice too small to be observable), and the larger (in practice infinitely larger)  the ``quantum Poincaré recurrence  time'' (where one might expect to get again $|\langle F_+ | F_-\rangle| \simeq  1$) is.

Of course, as already mentionned, this is just the first step in the discussion of the dynamics of a quantum measurement. One has in particular to check and to explain how, and under which conditions, the pointer states are quantum microstates which correspond to macroscopic classical-like macrostates, which can be manipulated, observed, stored in an efficient way. 
At that stage, I just paraphrase J. von Neumann (in the famous  chapter VI  ``Der Me\ss proze\ss '' of  \cite{vonNeumann32G})
\begin{center}
\emph{``Die weitere Frage (...) soll uns dagegen nicht beschäftigen.''}
\end{center}

Decoherence is a typical quantum phenomenon. It explains how, in most situations and systems, quantum correlations in small (or big) multipartite systems are ``washed out" and disappear through the interaction of the system with other systems, with its environment or its microscopic  internal degrees of freedom. 
Standard references on decoherence and the general problem of the quantum to classical transitions are  \cite{Zurek:1990fk} and\cite{Schlosshauer07}.

However, the underlying mechanism for decoherence has a well know classical analog: it is the (quite generic) phenomenon of \emph{ergodicity}, or more precisely the \emph{mixing property} of classical dynamical systems.
\index{Ergodicity}
I refer to textbooks such as \cite{ArnoldAvez-book68} and \cite{LichLieb-book92} for precise mathematical definitions, proofs and details. 
Again I give here an oversimplified presentation.

Let us consider a classical Hamiltonian system. One considers its dynamics on (a fixed energy slice $H=E$ of) the phase space $\Omega$ , assumed to have a finite volume $V=\mu(\Omega)$ normalized to $V=1$,  where $\mu$ is the Liouville measure. 
We denote $T$ the volume preserving  map $\Omega\to\Omega$ corresponding to the integration of the Hamiltonian flow during some reference time $t_0$. $T^k$ is the iterated map (evolution during time $t=k t_0$). This discrete time dynamical mapping given by $T$ is said to have the \emph{weak mixing property} if for any two (measurable) subsets $A$ and $B$ of $\Omega$ one has
\index{Mixing property}
\begin{equation}
\label{ }
\lim_{n\to\infty} \  {1\over n}\sum_{k=0}^{n-1} \mu(B\cap T^k A)  \ =\ \mu(B)\mu(A)
\end{equation}
The (weak) mixing properties means (roughly speaking)  that, if we take a random point $a$  in phase space, its iterations $a_k=T^k a$ are at large time and ``on the average'' uniformly distributed on phase space, with a probability $\mu(B)/\mu(\Omega)$ to be contained inside any subset $B\in\Omega$.
See fig.~\ref{fWMixing}
\begin{figure}[h]
\begin{center}
\includegraphics[height=2.in]{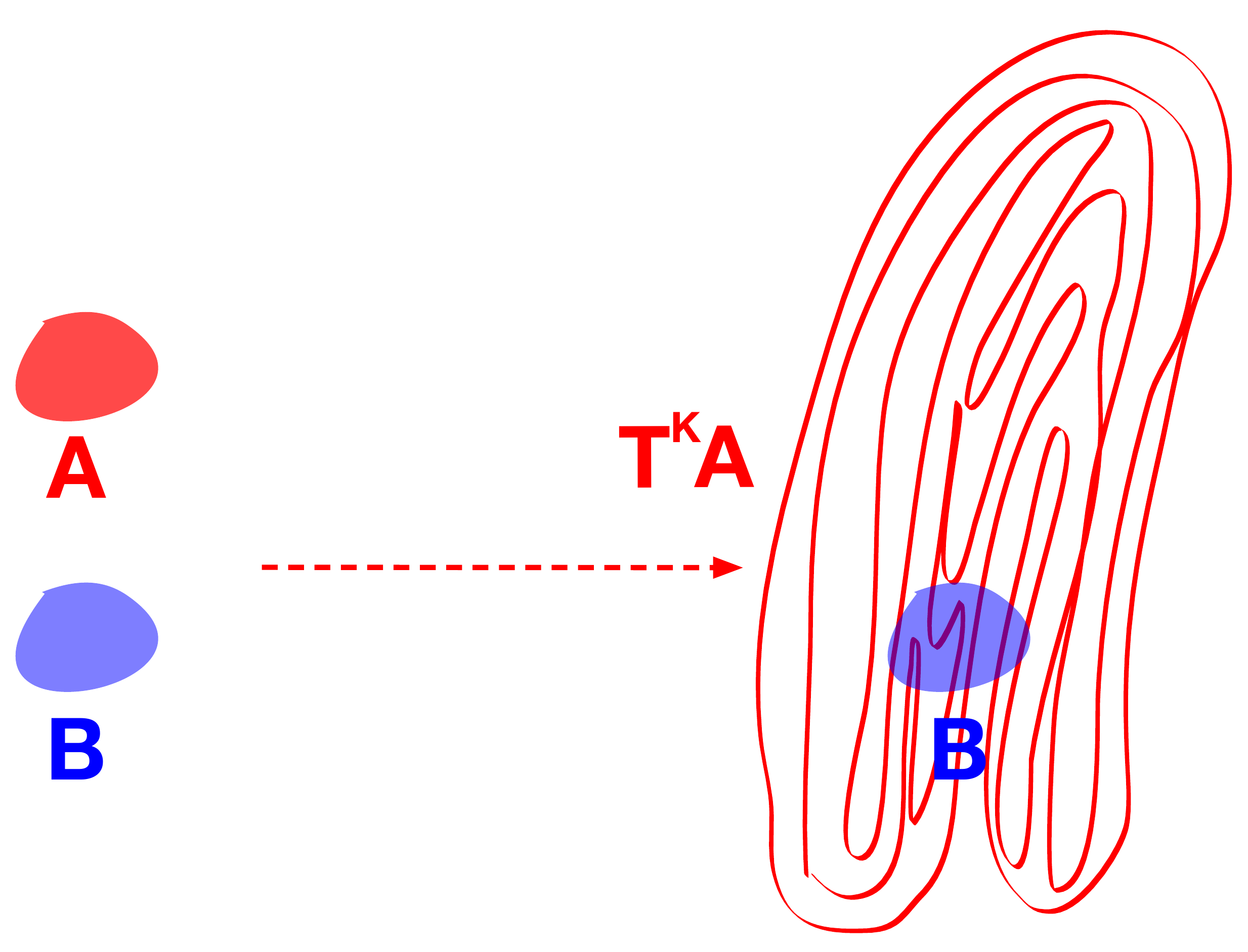}
\caption{Graphical representation of the mixing property (very crude)}
\label{fWMixing}
\end{center}
\end{figure}

Weak mixing is one of the weakest form of ``ergodicity'' (in a loose sense, there is a precise mathematical concept of ergodicity).

Now in semiclassical quantization (for instance using Bohr-Sommerfeld quantization rules)  if a classical system has $M$ independent degrees of freedom (hence its classical phase space $\Omega$ has dimension $2M$), the ``quantum element of phase space'' $\delta\Omega$ has volume $\delta V=\mu(\delta\Omega)= h^{M}$ with $h=2\pi\hbar$ the Planck's constant.
If the phase space is compact with volume $\mu(\Omega)<\infty$ the number of ``independent quantum states'' accessible to the system is of order $N=\mu(\Omega)/\mu(\delta\Omega)$ and should correspond to the dimension of the Hilbert space $N=\dim(\mathcal{H})$.
In this crude semiclassical picture, if we consider two pure quantum states $|a\rangle$ and $|b\rangle$ and associate to them two minimal semiclassical subsets $A$ and $B$ of the semiclassical phase space $\Omega$, of quantum volume $\delta V$, the semiclassical volume $\mu(A\cap B)$ corresponds to the overlap between the two quantum pure states through
\begin{equation}
\label{ }
\mu(A\cap B)\  \simeq \ {1 \over N} | \langle a |b\rangle |^2
\end{equation}
More generally if we associate to any (non minimal) subset $A$ of $\Omega$ a mixed state given by a quantum density matrix $\rho_A$ we have the semiclassical correspondence
\begin{equation}
\label{ }
{\mu(A\cap B)\over \mu(A)\mu(B)}\  \simeq\  {N}\   \tr(\rho_A\, \rho_B)
\end{equation}
With this semiclassical picture in mind (Warning! It does not work for all states, only for states which have a semiclassical interpretation! But pointer states usually do.) the measurement/interaction process discussed above has a simple semiclassical interpretation, illustrated on fig.~\ref{fFFMixing}. 

\begin{figure}[h]
\begin{center}
\includegraphics[height=2in]{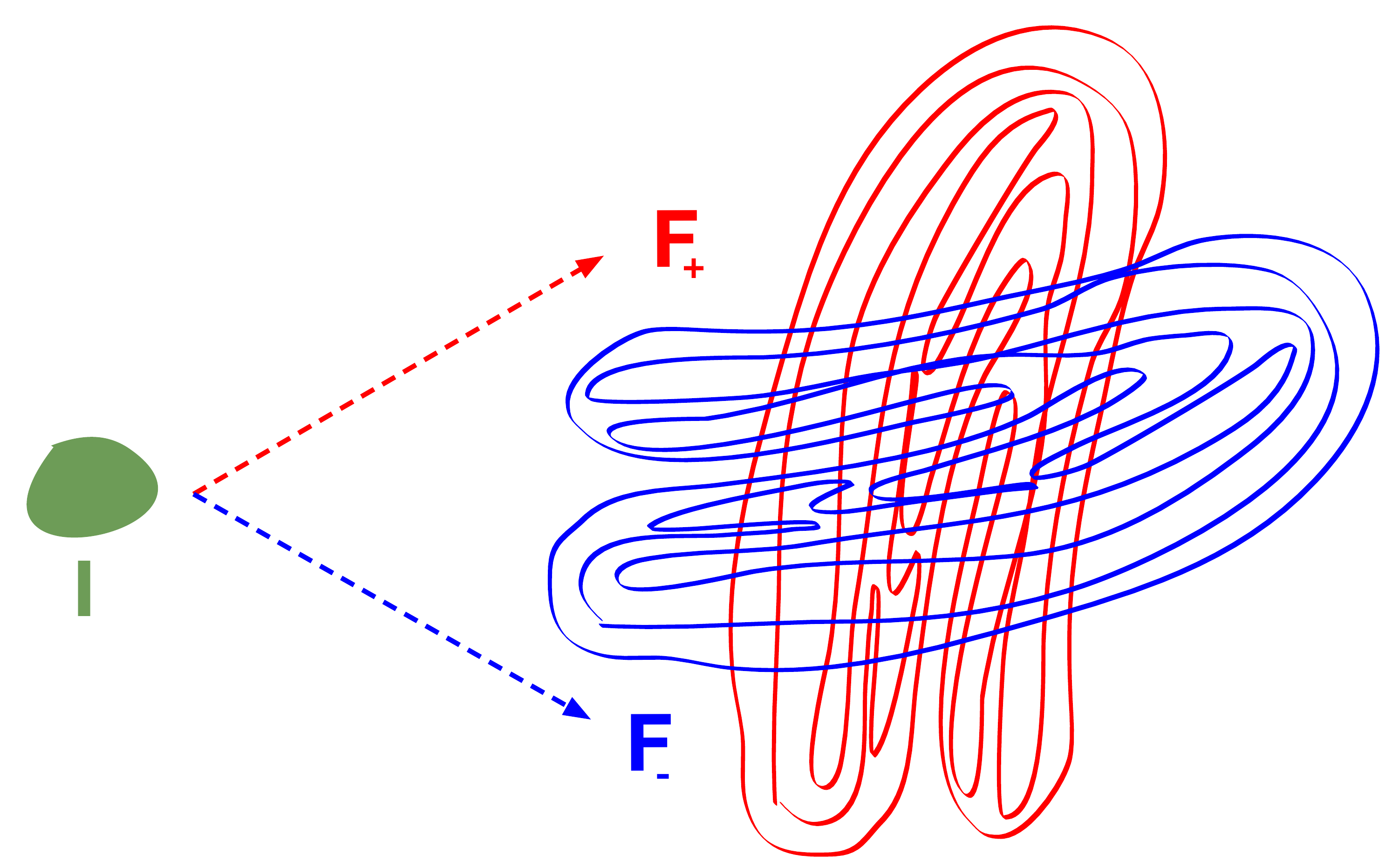}\qquad\qquad\includegraphics[height=2in]{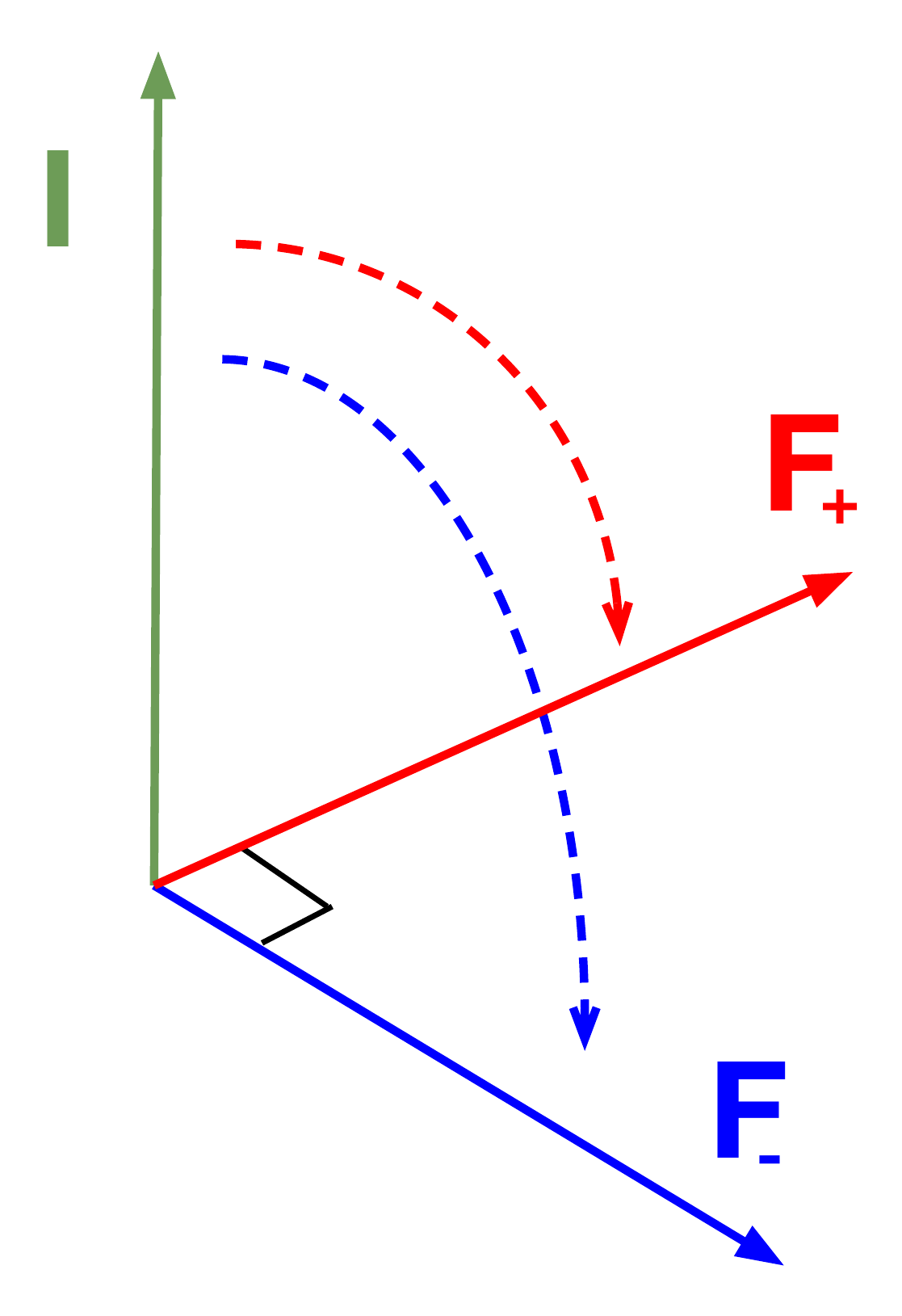}
\caption{Crude semiclassical and quantum pictures of the decoherence process \ref{F+F-Exp}-\ref{F+F-OSm}}
\label{fFFMixing}
\end{center}
\end{figure}

The big system $\mathcal{M}$ starts from an initial state $|I\rangle$  described by a semiclassical element $I$. 
If the system $\mathcal{S}$ is in the state $|\uparrow\rangle$, $\mathcal{M}$ evolves to a state $| F_+\rangle$ corresponding to $F_+$.
If it is in the state $|\uparrow\rangle$, $\mathcal{M}$ evolves to a state $| F_-\rangle$ corresponding to $F_-$.
For well chosen, but quite generic Hamiltonians $H_+$ and $H_-$, the dynamics is mixing, so that, while $\mu(F_+)=\mu(F_-)=1/N$, typically one has $\mu(F_+\cap F_-)=\mu(F_+) \mu(F_-) = 1/N^2\ll 1/N$.
Thus it is enough for the quantum dynamics generated by $H_+$ and $H_-$ to have a quantum analog the classical property of mixing, which is quite generic, to ``explain'' why the two final states $|F_+\rangle$ and $|F_-\rangle$ are generically (almost) orthogonal.

\subsection{Discussion}
As already stated, the points that I tried to discuss in this section represent only a small subset of the questions about measurements in quantum mechanics. Again, I refer for instance to \cite{AlBaNi2012} and \cite{Laloe-book} (among many other reviews) for a serious discussion and bibliography.

I have not discussed more realistic measurement processes, in particular the so called ``indirect measurements procedures'', or ``weak measurements'', where the observations on the system are performed through successive interactions with other quantum systems (the probes) which are so devised as to perturb as less as possible the observed system, followed by stronger direct (in general destructive) measurements of the state of the probes.
Such measurement processes, as well as many interesting questions and experiments in quantum physics, quantum information sciences, etc. are described by the general formalism of POVM's (Positive Operator Valued Measure). I do not discuss these questions here.
\index{Indirect measurements}
\index{POVM}

In any case,  important aspects of quantum measurements belong to the general class of problems of the emergence and the meaning of irreversibility out of reversible microscopic laws in physics (quantum as well as classical). See for instance 
\cite{Halliwell96}.

The quantum formalism as it is presented in these lectures starts (amongst other things) from explicit assumptions on the properties of measurements. The best one can hope is to show that the quantum formalism is consistent: the characteristics and behavior of (highly idealized) physical measurement devices, constructed and operated according to the laws of quantum mechanics, should be consistent with the initials axioms. 

One must be careful however, when trying to justify or rederive some of the axioms of quantum mechanics from the behavior of measurement apparatus and their interactions with the observed system and the rest of the world, not to make circular reasoning.

\section{Interpretations versus Alternative Theories }

In these notes I have been careful not to discuss the interpretation issues of quantum mechanics.
There are at least two reasons.
\begin{enumerate}
  \item These notes are focused on the mathematical formalism of ``standard quantum mechanics". Thus I adopt the ``operational'' point of view
\footnote{This is probably the point of view adopted by most physicists, chemists, mathematicians, computer scientists, engineers, ...  who deal with the quantum world.}
that quantum mechanics is a theoretical framework which provides rules to compute the probabilities to obtain a given result when measuring some observable of a system in a given state.  The concepts of ``observables'', ``states'' and ``probabilities'' being defined through the principles (axioms in a non-mathematical sense) of the formalisms considered.
  \item I do not feel qualified enough to discuss  all the interpretations that have been proposed  and all the philosophical questions raised by quantum physics since its birth. This does not mean that these question are unimportant.
 \end{enumerate}
 However, let me just make a few simple, and probably naïve,  remarks.

Many interpretations of quantum mechanics do not challenge the present standard mathematical formulations of the theory.
They rather insist on a particular point of view or a particular formulation of quantum mechanics as the best suited or the preferable one to consider and study quantum systems, and the quantum world.
They may be considered as particular choices
 \footnote{This does not mean that I am an adept of some post-modern relativism...}
 of  point of view and of  philosophical option to think about quantum mechanics and practice it.

 This is clearly the case for the so-called Copenhagen interpretations. 
 \index{Copenhagen interpretation}
 They insist on the fact that QM deals only with predictions for results of operations, and they can be considered as ``quantum mechanics from a strong pragmatist 
 \footnote{In the philosophical sense of pragmatism}
 point of view".
Remember however that there is no clear cut definition of what a Copenhagen interpretation is. The term was introduced only in 1955 by Heisenberg. I refer to the paper by Howard \cite{Howard2004} for an historical and critical review of the history, uses and misuses of the concept.
\index{Many world interpretation}
 This is also  the case for the ``many worlds interpretations'', that tries to take seriously the concept of ``wave function of the universe''. 
They can be considered (when used reasonably for physics) as the other extreme of ``quantum mechanics from a strong realist 
 \footnote{In the philosophical sense of realism}
 point of view".
 Again there are many variants of these kind of interpretations. I refer to \cite{MWDeWittGraham73} for  the original papers, and to \cite{MW2010} for a recent  presentation of the subject and contradictory discussions.
 
 There is a whole spectrum of  proposed interpretations, for instance the ``coherent history formulations" and the ``modal interpretations'' . I do not  discuss these interpretations here.
 
 The interpretations that rely on the mathematical formulations of quantum mechanics should be clearly distinguished
 \footnote{This is unfortunately not always the case in popular -- and even in some advanced -- presentations and discussions of quantum physics.}
  from another class of  proposals to explain quantum physics that rely on modifications of the rules and are different physical theories. 
These modified or alternative quantum theories  deviate from ``standard'' quantum mechanics and should be experimentally falsifiable (and sometimes are already falsified).

\index{Hidden variables}
This is the case of the various non-local hidden-variables proposals, such as the de Broglie-Bohm theory, which contain some variables (degrees of freedom) which do not obey the laws of QM, and which cannot be observed directly. One might think that they are not falsifiable, but remember that there are serious problems from contextuality, which means that in general, if one want to keep non-contextuality, not all physical (i.e. that can be measured) observables are expected to behave as QM predicts.

\index{Collapse models}
This is also the case for the class of models known as ``collapse models''. 
See \cite{GhRiWe1985,GhRiWe1986} for the first models.
In these models the quantum dynamics is modified (for instance by non-linear terms) so that the evolution of the wave functions is not unitary any more (while the probabilities are conserved of course), and the ``collapse of the wave function'' is a dynamical phenomenon. 
 These models are somehow phenomenological and of course not (yet?) fully  internally consistent, since the origin of these non linear dynamics is quite ad hoc. 
 They predict a breakdown of the law of QM for the evolution of quantum coherences and decoherence phenomenon at large times, large distances, or in particular for big quantum systems (for instance large molecules or atomic clusters).  
At the present day, despite the impressive experimental progresses in the control of quantum coherences, quantum measurements, study of decoherence phenomenon, manipulation of information in quantum systems, etc. , no such violations of the predictions of standard QM and of unitary dynamics have been observed.
 
\section{What about gravity?}
Another really big subject that I do not discuss in these lecture is quantum gravity. Again just a few trivial remarks.
\index{Quantum gravity}

It is clear that the principles of quantum mechanics are challenged by the question of quantizing gravity. 
The challenges are not only technical. General relativity (GR) is indeed a non-renormalizable theory, and from that point of view a first and natural idea is to consider it as an effective low energy theory. After all, in the development of nuclear and particle physics (in the 30', the 40', the 60'...)  there have been several theoretical false alerts and clashes between experimental discoveries and the theoretical understanding that led many great minds to question the principles of quantum mechanics. However QM came out unscathed and even stronger, and since the 70' its principles are not challenged any more.

However with gravity the situation is different.  For instance the discovery of the Bekenstein-Hawking entropy of black holes, of the Hawking radiation, and of the ``information paradox'' shows that  fundamental questions remain to be understood about the relation between  quantum mechanics and the GR concepts of space and time. Indeed even the most advanced quantum theories available, quantum field theories such as non-abelian gauge theories
the standard model, its supersymmetric and/or grand unified extensions, still rely on the special relativity concept of space-time, or to some extend to the dynamical but still classical concept of curved space-time of GR.
It is clear that a quantum theory of space time will deeply modify, and even abolish, the classical concept of space-time as we are used to.
One should note two things. 

Firstly, the presently most advanced attempts to build a quantum theory incorporating gravity, namely string theory and its modern extensions, as well as the alternative approaches to build a quantum theory of space-time such as loop quantum gravity (LQG) and spin-foam models (SF), rely mostly on the quantum formalism as we know it, but change the fundamental degrees of freedom (drastically and quite widely for string theories, in a more conservative way for LQG/SF).
The fact that  string theories offers some serious hints of solutions of the information paradox, and some explicit solutions and ideas, like holography and AdS/CFT dualities, for viewing space-time as emergent, is a very encouraging fact.

Secondly, in the two formalisms presented here, the algebraic formalism and the quantum logic formulations, it should be noted that space and time (as  continuous entities)  play a secondary role with respect to the concept of causality and locality/separability. I hope this is clear in the way I choose to present the algebraic formalism in section \ref{s:AlgQM} and quantum logic in section \ref{s:QuanLog}. 
Of course space and time are essential for constructing physical theories out of the formalism.
Nevetheless, the fact that it is causal relations and causal independence between physical measurement operations that are essential for the formulation of the theory is also a very encouraging fact.

Nevertheless, if for instance the information paradox is not solved by a quantum theory of gravity, or if the concepts of causality and separability have to be rejected (for instance if no repeatable measurements are possible, and if no two sub-systems/sub-ensembles-of-degrees-of-freedom can be considered as really separated/independent), then one might expect that the basic principles of quantum mechanics will not survive (and, according to the common lore, should  be replaced by something even more bizarre and inexplicable...).

Well! It is time to end this bar room discussion.

\newcommand{\etalchar}[1]{$^{#1}$}

%
%

\printindex
\end{document}